%% file: Clean.tex
\documentclass[a4paper,fleqn]{cas-sc}

\usepackage[square,numbers]{natbib}
\usepackage{csquotes}
\usepackage{xcolor}
\usepackage{tablefootnote}
\usepackage[colorinlistoftodos,prependcaption]{todonotes}
\usepackage[inline]{enumitem}
\usepackage{xspace}
\usepackage{hyperref}
\usepackage[nolist]{acronym}
\usepackage{longtable}
\usepackage{multicol}
\usepackage[utf8]{inputenc}
\usepackage{subcaption}
\usepackage{float}
\usepackage[usenames,dvipsnames]{pstricks}
\usepackage{pstricks-add}
\usepackage{amssymb}
\usepackage{pifont}

\usepackage[export]{adjustbox}

\def\tsc#1{\csdef{#1}{\textsc{\lowercase{#1}}\xspace}}
\tsc{WGM}
\tsc{QE}
\tsc{EP}
\tsc{PMS}
\tsc{BEC}
\tsc{DE}

\begin{document}
\let\WriteBookmarks\relax
\def\floatpagepagefraction{1}
\def\textpagefraction{.001}

\shorttitle{Advanced Deep Learning  and  Large Language Models for Cancer Detection}

\shortauthors{Dr. Y Habchi et~al.}

\title [mode = title]{Advanced Deep Learning  and Large Language Models: Comprehensive Insights  for Cancer Detection }                      

\vskip2mm

\author[1]{Yassine Habchi}
\ead{habchi@cuniv-naama.dz}
\credit{Conceptualization; Methodology; Resources; Investigation; Writing original draft; Writing, review, and editing}

\author[2]{Hamza Kheddar}
\ead{kheddar.hamza@univ-medea.dz}
\credit{Conceptualization; Methodology; Resources; Investigation; Writing original draft; Writing, review, and editing}

\author[3]{Yassine Himeur}
\ead{yhimeur@ud.ac.ae}
\credit{Conceptualization; Methodology; Resources; Investigation; Writing original draft; Writing, review, and editing}

\author[4]{Adel Belouchrani}
\ead{adel.belouchrani@g.enp.edu.dz}
\credit{Conceptualization; Methodology; Resources; Investigation; Writing original draft; Writing, review, and editing}

\author[5]{Erchin Serpedin}
\ead{eserpedin@tamu.edu}
\credit{Methodology; Resources; Visualization; Investigation; Writing – review and editing, Supervision}

\author[6]{Fouad Khelifi}
\ead{fouad.khelifi@northumbria.ac.uk}
\credit{Methodology; Resources; Visualization; Investigation; Writing – review and editing, Supervision}

\author[7]{Muhammad E.H. Chowdhury}
\ead{chowdhury@qu.edu.qa}
\credit{Methodology; Resources; Visualization; Investigation; Writing – review and editing, Supervision}

\address[1]{Institute of Technology, University Center Salhi Ahmed, Naama, Algeria}

\address[2]{LSEA Laboratory, Electrical Engineering Department, Faculty of Technology, University of Medea, 26000, Algeria}

\address[3]{College of Engineering and Information Technology, University of Dubai, Dubai, UAE}

\address[4]{Ecole Nationale Polytechnique/ LDCCP lab., El Harrach, Algiers, Algeria}

\address[5]{ECEN Department, Texas A\&M University, College Station TX 77843-3128 USA}

\address[6]{Department of Computer Science and Digital Technologies, Engineering and Environment,
Northumbria University at Newcastle}

\address[7]{Department of Electrical Engineering, Qatar University, Doha 2713, Qatar}

\tnotetext[1]{Dr. Hamza Kheddar is the corresponding author.}

\begin{abstract}
In recent years, the rapid advancement of machine learning (ML), particularly deep learning (DL), has revolutionized various fields, with healthcare being one of the most notable beneficiaries. DL has demonstrated exceptional capabilities in addressing complex medical challenges, including the early detection and diagnosis of cancer. Its superior performance, surpassing both traditional ML methods and human accuracy, has made it a critical tool in identifying and diagnosing diseases such as cancer. Despite the availability of numerous reviews on DL applications in healthcare, a comprehensive and detailed understanding of DL's role in cancer detection remains lacking. Most existing studies focus on specific aspects of DL, leaving significant gaps in the broader knowledge base. This paper aims to bridge these gaps by offering a thorough review of advanced DL techniques, namely transfer learning (TL), reinforcement learning (RL), federated learning (FL), Transformers, and large language models (LLMs). These cutting-edge approaches are pushing the boundaries of cancer detection by enhancing model accuracy, addressing data scarcity, and enabling decentralized learning across institutions while maintaining data privacy. TL enables the adaptation of pre-trained models to new cancer datasets, significantly improving performance with limited labeled data. RL is emerging as a promising method for optimizing diagnostic pathways and treatment strategies, while FL ensures collaborative model development without sharing sensitive patient data. Furthermore, Transformers and LLMs, traditionally utilized in natural language processing (NLP), are now being applied to medical data for enhanced interpretability and context-based predictions. In addition, this review explores the efficiency of the aforementioned techniques in cancer diagnosis, it addresses key challenges such as data imbalance, and proposes potential solutions. It aims to be a valuable resource for researchers and practitioners, offering insights into current trends and guiding future research in the application of advanced DL techniques for cancer detection.
\end{abstract}


\maketitle

\begin{table}[]
    \centering
\begin{multicols}{3}
\footnotesize
\input{acro_list}
\end{multicols}
\end{table}

\section{Introduction}
\label{sec1}

\begin{figure*}[tbph]
\begin{center}
\includegraphics[width=1\columnwidth]{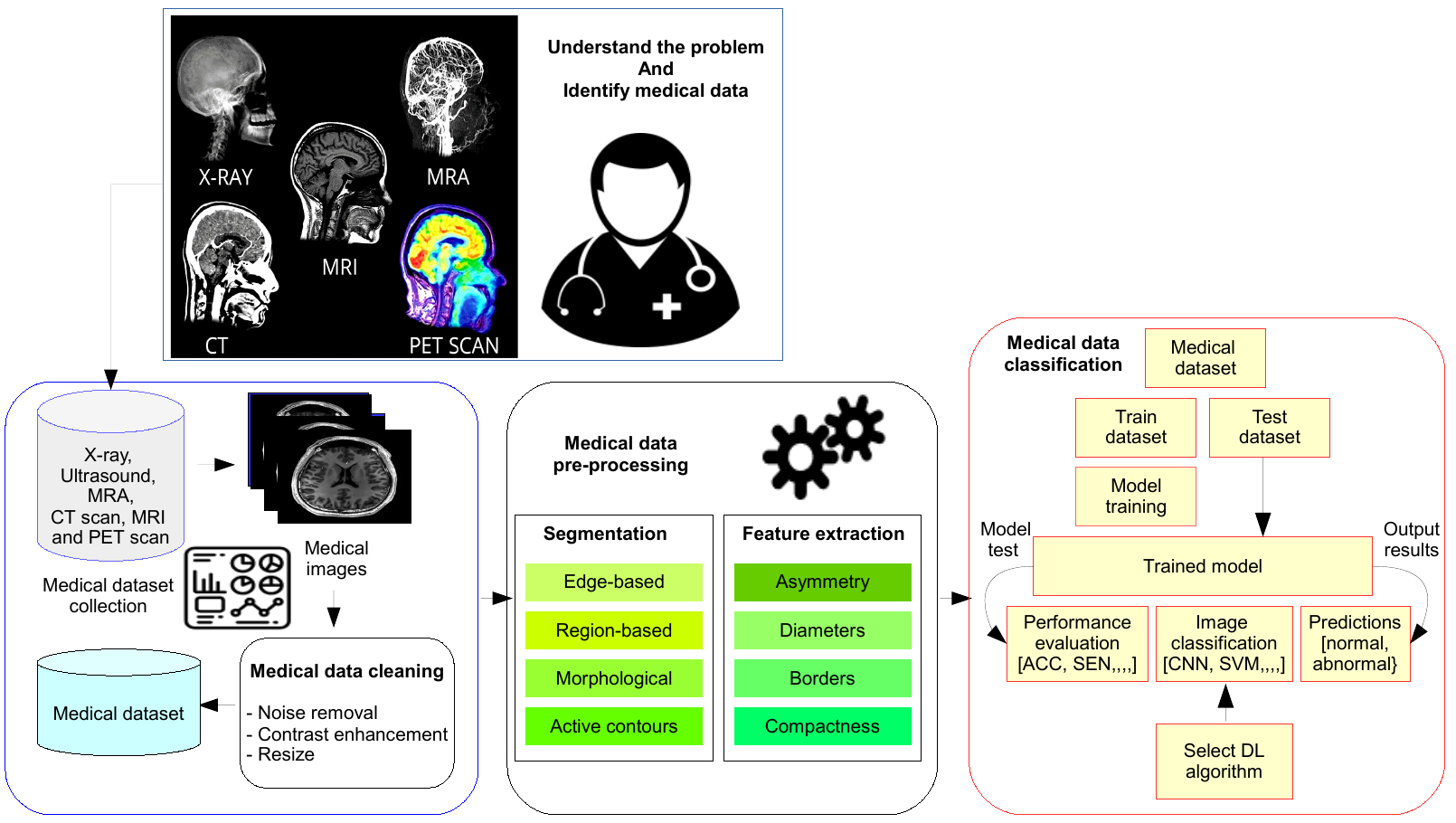}
\end{center}
\caption{The general process of cancer detection via \ac{DL}}
\label{fig1}
\end{figure*}

Researchers and clinicians face the formidable challenge of combating cancer, a leading cause of death globally. The World Health Organization has issued warnings about the anticipated increase in cancer-related deaths if effective solutions are not developed \cite{lam2020squamous}. Consequently, early detection of cancer has become crucial for saving numerous lives. However, the traditional process of diagnosing cancer through manual inspection of medical images is error-prone and time-consuming, highlighting the urgent need for more accurate and efficient detection methods. The \ac{CAD} systems have aided physicians by enhancing the accuracy and efficiency of medical image analysis, a task where feature extraction is crucial and heavily reliant on \ac{ML} techniques \cite{kwon2024diagnostic}. While various feature extraction methods have been explored for different types of cancer, these approaches face inherent limitations that \ac{ML} strives to overcome. \Ac{DL}, a subset of \ac{ML} algorithms, has gained significant prominence due to its layered structure, which makes it highly effective in medical fields, particularly in response to the increasing patient numbers, rapid technological advancements, and the exponential growth of medical data \cite{tran2021deep}. The use of \ac{DL} in medical diagnostics has broadened to cover a wide range of cancer types, including breast \cite{jiang2024deep}, lung  \cite{xie2021early}, kidney \cite{lal2024fpga}, skin \cite{das2021machine}, leukemia \cite{jawahar2024attention}, thyroid \cite{habchi2023ai}, brain cancer \cite{mathivanan2024employing}, among others.  By leveraging its multi-layered architecture, \ac{DL} outperforms traditional neural networks (Figure \ref{fig1}). The capability of \ac{DL} algorithms to effectively extract diverse features from data underscores their importance, particularly in cancer detection. The architecture of \ac{CNN}, a prominent class of \ac{DL}, mirrors the structure of the human visual cortex and is composed of multiple layers, including convolutional, pooling, and \ac{FC} layers. \ac{CNN}s excel at learning features directly from raw data without requiring human intervention, allowing them to develop hierarchical data representations that are highly valuable in image analysis \cite{ajit2020review}.

{Many methods are designed to efficiently classify diseases despite data imbalance. For example, the ENTAIL framework \cite{auriemma2022entail} demonstrates strong robustness on an unbalanced dataset, achieving high accuracy and sensitivity.} However, limited dataset sizes and imbalanced data are particularly significant challenges in the field of medical cancer research. Collecting comprehensive cancer datasets is often constrained by privacy regulations, ethical considerations, and the high cost and complexity of medical procedures. Consequently, the available datasets tend to be small, especially for rare types of cancer, which can hinder the development of robust and generalizable \ac{ML} models. This limitation can lead to overfitting, where models perform well on training data but fail to generalize to new, unseen data. Additionally, imbalanced data is a prevalent issue, as certain types of cancer are much more common than others. This imbalance can cause models to become biased towards the majority class, reducing their ability to accurately detect and diagnose less common cancers.

To address these issues, techniques such as data augmentation where synthetic data generation can be utilized to expand the size of training datasets, and  \ac{TL} can be used \cite{mazari2023deep,himeur2023video}. Moreover, \ac{FL} offers a promising solution by enabling the training of \ac{ML} models on decentralized datasets distributed across various institutions without the need to share sensitive patient data \cite{himeur2023federated}, This will encourage patients to accept participating in gathering their own private data. This approach not only preserves privacy but also increases the diversity of data used in model training, which enhances the model robustness and generalizability. In addition, \ac{RL} can be employed in clinical decision-making, where models are trained to optimize treatment plans and diagnostic pathways by learning from their interactions with the environment \cite{kheddar2024automatic,zhou2021deep}. This adaptive learning strategy allows the system to continuously improve and personalize recommendations based on real-time data.

The emergence of Transformer-based models and \acp{LLM} has further advanced the fields of computer vision and \ac{NLP} \cite{djeffal2023automatic}, and found applications in the analysis of medical data \cite{panagoulias2024evaluating}. Unlike conventional \ac{DL} models such as \acp{CNN} and \acp{RNN}, which rely on localized feature extraction and sequential dependencies, Transformers introduce self-attention mechanisms that enable them to capture long-range dependencies and global contextual relationships across large datasets. This capability is particularly crucial in cancer diagnostics, where heterogeneous data sources--including medical images, electronic health records, and genomic sequences--must be analyzed holistically to improve diagnostic precision.

Transformers excel at processing sequential data and can be applied to medical records, imaging data, and genomics, providing enhanced interpretability and context-aware predictions \cite{habchi2024machine}. Furthermore, \acp{LLM}, with their ability to process vast amounts of unstructured medical text, extend these capabilities by enabling automated medical reasoning and decision support. By leveraging large-scale textual and multimodal data, \acp{LLM} facilitate tasks such as medical literature review, patient history analysis, and clinical decision-making, surpassing traditional methods that primarily rely on structured data inputs. On the other side, traditional resampling methods, cost-sensitive learning, and ensemble techniques can help mitigate the effects of data imbalance, ensuring that \ac{ML} models are more accurate and reliable across different types of cancer. By effectively integrating advanced techniques such as \ac{FL}, \ac{RL}, Transformers, and \acp{LLM} alongside traditional methods, the performance and generalizability of \ac{AI} models in cancer diagnosis and treatment can be significantly improved.

Despite numerous reviews \cite{tran2021deep, rai2023comprehensive}, this coverage of \ac{ML} in cancer detection often lacks depth, leading to an incomplete understanding of its challenges. This paper proposes to cover this gap by offering a comprehensive review of \ac{DL} in cancer detection, discussing the key challenges, current applications, and most recent \ac{AI} models. Our contribution includes an in-depth examination of the aforementioned algorithms \ac{TL}, \ac{RL}, \ac{FL}, and Transformers, which are considered as the state-of-the-art DL tools, and highlight their key features. In addition, this paper overviews the contributions of DL in early detection of cancer as well as in improving the diagnosis and outcome prediction. Unlike other reviews that focus on specific aspects such as \ac{FL} or \ac{TL} \cite{sharma2023comprehensive, atasever2023comprehensive}, this paper delves into a broader spectrum, including emerging techniques like \ac{RL}, \acp{ViT}, and \acp{LLM}. While some reviews limit their discussion to particular cancers, such as lung or breast cancer \cite{gayap2024deep, carriero2024deep}, this review offers a thorough analysis of multiple cancer types, showcasing the versatility of \ac{DL} models across various domains.  Additionally, other reviews such as \cite{nerella2024Transformers} focuses solely on Transformers or omit certain advanced algorithms. In contrast, our review integrates these advanced algorithms alongside \ac{FL}, \ac{RL}, and \acp{LLM}, providing a more holistic view of their contributions to the detection of many types of cancer, including \textit{skin}, \textit{brain}, \textit{thyroid}, \textit{liver}, \textit{kidney}, \textit{pancreas}, \textit{lung}, \textit{leukemia}, \textit{breast}, \textit{cervical}, \textit{ovarian}, \textit{stomach}, \textit{bladder}, and \textit{colorectal} cancer.  Furthermore, this review addresses critical gaps by exploring challenges such as data privacy, model scalability, and dataset limitations, topics that other reviews often overlook. By covering evaluation metrics, model performance, and the role of advanced architectures, our survey bridges the gaps in existing literature and offers valuable insights into the future potential of \ac{DL} for improving cancer diagnosis and treatment planning.  Table \ref{table:01} presents a summary of the topics covered, highlighting the similarities and differences compared to existing reviews. It also identifies important topics often overlooked in the literature by current reviews, which our paper seeks to cover.

\begin{center}
\scriptsize
\begin{longtable}{m{0.5cm} m{0.5cm}m{5cm} m{6cm} m{0.3cm} m{0.3cm} m{0.3cm} m{0.3cm} m{0.3cm} m{0.3cm} }
\caption{Comparison with existing advanced \ac{DL}-based cancer diagnosis. The markers $\mdblkdiamond$ and $\mdwhtdiamond$ signify that a specific topic has been addressed or ignored, respectively.} 
\label{table:01} \\
\hline
\textbf{Ref} & \textbf{Year} & \textbf{ The similarity with our review} & \textbf{Differentiate from our review} & \textbf{DL} & \textbf{RL} & \textbf{FL} & \textbf{TL} & \textbf{ViT} & \textbf{LLM}\\
\hline
\endfirsthead

\hline
\textbf{Ref} & \textbf{Year} & \textbf{ The similarity with our review} & \textbf{Differentiate from our review} & \ac{DL} & \ac{RL} & \ac{FL} & \ac{TL} & \ac{ViT} & \ac{LLM}\\
\hline
\endhead

\hline
\multicolumn{3}{r}{\textit{Continued on next page}} \\
\endfoot

\hline
\endlastfoot

\cite{sharma2023comprehensive} &  2023 & The work highlights \ac{FL} for disease detection, emphasizing its impact on diagnostic accuracy, data privacy, and model effectiveness. &  This paper explores, in a general manner, various models for disease detection using \ac{FL} alone. However in our paper, the disease type is identified as cancer. & $\mdblkdiamond$ & $\mdwhtdiamond$ & $\mdblkdiamond$& $\mdwhtdiamond$ & $\mdwhtdiamond$ & $\mdwhtdiamond$\\ \hline

\cite{atasever2023comprehensive} &  2023 &  This study examines \ac{TL} in medical image analysis, detailing its methods, applications, source and target data, and the use of public or private imaging datasets. &   Compared to our work, it lacks exploration of other advanced \ac{DL} techniques and fails to highlight the significance of these methods in cancer image analysis. & $\mdblkdiamond$ & $\mdwhtdiamond$ & $\mdwhtdiamond$& $\mdblkdiamond$ & $\mdwhtdiamond$ & $\mdwhtdiamond$\\ \hline

\cite{al2024reinforcement} & 2024 & In this work, \ac{RL} has emerged as a dynamic and transformative paradigm in the field of \ac{AI}, offering the promise of intelligent decision making, especially in robotics and healthcare.  & Although \ac{RL} is discussed, this work does not review its application in the context of cancer diagnosis. Moreover, it only presents a comparative study of \ac{RL} algorithms, excluding others like \ac{LLM}s and \ac{ViT}s.& $\mdblkdiamond$& $\mdblkdiamond$& $\mdwhtdiamond$ & $\mdwhtdiamond$& $\mdwhtdiamond$ &$\mdwhtdiamond$.\\\hline

\cite{nazi2024large}&  2024 &  The review offers a comprehensive analysis of the use of advanced \ac{AI} models in healthcare. & The survey emphasizes the evolution of \ac{LLM}s and their performance metrics in the biomedical domain, without mentioning the other methods used to detect cancer.& $\mdblkdiamond$& $\mdwhtdiamond$ & $\mdwhtdiamond$ & $\mdwhtdiamond$ & $\mdwhtdiamond$ & $\mdblkdiamond$\\ \hline

\cite{gayap2024deep} & 2024 & This review highlights the \ac{DL} techniques for lung cancer diagnosis. This study emphasizes the high performance of these models compared to traditional methods. & The review focuses on \ac{DL} techniques such as \ac{CNN} for lung cancer. It covers public datasets and addresses challenges in deploying models clinically, but did not examine the role of advanced \ac{DL} in the diagnosis of several type of cancers.& $\mdblkdiamond$& $\mdwhtdiamond$ & $\mdwhtdiamond$&$\mdwhtdiamond$&$\mdwhtdiamond$&$\mdwhtdiamond$\\ \hline

\cite{nerella2024Transformers} & 2024 & This work explores Transformer models in healthcare, focusing on their application to complex data. &  The provided work offers a broad overview types of Transformer applications in various healthcare settings, including surgical outcomes and drug synthesis, without delving into specific models or datasets related to cancer.& $\mdblkdiamond$&$\mdwhtdiamond$&$\mdwhtdiamond$&$\mdwhtdiamond$&$\mdblkdiamond$&$\mdwhtdiamond$\\ \hline

\cite{carriero2024deep} & 2024 & The focus is on the application of \ac{DL} techniques to breast cancer imaging. &  This work explores discusses only \ac{DL} techniques for cancer detection and focuses exclusively on applications for diagnosing of the breast  cancer. &$\mdblkdiamond$&$\mdwhtdiamond$&$\mdwhtdiamond$&$\mdwhtdiamond$&$\mdwhtdiamond$&$\mdwhtdiamond$\\ \hline

\cite{jiang2024vision} & 2024 &  This article illustrates the importance of \ac{ViT} for cancer diagnosis. & 
A comprehensive study on the application of \ac{ViT}s in addressing intractable diseases is lacking, and other relevant algorithms have not been thoroughly explored either. &$\mdblkdiamond$&$\mdwhtdiamond$&$\mdwhtdiamond$&$\mdwhtdiamond$&$\mdblkdiamond$&$\mdwhtdiamond$\\\hline

Ours & 2024 & This paper reviews the \ac{DL} tools for cancer diagnosis by highlighting their key features and applications. & This survey first presents the conventional \ac{DL} tools. Our survey first presents the background of conventional \ac{DL} techniques, followed by advanced approaches for cancer diagnosis, including \ac{RL}, \ac{FL}, \ac{TL}, \ac{LLM}s, and \ac{ViT}s. It examines their effectiveness, challenges, datasets, evaluation metrics, and potential to enhance detection, classification, and treatment. &$\mdblkdiamond$&$\mdblkdiamond$&$\mdblkdiamond$&$\mdblkdiamond$&$\mdblkdiamond$&$\mdblkdiamond$\\
\hline
\end{longtable}
\end{center}

\begin{figure*}[tbph]
\begin{center}
\includegraphics[width=0.8\columnwidth]{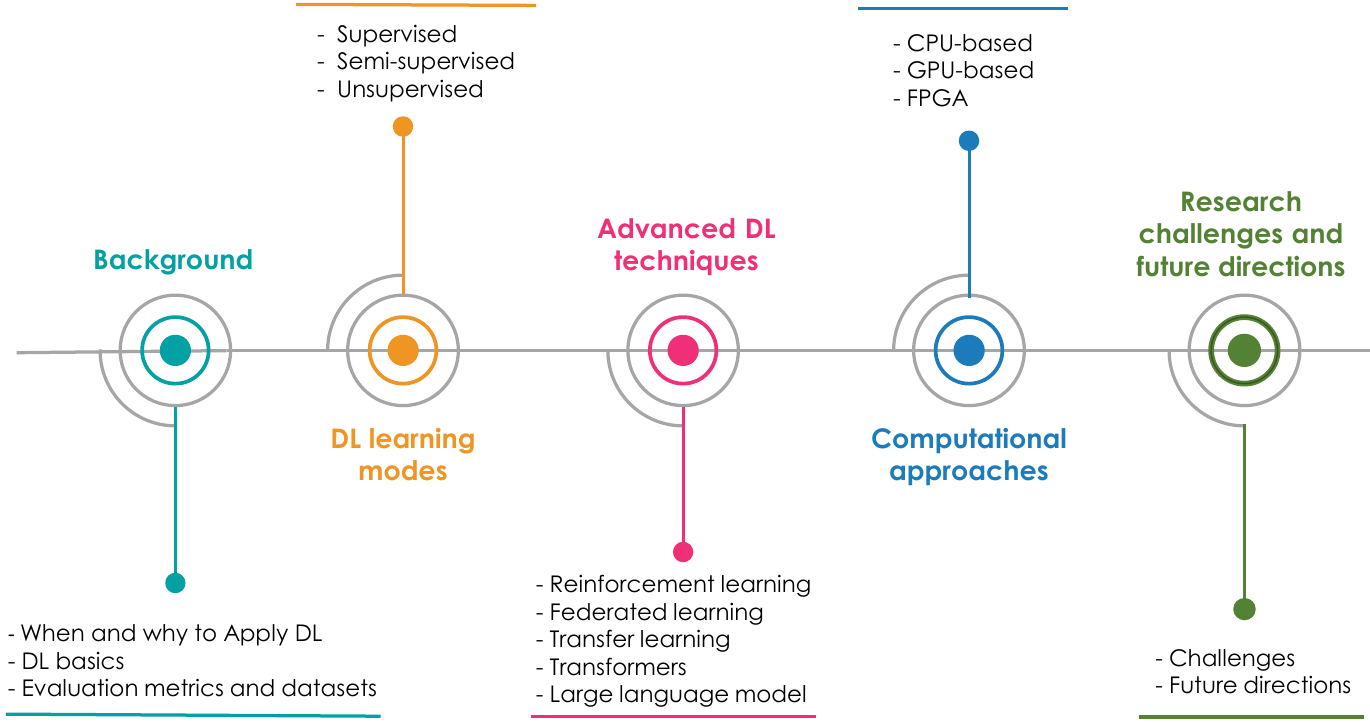}
\end{center}
\caption{Road-map outlining the structure of the review.}
\label{fig2}
\end{figure*}

The remaining of the paper is structured to provide a comprehensive review of \ac{DL} in cancer detection. Section \ref{sec2} offers background information on \ac{DL} fundamentals. Section \ref{sec3} overviews various \ac{DL} training modes. Section \ref{sec4} delves into the most commonly used advanced \ac{DL} networks for cancer detection, namely \ac{TL}, \ac{RL}, \ac{FL}, and Transformers. Section \ref{sec5} discusses existing computational approaches. Section \ref{sec6} addresses research challenges and future directions. Finally, Section \ref{sec7} concludes this paper. Figure \ref{fig2} illustrates the road-map structure of this review in terms of sections and subsections.

\section{Background} 
\label{sec2}

This section presents the basics of \ac{DL}, covering the necessary layers for building models in medical image recognition, such as convolutional, pooling, and \ac{FC} layers. It also discusses the concepts of activation function \acp{AF}, regularization, hyper-parameter optimization, training, and other essential components required to construct efficient \ac{DL} models. 

\subsection{When and why applying \ac{DL} in the medical domain}

\ac{DL} is applied in the medical field when addressing complex tasks that require advanced data processing and interpretation. One key area is medical imaging, where \ac{DL} models are used for classifying, segmenting, and detecting abnormalities in images, such as detecting tumors and diagnosing diseases from radiological scans. It plays a critical role in interpreting \ac{MRI} and \ac{CT} scans to identify irregularities \cite{ryu2025radiography}. \ac{DL} is also instrumental in disease diagnosis and prognosis by being capable to analyze vast datasets of medical records and images, providing more precise predictions and enabling personalized treatments for patients. In drug discovery and development, \ac{DL} accelerates the process by analyzing molecular data to identify potential drug candidates, predict their effectiveness, and model their interactions with biological systems. Additionally, \ac{DL} supports \ac{EHR} analysis by parsing large datasets to generate clinical insights, predict outcomes, and identify disease patterns. Genomics and precision medicine benefit from \ac{DL} by analyzing genetic data to identify anomalies and predict disease risks, while \ac{NLP} applications in healthcare aid in extracting information from unstructured medical texts \cite{piccialli2021survey}. 

\begin{figure*}[tbph]
\begin{center}
\includegraphics[width=1\columnwidth]{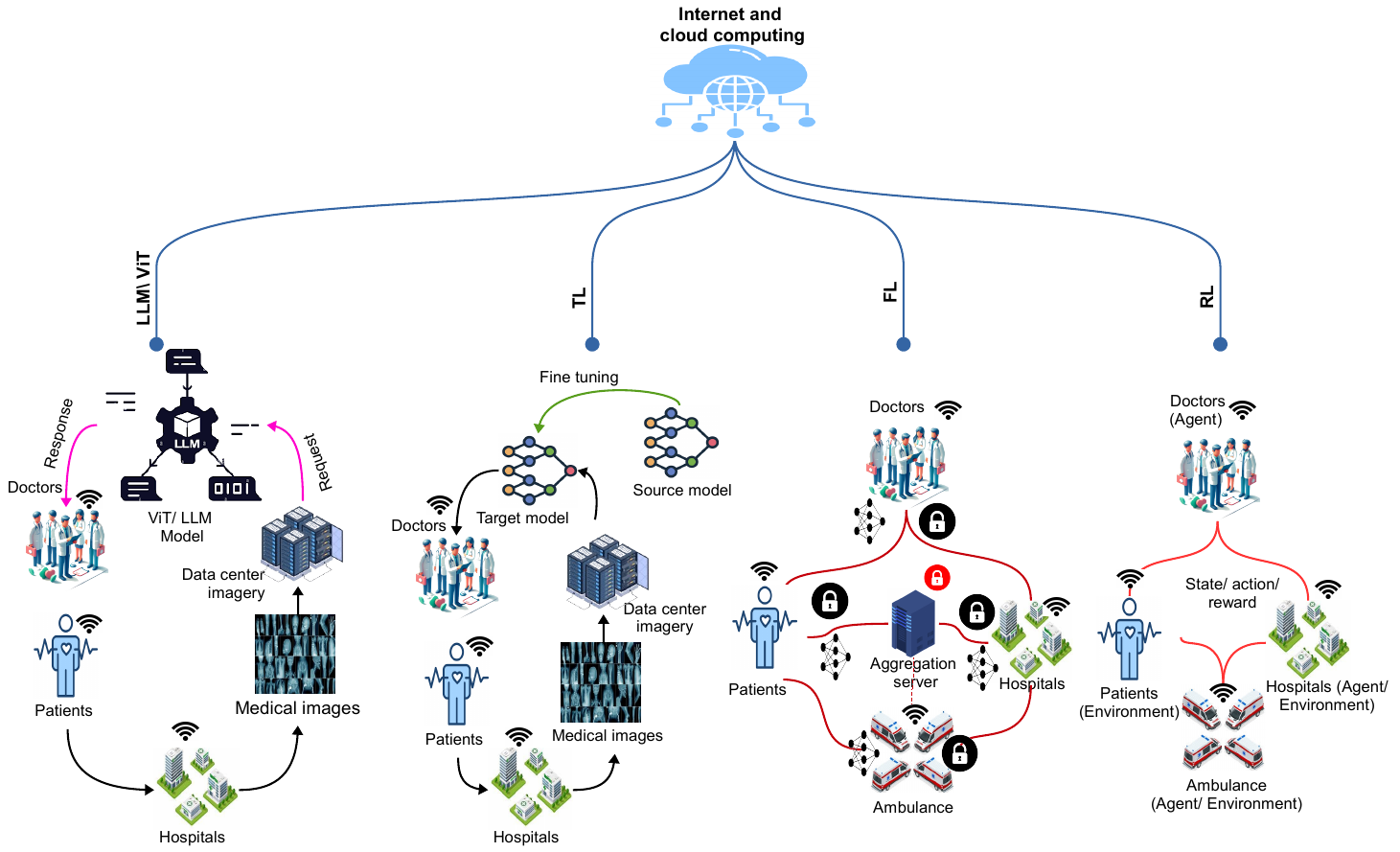}
\end{center}
\caption{Application of advanced \ac{DL} methods in healthcare, showcasing technologies such as \ac{RL}, \ac{FL}, \ac{TL}, \ac{ViT}, and \acp{LLM}. It highlights scenarios like real-time decision-making, privacy-preserving model training, and personalized treatment. The ecosystem integrates agents like hospitals, doctors, and ambulances with cloud computing, pretrained models, and aggregation servers. Key use cases include optimizing resource allocation, enhancing diagnostic accuracy, and improving scalability. Advanced DL solutions address challenges such as data privacy, limited labeled datasets, and computational efficiency, ultimately supporting accurate, efficient, and personalized patient care \cite{zhang2023applying}.}
\label{fig3}
\end{figure*}

\ac{DL} is favored in the medical domain because of its ability to handle large-scale and complex data, such as medical images, genomic data, and electronic health records. Its capacity for feature extraction and representation learning allows it to automatically learn meaningful features from raw medical data, which is particularly useful for tasks like medical image analysis. \ac{DL} models also provide improved accuracy and predictive power, often outperforming traditional methods and even surpassing human performance in certain diagnostic tasks. Another advantage is in personalized medicine, where \ac{DL} leverages extensive patient data to customize treatment plans, improving disease management and treatment outcomes. Moreover, \ac{DL} enhances efficiency by automating routine tasks, thus alleviating the workload on medical professionals and allowing them to focus on patient care. In research and drug discovery, \ac{DL} accelerates the discovery process and helps reduce the cost and time involved in clinical trials. Finally, \ac{DL} models continually improve their performance with more data, making them highly adaptable to the evolving needs of the medical field. Figure \ref{fig3} summarizes the benefits of \ac{DL} in the healthcare domain \cite{zhang2023applying}. 

\subsection{{\ac{DL} and \ac{CNN} basics}}

\ac{DL} has revolutionized medical image analysis by enabling automated feature extraction and hierarchical representation learning. Unlike traditional \ac{ML} approaches that require handcrafted feature engineering, \ac{DL} models, particularly \acp{CNN}, learn patterns directly from raw medical images, improving accuracy and diagnostic reliability. These models leverage multiple layers, including convolutional layers for spatial feature extraction, pooling layers for dimensionality reduction, and fully connected layers for classification. Regularization techniques and optimization strategies further enhance model generalization and performance.

\acp{CNN} have been widely applied in \textit{cancer detection}, excelling in tumor segmentation, classification, and localization. Their hierarchical structure allows them to capture intricate patterns within medical images, making them highly effective for tasks such as identifying malignant lesions in radiology scans. Table~\ref{tab:cnn_components} summarizes the main CNN components and their roles in cancer detection. Unlike traditional \ac{ML}-based \ac{CAD} systems, CNNs automatically extract tumor-related features without requiring manual selection, streamlining the diagnostic process and improving efficiency. Table \ref{table:3} summarizes the most frequently used activation, loss, and pooling functions in \ac{DL}, particularly in \ac{CNN}-based approaches.

\begin{table}[h]
    \centering
    \caption{{Key CNN components for cancer detection.}}
    \label{tab:cnn_components}
    \begin{tabular}{m{3.5cm}m{5cm}m{6cm}}
        \hline
        \textbf{{Component}} & \textbf{{Function}} & \textbf{{Relevance to cancer detection}} \\
        \hline
        \textbf{{Convolutional layer}} & {Extracts spatial features by applying filters to input images.} & {Detects tumor patterns and abnormal tissue structures.} \\
        \hline
        \textbf{{Pooling layer}} & {Reduces feature map size, preserving key information.} & {Enhances computational efficiency without losing critical features.} \\
        \hline
        \textbf{{Activation function}} & {Introduces non-linearity to improve model learning.} & {Helps differentiate between cancerous and non-cancerous regions.} \\
        \hline
        \textbf{{Fully connected layer}} & {Aggregates extracted features for final classification.} & {Used to distinguish between cancer types or stages.} \\
        \hline
        \textbf{{Regularization (Dropout, Batch norm.)}} & {Prevents overfitting by stabilizing learning.} & {Improves model generalization to unseen medical data.} \\
        \hline
    \end{tabular}
\end{table}

\begin{table*}[tbph]
\caption{{Types of functions used in \ac{CNN}-based cancer detection.}}
\label{table:3}
\begin{center}

\begin{tabular}{m{0.3cm}m{3cm}m{3.5cm}m{10cm}}
\hline
& Function & Mathematical formula & Role overview \\ \hline
& Sigmoid & \(\displaystyle f(x)=\frac{1}{1+e^{-x}}\) &  The input is real numbers, while the output is between "0" and "1". \\

\multirow{6}{*}{\rotatebox{90}{{Activation functions}}}& Tanh & \(\displaystyle f(x)=\frac{e^{x}-e^{-x}}{e^{x}+e^{-x}} \) & Its input is real numbers, and the output is between -1 and 1.\\
& ReLU & \(\displaystyle f(x)=max(0,x) \) & It converts the input values to positive numbers and presents low computational complexity. Is characterized with lower computational load.\\

& Leaky ReLU & 

$f(x) = \left\{
    \begin{array}{ll}
        x, &  x>0 \\
        m.x, &  x \leq 0
    \end{array}
\right.$

& It has a small slope for negative values. It is employed to overcome the Dying ReLU problem. m: the leak factor.\\

& Noisy ReLU &$f(x)=max(x+Y)$ & It employs a Gaussian distribution $Y \sim N(0,\sigma(x))$ to create ReLU noisy.\\

& Parametric linear units & 

$f(x) = \left\{
    \begin{array}{ll}
        x, &  x>0 \\
        a.x, &  x \leq 0
    \end{array}
\right.$

& It is similar to Leaky ReLU, but it differs in that the leak factor is updated during the model training process. a:learnable weight.\\

\hline

\multirow{3}{*}{\rotatebox{90}{{Loss functions}}}& SLF & $H(p,y)=-\sum_{i} y_ilog(p_i),$ & Is employed for measuring the performance. $P_i=\frac{e^{a_i}}{\sum_{k=1}^{N} e_{k}^{a}}, i\in[1,N]$ Its output is the probability $p\in[0,1]$.\\

& ELF & $H(p,y)=\frac{1}{2N} \sum_{i=1}^{N} (p_i-y_i)^2$ & It is called mean square error and is used in regression problems.\\

& HLF & $H(p,y)=\sum_{i=1}^{N} max(0, m-(2y_i-1)p_i)$ & It is employed in binary classification
 problems; this is important for SVMs.\\ \\
 \hline

 & Max pooling & 
$y = \max(x_i)$ & 
Retains the most significant feature by selecting the maximum activation in each pooling region. Preserves texture details. \\

\multirow{3}{*}{\rotatebox{90}{{Pooling functions}}} & Average pooling & 
$y = \frac{1}{n} \sum_{i=1}^{n} x_i$ & 
Averages the activations in each pooling region. Helps retain background information. \\

& Global pooling & 
$y = \frac{1}{N} \sum_{i=1}^{N} x_i$ & 
Reduces feature maps to a single value per channel, helping in model regularization. Often used in fully convolutional architectures. \\

 & Stochastic pooling & 
$P(x_i) = \frac{x_i}{\sum_{j} x_j}$ & 
Randomly selects activation based on a probability distribution. Helps prevent overfitting. \\

& Rank-based pooling & 
$y = \operatorname{Rank}(x_i)$ & 
Ranks the activations instead of selecting absolute values. Offers stronger robustness in noisy environments. \\

& Ordinal pooling & 
$\text{Sort}(x_1, x_2, ..., x_n)$ & 
Sorts activations in ascending or descending order to retain structured information. Speeds up training. \\ 
\hline

\end{tabular}

\end{center}
\end{table*}

\subsection{Evaluation metrics and datasets}

When evaluating the performance of \ac{DL}, Transformers, and \ac{LLM} models, selecting an appropriate metric is of paramount importance. Numerous metrics have been proposed and employed across various DL applications \cite{kheddar2024deepSteg,hemis2024deep}. Table \ref{table:12} presents a concise summary of commonly used metrics that are particularly suited for assessing the performance of \ac{AI} algorithms in cancer detection. Additionally, several datasets have been employed in recent years across various medical domains to evaluate performance and accelerate progress. Table \ref{table:13} lists the most widely used datasets in different medical applications.

\begin{table*}[ht!]
\caption{Summary of metrics used in evaluating \ac{DL}, Transformers, and \ac{LLM}-based cancer detection.}
\label{table:12}
\begin{center}
\begin{tabular}{m{3cm}m{7cm}m{7cm}} 
\hline
Task type &  Metric  & Description\\  \hline

Classification   & \(\displaystyle\mathrm{ Acc=\frac{T_{P}+T_{N}}{T_{P}+F_{P}+T_{N}+F_{N}}} \)  & Gives the correct percent of the total number of positive and negative predictions. \\

Classification   & \(\displaystyle Spe=\frac{T_{N}}{T_{N}+F_{P}} \)  & It is the ratio of correctly predicted negative samples to the total negative samples.  \\

Classification    & \(\displaystyle \mathrm{Sen= \frac{T_{P}}{T_{P}+F_{N}}} \)   &It is a quantifiable measure of real positive cases predicted as true positive cases.  \\

Classification   & \(\displaystyle\mathrm{P=\frac{T_{P}}{T_{P}+F_{P}}100\%} \)  &Measures the proportion of true positive predictions made by the model, out of all positive predictions.  \\

Classification  & \(\displaystyle \mathrm{F1 = 2\times \frac{Precision \times Recall}{Precision + Recall}} \)  & It is the harmonic mean of precision and sensitivity of the classification.  \\

Classification   & \(\displaystyle \mathrm{NPV=\frac{T_{N}}{T_{N}+F_{N}}} \)  & The proportion of negative results in diagnostic tests; a higher value indicates better diagnostic accuracy.   \\

Similarity    & \(\displaystyle\mathrm{JSC=\frac{\left\vert A\cap B\right\vert }{\left\vert A\cup B\right\vert }=
\frac{T_{P}}{T_{P}+F_{P}+F_{N}}}\) & Proposed by Paul Jaccard to gauge the similarity and variety in samples. \\

Error rate  & \(\displaystyle \mathrm{FPR=\frac{F_{P}}{F_{P}+T_{N}}=1-SP}\)  &Measures the proportion of negative samples incorrectly classified as positive by the model.  \\

Correlation \newline (\ac{ViT} and \ac{LLM})   & \(\displaystyle \mathrm{MCC=\frac{T_{P}.T_{N}-F_{P}.F_{N}}{\sqrt{(T_{P}+F_{P})(T_{P}+F_{N})(T_{N}+F_{P})(T_{N}+F_{N})}}}\) & Assesses binary classification by balancing true/false positives and negatives across varying class sizes.\\

Adversarial robust \newline (\ac{ViT} and \ac{LLM})  & \(\displaystyle\mathrm{FR=\frac{\text{Number of samples with changed predictions}}{\text{Total number of adversarial samples}}}\) & Measures misclassified samples after adversarial manipulation, crucial for attack evaluation.\\

Anomaly detection \newline (\ac{ViT} and \ac{LLM}) 
& \(\displaystyle\mathrm{AS=\frac{\text{Score - Max  baseline  score}}{\text{Baseline  standard  deviation}}}\) & 
Assesses anomaly deviations from normal activity, adjusting thresholds to balance false positives and negatives.\\

\hline
\end{tabular}
\end{center}
\begin{flushleft}
\scriptsize{Abbreviations: Accuracy (Acc); Alert score (AS); Specificity (Spe); Sensitivity (Sen); Fooling rate (FR); Matthew’s correlation coefficient (MCC); Fallout or False positive rate (FPR); Jaccard similarity index (JSI); Negative predictive value (NPV); F1 score (F1); Precision (P).}
\end{flushleft}
\end{table*}

\begin{table}[tbph]
\caption{Summary of the most commonly  used datasets in different  cancer type.}
\label{table:13}
\begin{center}
\scalebox{0.8}{
\begin{tabular}{ccccccccc}
\hline
Datasets & Available? & Format&Bit per pixel&Resolution&N of images & Type of images &Application & Related work\\  \hline 

DDTI & Yes\tablefootnote{http://cimalab.intec.co/?lang=en\&mod=project\&id=31} &JPG&8-bit&-&134 &Ultrasound images& Thyroid & {\cite{bal2023approaching}} \\ 

{ALL-IDB} & Yes\tablefootnote{http://homes.di.unimi.it/scotti/all/} &JPG&24-bit&2592x1944& 108& Microscope images & {Leukemia} & {\cite{bibi2020iomt}} \\ 

{TCGA} & Yes\tablefootnote{https://archive.ics.uci.edu/ml/datasets/chronic\_kidney\_disease}&TIFF, DICOM&16-bit&1000x1000 &-&Histopathological, radiology& {Kidney} & {\cite{mohammed2021stacking}} \\%

{BreakHis} & Yes\tablefootnote{https://web.inf.ufpr.br/vri/databases/breast-cancer-histopathological-database-breakhis/}&JPEG&8-bit&700x460&7,909&Histopathological& {Breast} & {\cite{agarwal2022breast}} \\

{HAM10000} & Yes\tablefootnote{https://dataverse.harvard.edu/dataset.xhtml?persistentId=doi:10.7910/DVN/DBW86T}&JPEG&8-bit&600x450&10,015&Dermatoscopic& {Skin} & {\cite{gomathi2023skin}} \\ 

{LUNA16} & Yes\tablefootnote{https://luna16.grand-challenge.org/} &-&16-bit&512x512&~1,200 &Lung \ac{CT}& {Lung} & {\cite{gharaibeh2024automated}} \\

{ProstateX} & Yes\tablefootnote{https://www.cancerimagingarchive.net/collection/prostatex/}  &DICOM&16-bit&384x384&~1,000  &Prostate \ac{MRI}& {Prostate} & {\cite{cai2024fully}} \\

{TCIA - Colorectal Histology} & Yes\tablefootnote{https://www.cancerimagingarchive.net/collection/stageii-colorectal-ct/} &PNG&8-bit&150x150&~5,000  &Histopathological& {Colorectal } & {\cite{santhoshi2024optimizing}} \\

{PAIP 2019} & Yes\tablefootnote{https://paip2019.grand-challenge.org/Dataset/}&TIFF&16-bit&20kx20k&100  &Histopathological& {Liver} & {\cite{wang2024multi}} \\

{BraTS 2023 dataset} & Yes\tablefootnote{https://www.synapse.org/Synapse:syn51156910} &NIfTI &16-bit&240x240x155 &~15,000 &MRI& {Brain} & {\cite{mostafa2023brain}} \\

{NIH Pancreatic} & Yes\tablefootnote{https://cdas.cancer.gov/datasets/plco/10/}&DICOM &16-bit&512x512 &~100+ &MRI& {Pancreatic} & {\cite{dutta2023using}} \\

\hline
\end{tabular}
}
\end{center}
\end{table}

\section{Training modes in \ac{DL} }
\label{sec3}

\ac{DL} techniques can be trained using three primary modes: \ac{SL}, \ac{SSL}, and \ac{USL}. A detailed explanation of each mode is provided below, along with examples relevant to their respective category. Table \ref{table:modes} provides an overview of state-of-the-art studies employing these \ac{DL} training modes for various cancer detection applications. {Table~\ref{tab:dl_training_modes} provides an overview of these modes, their applications in cancer detection, 
and their key limitations.}

\subsection{Deep \ac{SL}}
Deep \ac{SL} is a specialized branch of \ac{ML} that trains algorithms to predict or make decisions based on labeled data sets.  The primary objective of deep \ac{SL} is to develop a function that maps inputs to outputs by identifying patterns within the data.  The "deep" aspect of \ac{SL} pertains to the application of \ac{DNN}, which are characterized by multiple hidden layers.  Training involves iterative adjustments of the network's weights and biases to reduce the disparity between its predictions and the actual labels, usually employing a \ac{LF} to measure this difference. Back-propagation is used to update parameters and incrementally enhance performance.

\renewcommand{\arraystretch}{1.5}
\begin{table*}[ht!]
\centering
\caption{{Overview of \ac{DL} training modes for cancer diagnosis.}}
\label{tab:dl_training_modes}
\scalebox{0.8}{
\begin{tabular}{m{2.5cm}m{3cm}m{8cm}m{6cm}}
\hline
\textbf{{Training mode}} & \textbf{{Key characteristic}} & \textbf{{Application in cancer diagnosis}} & \textbf{{Limitations}} \\
\hline
{\textbf{\Ac{SL}}} & {Requires labeled data for training} & {Used in cancer image classification, tumor segmentation, and disease diagnosis with well-annotated datasets \cite{johnson2019survey}.} & {High labeling cost; performance depends on dataset quality; limited generalization to unseen data.} \\
\hline
{\textbf{\Ac{SSL}}} & {Uses a mix of labeled and unlabeled data} & {Reduces the reliance on labeled data, making it useful for medical imaging where obtaining labeled samples is costly. Applied in histopathology and radiology \cite{habchi2024machine}.} & {Performance is sensitive to the proportion of labeled data; risk of learning incorrect patterns from noisy unlabeled data.} \\
\hline
{\textbf{\Ac{USL}}} & {Extracts patterns from unlabeled data} & {Used in clustering cancer subtypes, anomaly detection in medical images, and identifying novel biomarkers \cite{panagoulias2024evaluating}.} & {Lack of ground truth validation; difficult to interpret results; risk of overfitting to irrelevant patterns.} \\
\hline
\end{tabular}
}
\end{table*}

\begin{figure}[tbph]
\begin{center}
\includegraphics[width=0.7\columnwidth]{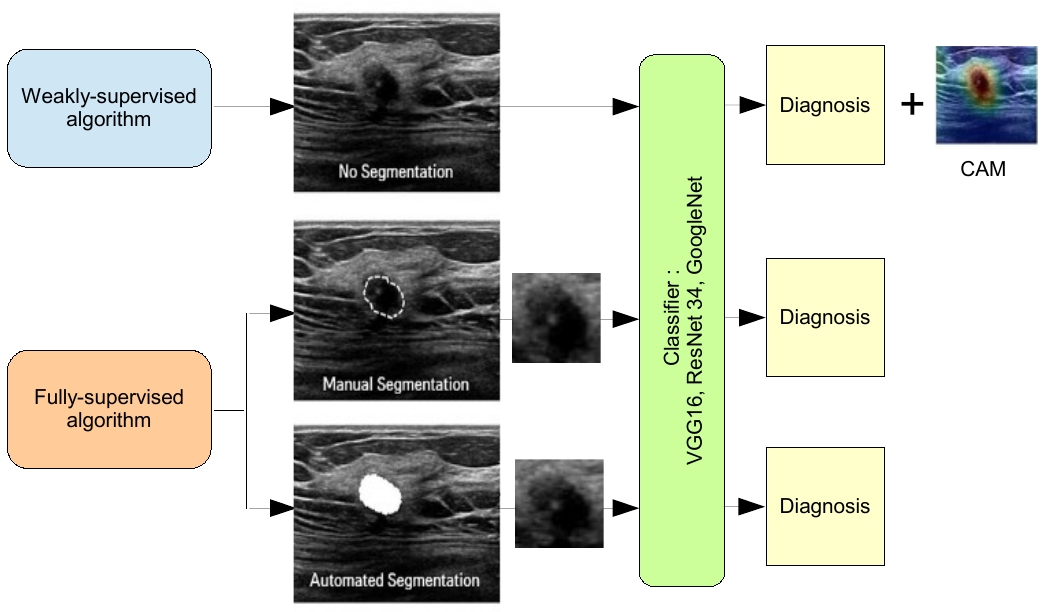}
\end{center}
\caption{ \ac{SL} for ultrasound diagnosis of breast cancer images using both fully-supervised and weakly-supervised \ac{DL} algorithms \cite{han2020semi}.}
\label{fig6}
\end{figure}

 In the context of image processing, \ac{SL} can be further divided into two sub-categories: fully-supervised \ac{DL} and weakly-supervised \ac{DL} algorithms. Fully-supervised \ac{DL} algorithms require  segmentation—either manual or automated and cropping of the \ac{ROI} before inputting the data into classifiers. In contrast, weakly-supervised \ac{DL} algorithms do not require image annotation of the lesion's \ac{ROI}. Instead, they leverage class activation maps generated by incorporating a \ac{GAP} layer into the final convolutional layer to visualize detected regions. Figure \ref{fig6} illustrates a comparison between weakly-supervised and fully-supervised \ac{DL} algorithms for breast mass classification and localization.
 
Many studies employing advanced \ac{DL} rely on \ac{SL}. For example, in \cite{bhuiyan2019transfer}, the authors explore the use of \ac{TL} and \ac{SL} techniques to enhance the classification of breast cancer histopathology images. The proposed approach aims to automate the differentiation between benign and malignant tissue by employing various feature extractors and classifiers, addressing challenges associated with image analysis. Moreover, in \cite{pati2022federated}, the authors examine how \ac{FL} can facilitate the detection of boundaries in rare cancers using big data. This research highlights the integration of \ac{FL} with \ac{SL} techniques to enhance the accuracy and effectiveness of cancer boundary detection while protecting patient privacy. At the same time, it enables the use of diverse datasets to improve the model's performance in detecting rare cancer types. Moving forward, the study \cite{an2024transformer}, proposes a two-stage transformer-based weakly \ac{SL} framework, called SSRViT, to support histo-pathological diagnosis of lung cancer. Due to the large size of whole-slide images  and the difficulty of obtaining precise annotations, SSRViT aims to leverage weak labels for efficient learning. The framework uses a Shuffle-remix \ac{ViT} to extract discriminative local features, which are then aggregated for slide-level classification via a simple transformer-based classifier. The method demonstrates superior performance in distinguishing between lung adenocarcinoma, pulmonary sclerosing pneumocytoma, and normal lung tissue. Sushil et al. \cite{sushil2024comparative} explored the use of \acp{LLM} to reduce the need for extensive data annotation in breast cancer pathology. A manually labeled dataset of 769 reports with 13 categories was used to compare the zero-shot classification performance of \ac{GPT}-4 and \ac{GPT}-3.5 with traditional supervised models like random forests, \ac{LSTM}-Attention, and UCSF-\ac{BERT}. \ac{GPT}-4 performed as well as or better than the best \ac{SL} models, particularly in handling imbalanced label tasks. The findings suggest that \acp{LLM} can significantly reduce the labor-intensive data labeling process, advancing cancer diagnosis without sacrificing accuracy.

\subsection{Deep \ac{USL}}

Deep \ac{USL} is a branch of \ac{ML} where neural networks are trained on large datasets without specific guidance or supervision. Unlike \ac{SL}, which use labeled data, \ac{USL} aims to identify inherent patterns and structures in the data autonomously. A prevalent technique in deep \ac{USL} is the use of \ac{AE}s, which are designed to compress data into a lower-dimensional space before reconstructing it to its original form, thereby learning the essential structures and features without labels. Other significant methods in deep \ac{USL} include \acp{GAN}, which learn to create new data mimicking the original by understanding its distribution. Clustering is another common unsupervised strategy. Deep \ac{USL}'s primary benefit is its ability to reveal critical data features that are not readily visible, aiding in tasks such as image recognition where it can uncover crucial visual patterns like edges and textures. This insight can improve the accuracy and robustness of models for tasks like object recognition. However, deep \ac{USL} presents several challenges: (1) \textit{Interpretability}: The complexity of deep \ac{USL} models makes it difficult to interpret how decisions are made or to diagnose errors effectively. (2) \textit{Data requirements}: Deep \ac{USL} requires large datasets for effective learning; insufficient data may hinder accurate pattern recognition. (3) \textit{Model complexity}: Deep \ac{USL} models, often consisting of numerous layers and parameters, can complicate training and demand significant computational resources and time. (4) \textit{Fine-tuning issues}: While deep \ac{USL} excels at learning general features, adapting these models to specific tasks remains challenging. (5) \textit{Noise sensitivity}: These models are vulnerable to noise, which can mislead the learning process and obscure relevant patterns. (6) \textit{Pattern bias}: Deep \ac{USL} may introduce bias towards certain patterns or data points, potentially reducing diversity in the learned features \cite{liu2022deep, raza2021tour}.

\begin{figure*}[tbph]
\begin{center}
\includegraphics[width=0.9\columnwidth]{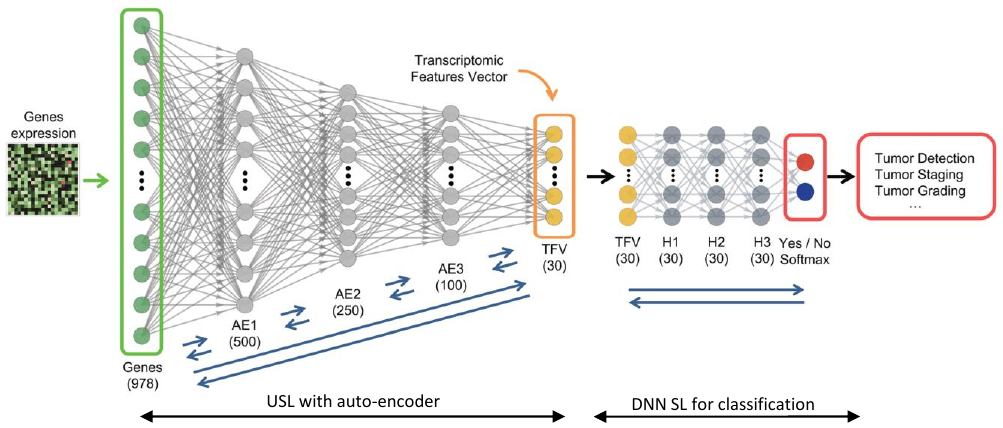}
\end{center}
\caption{An example on \ac{USL} for cancer diagnosis using \ac{AE}  \cite{yuan2020unsupervised}. A 5-layer \ac{AE}  is trained in an \ac{USL} manner to extract features from high-dimensional gene expression data, compressing them into a 30-dimensional transcriptomic feature vector. These feature vectors are subsequently used for supervised training in a fully connected deep softmax classifier, which distinguishes between normal and tumor samples and classify tumors by grade and stage.}
\label{figAE}
\end{figure*}

In this context, numerous studies have been proposed in the literature, with many schemes adopting \ac{AE} as the primary unsupervised approach, as depicted in Figure~\ref{figAE}. For example, the study \cite{silva2021egfr} employed \ac{USL} and \ac{TL} to assess epidermal growth factor receptor mutation status in lung cancer using \ac{CT} images. A convolutional \ac{AE} was developed to reconstruct images and extract features from three \ac{ROI}: the nodule, the lung containing the nodule, and both lungs. By leveraging \ac{TL}, the model improved feature extraction and analysis. The study found that analyzing beyond the nodule captured more relevant information, enhancing prediction accuracy.  Bercea et al. \cite{bercea2021feddis} introduce FedDis, a novel \ac{FL} approach that integrates \ac{USL} to address data heterogeneity in medical imaging. FedDis disentangles model parameters into shape and appearance components, sharing only the shape parameters among clients. This method leverages the assumption that anatomical structures in brain \ac{MRI} images are consistent across institutions, improving anomaly detection, such as cancer identification. By utilizing healthy brain scans from various sources, FedDis segments abnormal structures in pathological databases, including Glioblastoma cases. Similarly, Stember et al. \cite{stember2022unsupervised} propose a method that combines \ac{USL} (clustering) with \ac{RL} to segment brain lesions in \ac{MRI} scans. Initially, clustering   generates candidate lesion masks for each image. Users then select the best mask for a subset of images, which is subsequently used to train an \ac{RL} algorithm to identify the optimal masks. This approach was compared to a U-net \ac{SL} network. While the \ac{SL} model suffered from overfitting and performed poorly, the combined \ac{USL} and \ac{RL} approach achieved a high Dice score, demonstrating its effectiveness in lesion segmentation with minimal radiologist input. Additionally, Pina's study \cite{pina2024unsupervised} addresses the challenges of digital pathology image analysis for breast cancer, focusing on the variability in histopathological slides and differences in staining techniques. The manual annotation of large datasets across different stains (Estrogen receptor, progesterone receptor, human epidermal growth factor receptor 2, and Ki-67 protein) is highly labor-intensive. To overcome this, the study applies \ac{USL} combined with domain adaptation techniques and \ac{ViT} to enhance cell detection tasks. By leveraging adversarial feature learning, the research improves pipelines for \ac{CAD}, significantly boosting diagnostic accuracy across various staining methods.  Moving forward, the study \cite{jung2024expansive} presents a comprehensive exploration of \ac{USL} techniques, which were employed using topic modeling methods such as \ac{LDA}, \ac{NMF}, and \ac{CMT}. These unsupervised methods were applied to analyze and categorize large volumes of text data from the Web of Science and LexisNexis concerning discussions on \ac{LLM}s. The use of \ac{USL} enabled the study to evaluate topic coherence and diversity across different models, with BERTopic emerging as a top performer in both metrics \cite{grootendorst2022bertopic}.

\begin{table}[ht!]
\caption{Summary of classification examples employing various training modes in different medical domains.}
\label{table:modes}
\centering
\scalebox{0.85}{
\begin{tabular}{lllllm{3cm}lcccccc}
\hline
\multicolumn{1}{c}{\multirow{2}{*}{Ref.}} & 
\multicolumn{1}{c}{\multirow{2}{*}{Year}} & 
\multicolumn{1}{c}{\multirow{2}{*}{C}} & 
\multicolumn{1}{c}{\multirow{2}{*}{Algo.}} & 
\multicolumn{1}{c}{\multirow{2}{*}{Disease}} & 
\multicolumn{1}{c}{\multirow{2}{*}{Dataset}} & 
\multicolumn{1}{c}{\multirow{2}{*}{Modality}} & 
\multicolumn{6}{c}{{Metrics}} \\ 
\cline{8-13} 
& & & & & & & {Acc.} &  {F1} & {FPR} & {AUC} & {Pre.} & {Sens.} \\ 
\hline 
\cite{an2024transformer} & 2024 & DSL & ViT & Lung Cancer & TCGA-LUAD & Whole Slide Images & {96.90} & -- & -- & {99.60} & -- & -- \\ 

\cite{sushil2024comparative} & 2024  & DSL & \ac{LLM} & Breast  cancer & BreaKHis& Pathology Images & {--} & {83.00} & {--} & {--} & {--} & {--}\\

\cite{silva2021egfr} & 2021  & DUSL & \ac{TL} & Lung cancer &LIDC-IDRI & Chest CT Images & {--} & {--} & {--} & {68.00} & {--} & {--}\\

\cite{ferber2024context} & 2024  & DSL & \ac{LLM} & Multiple cancers & TCGA \newline PatchCamelyon & Histopathology & {90.00} & {--} & {80.00} & {--} & {--} & {--}\\

\cite{huang2022self} & 2023  & DSL & \ac{TL} & Lung cancer &LIDC-IDRI& DICOM & {96.00} & {--} & {--} & {95.84} & {--} & {--}\\

\cite{dadgar2022comparative} & 2022  & DSL & \ac{TL} & Lung cancer &LIDC-IDRI, LUNA16 & CT Scans & {91.10} & {81.50} & {--} & {95.80} & {84.9} & {--}\\

\cite{shaikh2020transfer} & 2020  & DSL & \ac{TL} & Breast cancer &INbreast, DDSM & Mammography & {89.77} & {88.91} & {--} & {--} & {--} & {83.78}\\

\cite{zhang2023pseudo} & 2023  & DSL & \ac{FL} & Multiple Cancers & Histopath. images &JPEG, PNG & {96.66} & {96.64} & {--} & {--} & {97.14} & {--}\\

\cite{liu2023federated} & 2023  & DSL & \ac{FL} & Lung Cancer & LUNA16 &CT Scans & {83.41} & {83.40} & {--} & {88.38} & {83.41} & {83.38}\\

\cite{kumbhare2023federated} & 2023  & DSL & \ac{FL} & Breast Cancer & VINDR-MAMMO, \newline  CMMD,  INbreast &Mammography & {95.00} & {--} & {--} & {--} & {--} & {--}\\

\cite{garia2023vision} & 2023  & DSL & \ac{ViT} & Breast  cancer &Thermal dataset & Thermal Images & {95.78} & {--} & {--} & {--} & {--} & {--}\\

\cite{shi2021semi} & 2021  & DSSL & \ac{TL} & Pulmonary nodules &LUNA16  & Chest CT Images & {88.30} & {--} & {--} & {91.00} & {--} & {--}\\

\cite{thungprue2022using} & 2023  & DSSL & \ac{TL} & Skin cancer &Custom dataset  & 2D Dermoscopy & {98.00} & {98.00} & {--} & {--} & {--} & {98.00}\\

\cite{leung2024deep} & 2023  & DSSL & \ac{TL} & Prostate cancers &Custom PET/CT dataset  & 3D PET/CT & {83.00} & {--} & {--} & {86.00} & {--} & {--}\\

\cite{wang2022semi} & 2024  & DSSL & \ac{ViT} & Breast cancer &BreakHis  & Histopath. Images & {98.12} & {98.41} & {--} & {--} & {98.17} & {--}\\

\hline
\end{tabular}
}
\begin{flushleft}
\small Abbreviations: Category (C); 
\end{flushleft}
\end{table}

\subsection{Deep \ac{SSL}}

\begin{figure*}[ht!]
\begin{center}
\includegraphics[width=1\columnwidth]{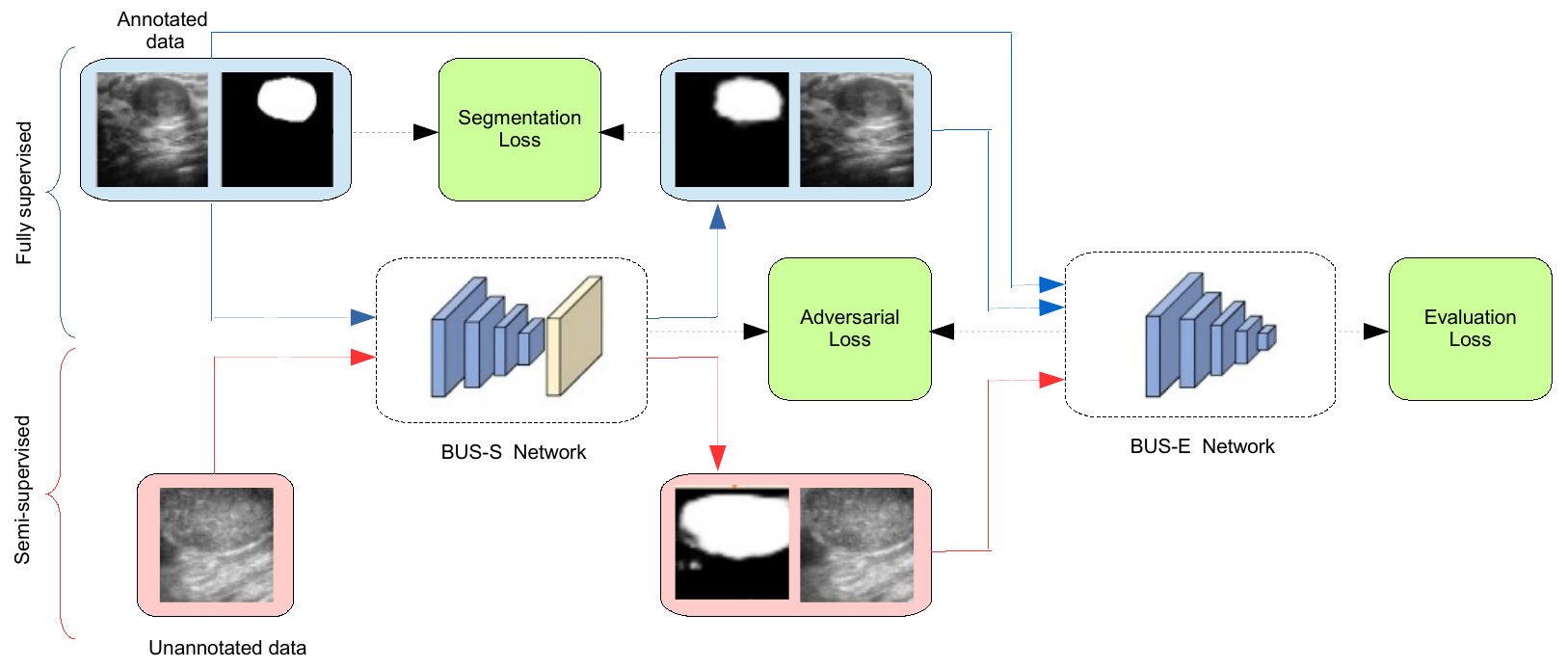}
\end{center}
\caption{An example on recognition of breast cancer with \ac{SSL} \cite{han2020semi}. }
\label{fig8}
\end{figure*}

Deep \ac{SSL} is a strategy that merges the strengths of \ac{SL} and \ac{USL}. In \ac{SL}, models are trained with labeled data, while \ac{USL} involves training with unlabeled data. Deep \ac{SSL} use a mix of a small set of labeled data with a significantly larger batch of unlabeled data, enhancing the training process under conditions where labeled data is scarce or costly to acquire. This hybrid approach allows the labeled data to steer the learning while the unlabeled data aids in discerning broader data attributes.  However, several challenges impact its effectiveness: (1) \textit{Data quality and quantity}: The efficacy of a deep \ac{SSL} model is highly reliant on the volume and quality of the available unlabeled data. Insufficient or non-representative unlabeled data can degrade model performance. (2) \textit{Data distribution:} Performance can also suffer from disparities in the distribution between labeled and unlabeled data, especially if labeled data represents limited classes. (3) \textit{Hyperparameter selection:} Deep \ac{SSL} requires careful tuning of numerous hyperparameters such as the amount of labeled versus unlabeled data, the strength of regularization, and learning rates, which can be complex and time-consuming to optimize. (4) \textit{Model interpretability:} The complexity of \ac{DNN} architectures in deep \ac{SSL} makes it challenging to interpret how decisions are made or to pinpoint which data features are most influential in those decisions. (5) \textit{Adversarial vulnerability:} Deep \ac{SSL} models are susceptible to adversarial attacks, where manipulated inputs can lead the model to incorrect outputs \cite{yang2022survey}. Figure \ref{fig8} illustrates a colorectal cancer study where \ac{SSL} and \ac{SL} are applied to labeled and unlabeled image patches from 70\% of Dataset-PATT, generating models.  Patient-level tests and human-AI competitions classify subjects as cancerous if clusters of positive patches are detected in whole slide images.

The study \cite{shi2021semi}, proposes SDTL framework for diagnosing benign and malignant pulmonary nodules. By utilizing \ac{TL} and an iterated feature-matching-based \ac{SSL} method, the model benefits from both a pre-trained classification network and a large dataset of unlabeled nodules. The SDTL framework gradually incorporates unlabeled samples to optimize the classification network. Experimental results show that \ac{TL} and \ac{SSL} significantly improve diagnostic accuracy, highlighting the framework's potential as an effective tool in clinical practice for lung nodule diagnosis. In addition, Bdair et al. \cite{bdair2021fedperl}, introduces FedPerl, a semi-supervised \ac{FL} method for skin cancer detection that addresses the challenge of limited annotated data. FedPerl integrates peer learning and ensemble averaging to improve pseudo label accuracy by fostering collaboration among distributed clients. Unlike previous methods, \cite{barata2023reinforcement} introduces DGARL, a novel approach for end-to-end semi-supervised segmentation of medical images, including cancer detection. DGARL combines deep \ac{RL} with \acp{GAN}, improving both tumor detection and segmentation tasks simultaneously. This method incorporates a task-joint \ac{GAN} with two discriminators to link detection outcomes with segmentation performance, enabling mutual optimization. Furthermore, a bidirectional exploration \ac{RL} technique is employed to address challenges associated with unlabeled data. Experiments conducted on datasets of brain, liver, and pancreas tumors demonstrate that DGARL enhances segmentation accuracy, underlining its potential effectiveness in cancer diagnosis. Wang et al. \cite{wang2022semi} propose a \ac{SSL} framework for breast cancer detection using the \ac{ViT}, which has demonstrated superior performance compared to traditional \acp{CNN} across various tasks. While \acp{CNN} have been extensively studied for breast cancer detection, the use of \ac{ViT} in this domain has been relatively limited. Nonetheless, validation on ultrasound and histopathology datasets reveals that this method consistently outperforms \ac{CNN} baselines across multiple tasks. Kumari et al. \cite{kumari2024leveraging} introduce \ac{LLM}-SegNet, a \ac{SSL} model for 3D medical image segmentation, including cancer imaging, that reduces the reliance on extensive voxel-level annotations. By incorporating a \ac{LLM} into its co-training framework, \ac{LLM}-SegNet improves learning from unannotated samples, enhancing the model's ability to identify cancerous regions.  Experimental results on multiple datasets demonstrate that \ac{LLM}-SegNet outperforms existing models in terms of segmentation accuracy.

\section{Advanced DL techniques}
\label{sec4}

{To improve cancer diagnosis, various advanced \ac{DL} techniques have been integrated into medical applications. This section summarizes key proposed techniques, emphasizing their effectiveness. Table~\ref{tab:dl_comparison} presents an overview, detailing their advantages and specific applications in cancer detection.}

\begin{table*}[ht!]
\centering
\caption{{Comprehensive comparison of advanced \ac{DL} techniques for cancer diagnosis.}}
\label{tab:dl_comparison}
\scalebox{0.8}{
\begin{tabular}{m{1cm}m{2.5cm}m{4cm}m{3cm}m{3cm}m{3cm}m{3cm}}
\hline
\textbf{{DL method}} & \textbf{{Scalability}} & \textbf{{Real-world implementation}} & \textbf{{Computational complexity}} & \textbf{{Accuracy}} & \textbf{{Privacy considerations}} & \textbf{{Research gaps}} \\
\hline
{\textbf{\Ac{RL}}} & {\textbf{Moderate} (limited scalability due to need for interactions with environment and reward structure complexity)} & {Used for tumor segmentation, automated lesion localization, and personalized treatment planning by continuously refining predictions based on rewards \cite{djeffal2023automatic}.} & {\textbf{High} (trial-and-error learning requires significant computational resources)} & {\textbf{High} (if well-defined reward functions are used)} & {\textbf{Low} (centralized data required for training)} & {Limited application in cancer detection; requires large datasets for effective learning.} \\
\hline
{\textbf{\Ac{FL}}} & {\textbf{High} (scalable across institutions without data sharing)} & {Trains models across decentralized datasets from different hospitals, enhancing generalizability while ensuring patient data confidentiality \cite{habchi2024machine}.} & {\textbf{Moderate} (network bandwidth and encryption overhead affect performance)} & {\textbf{Comparable} to centralized models (when sufficient data diversity exists)} & {\textbf{High} (data remains decentralized, improving security)} & {Model divergence, communication delays, and potential security risks in distributed learning.} \\
\hline
{\textbf{\Ac{TL}}} & {\textbf{High} (leverages pre-trained models for adaptation)} & {Fine-tunes pre-trained models on cancer imaging datasets, improving accuracy with minimal labeled data \cite{panagoulias2024evaluating}.} & {\textbf{Lo}w (reduces training time by reusing pre-trained models)} & {\textbf{High} (if source and target domains are well-matched)} & {\textbf{Moderate} (requires labeled data, but training can be privacy-preserving)} & {Domain shift challenges when applied to different datasets.} \\
\hline
{\textbf{Transf.}} & {\textbf{High} (well-suited for large-scale datasets)} & {Improves feature extraction in cancer histopathology, radiology, and genomics by leveraging self-attention mechanisms \cite{kheddar2024Transformers}.} & {\textbf{Very High} (self-attention computations are expensive)} & {\textbf{High} (state-of-the-art performance in medical imaging tasks)} & {\textbf{Low} (requires centralized data for training)} & {Overfitting in small medical datasets; high resource requirements.} \\
\hline
{\textbf{\Ac{LLM}}} & {\textbf{Moderate} (requires fine-tuning for specific medical applications)} & {Assists in clinical decision-making by analyzing patient histories, summarizing pathology reports, and automating medical literature reviews \cite{johnson2019survey}.} & {\textbf{Very High} (pretraining and inference are computationally demanding)} & {\textbf{High} (excellent for text-based analysis, but needs domain adaptation)} & {\textbf{Low} (requires centralized data for pretraining, posing privacy risks)} & {Ethical concerns, hallucination issues, and regulatory challenges in healthcare applications.} \\
\hline
\end{tabular}}
\end{table*}

\subsection{Reinforcement learning}

\ac{RL} is a \ac{ML} paradigm in which an agent learns to make decisions by interacting with an environment, aiming to maximize cumulative rewards over time. The agent selects actions based on a policy (\(\pi\)) that maps states (\(s\)) to actions (\(a\)), receiving feedback in the form of rewards (\(r\)) from the environment. \ac{RL} problems are typically modeled as Markov decision processes (MDPs), where the agent's objective is to learn an optimal policy that maximizes long-term returns. Various types of \ac{RL} have been explored in the literature (Figure \ref{fig9}), with a summary of their key findings presented in Table \ref{table:6}. While most \ac{RL} approaches have already been reviewed in \cite{kheddar2024reinforcement}, the most commonly employed types in cancer diagnosis include:

\begin{figure}[ht]
\begin{center}
\includegraphics[width=0.5\columnwidth]{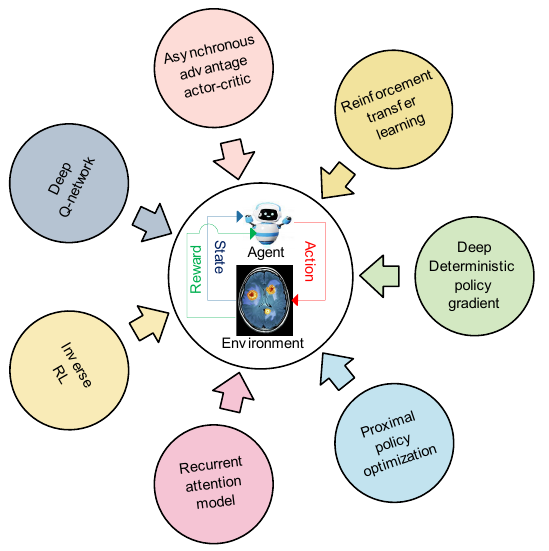}
\end{center}
\caption{\ac{RL} types used in cancer diagnosis. DQN \cite{khajuria2024active, luo2023lung, dahdouh2023new};  RAN \cite{pesce2019learning};  PPO \cite{zhao2023rlogist}; DDPG \cite{kumar2023attention};   IRL \cite{petousis2018generating,petousis2019using}; A3C \cite{renith2024automated};    RTL \cite{arora20243d,xu2023spatiotemporal}. }
\label{fig9}
\end{figure}

\subsubsection{Q-learning {and DQN}} Aims to find the optimal action-selection policy by maximizing cumulative rewards in a given environment. However, Q-learning struggles to handle large state spaces where storing a Q-table becomes inefficient. To address this, the \ac{DQN} algorithm extends Q-learning by using neural networks to approximate the Q-values, solving problems with large, discrete action spaces. \ac{DQN} introduces experience replay and target networks, which decouple the target and learned Q-values, stabilizing learning. This approach is particularly useful in environments where exploration is critical. The key functions associated with the \ac{DQN} are  \cite{barata2023reinforcement,dahdouh2023new}: (i) The Q-function approximates the expected cumulative reward of taking an action $a$ in a given state $s$ and following the policy afterward, \(\displaystyle Q(s, a) = \mathbb{E}[R_t | s_t = s, a_t = a] \), where, $Q(s, a)$ is the Q-value (action-value), $s$ is the state, and $a$ is the action and $R_t$ is the cumulative reward from time $t$ onward. (ii) The update rule for \ac{DQN} based on the Bellman equation is:
\begin{equation}
Q(s, a) \leftarrow Q(s, a) + \alpha \left( r + \gamma \max_{a'} Q(s', a') - Q(s, a) \right)
\end{equation}

\noindent Where $r$ is the reward received after taking action $a$ in state $s$, $s'$ is the next state and $\gamma$ is the discount factor, and $\alpha$ is the learning rate.  (iii) The loss function minimizes the difference between the target and predicted Q-values:
\begin{equation}
L(\theta) = \mathbb{E} \left[ \left( r + \gamma \max_{a'} Q(s', a'; \theta^{-}) - Q(s, a; \theta) \right)^2 \right]
\end{equation}

Where, $\theta$ and $\theta^{-}$ represent the parameters of the Q-network and target Q-network. Many studies related to \ac{DQN}-based cancer detection have been proposed in the literature. For example, \cite{khajuria2024active} focuses on enhancing the localization of malignant cervical cells, a key aspect of detecting tumors. It employs a \ac{DQN} algorithm  to fine-tune bounding boxes around cancerous nuclei to improve localization accuracy through reward-based reinforcement. This method addresses \ac{DL} overfitting issues by incorporating randomness, leading to successful and competitive localization performance compared to existing techniques. Similarly,  \cite{luo2023lung} addresses the inefficiencies of manual lung nodule detection methods and proposes the LLC-QE model, which integrates ensemble learning with \ac{DQN}. The model is pre-trained using the artificial bee colony algorithm to avoid local optima. It employs multiple \ac{CNN}s to extract and combine feature vectors for classification. Trained on the LIDC-IDRI dataset, the model tackles dataset imbalance by using \ac{RL} to prioritize accurate classification of underrepresented classes. Moving on,  Dahdouh in \cite{dahdouh2023new}, proposes an advanced model for early skin cancer detection by integrating \ac{DL} and \ac{DQN}. The approach uses the watershed algorithm for segmentation to isolate affected areas. A deep \ac{CNN} classifies the lesions into seven categories: actinic keratosis, basal cell carcinoma, benign keratosis, dermatofibroma, melanocytic nevi, melanoma, and vascular skin lessions. The model is further refined using the \ac{DQN} algorithm, which optimizes performance through \ac{RL}. Likewise, Tao \cite{tao2022seqseg}, contributes by proposing SeqSeg framework, a novel method for reliable carcinoma segmentation in \ac{MRI}. SeqSeg addresses background dominance issues at two scales: instance level and feature level. It uses a \ac{DQN}-based model to focus attention on the tumor and reduce segmentation background scale, and employs high-level semantic features to guide \ac{FL}. Evaluated on a large dataset, SeqSeg outperforms state-of-the-art methods and demonstrates superior performance across multi-device and multi-center datasets. Praneeth et al.  \cite{praneeth2022rl2ndgsnet}, presents RL2NdgsNet, a \ac{DL} network enhanced by \ac{RL} for diagnosing mediastinal lymph nodes and distinguishing between benign and malignant cases.  The proposed approach leverages radiological modalities like X-ray, ultrasound, \ac{CT}, and \ac{MRI}, which are non-invasive and painless. The RL2NdgsNet network incorporates a custom \ac{DQN} policy to optimize its performance. Various state-of-the-art \acp{AF} and exploration fractions were tested to refine the network’s capabilities. The results demonstrate that the RL2NdgsNet achieves superior diagnostic performance, improving upon existing \ac{DL}-based methods for medical imaging and offering a promising alternative to traditional invasive procedures. The reference \cite{liu2019deep} explores the potential of deep \ac{RL} in lung cancer detection, emphasizing its integration with medical big data from the medical \ac{IoT}. This work discusses the use of \ac{DQN} and its variants, such as double \ac{DQN} and hierarchical \ac{DQN}. \\

\subsubsection{\Ac{PPO}}  Is another popular type of \ac{RL} algorithm designed to balance exploration and exploitation. It improves training stability by updating policies using clipped objective functions within a trust region to avoid large updates, ensuring smoother policy updates. The key functions involved in \ac{PPO} are: (i) The policy $\pi_\theta(a | s)$ represents the probability of taking action $a$ given state $s$, parameterized by $\theta$. (ii) The advantage function $\hat{A}_t$ which represents how much better a particular action is compared to the average action taken at state $s_t$ where:
\begin{equation}
    \hat{A}_t = R_t - V(s_t)
\end{equation}

\noindent The $R_t$ is the cumulative reward at time step $t$ and $V(s_t)$ is the value function, representing the expected cumulative reward from state $s_t$. The objective is to maximize:

\begin{equation}
L(\theta) = \mathbb{E}_t \left[ \min \left( r_t(\theta) \hat{A}_t, \text{clip}(r_t(\theta), 1 - \epsilon, 1 + \epsilon) \hat{A}_t \right) \right]
\end{equation}

\noindent Where $r_t(\theta)$ is the probability ratio of the new policy to the old policy, $\hat{A}_t$ is the advantage estimate at time step $t$, $\epsilon$ is a small hyper-parameter that defines the clipping range and $\text{clip}(r_t(\theta), 1 - \epsilon, 1 + \epsilon)$ limits the policy update to be within a trusted region.\\

Within the class of \ac{PPO}-based cancer detection  methods, the study \cite{balaprakash2019scalable} introduces a \ac{RL}-based neural architecture search method to automate the development of \ac{DL} models for cancer data. This approach streamlines the creation of predictive models by incorporating domain-specific characteristics and reducing reliance on manual trial-and-error methods. Custom building blocks are designed to cater to the specific needs of cancer data, leading to the discovery of \ac{DNN} architectures with fewer trainable parameters and shorter training times, while achieving comparable or superior accuracy to manually designed models. Similarly, \cite{zhao2023rlogist} presents RLogist, a deep \ac{RL} method based on the \ac{PPO} algorithm designed to improve the efficiency of whole-slide image analysis in computational pathology. Unlike traditional methods that require extensive sampling of high-magnification patches, RLogist emulates the diagnostic process of human pathologists to strategically identify valuable regions for observation. This approach reduces the need for dense, high-magnification patch analysis by learning to select representative features from multiple resolution levels. Evaluated on tasks such as detecting metastases in lymph node sections and subtyping lung cancer, RLogist demonstrates competitive performance and provides interpretable decision-making pathways, potentially offering educational and assistive benefits for pathologists.

\subsubsection{\Ac{DDPG}} 

Is a type of \ac{RL} used to handle problems with continuous action spaces. It is an \ac{AC} method, which means it combines two networks: an actor network that suggests actions and a critic network that evaluates them. \ac{DDPG} is an off-policy algorithm, meaning it learns the value function from actions taken by a behavior policy different from the target policy \cite{usmani2021reinforcement}. {DDPG consists of two neural networks:}

\noindent \textbf{(i) Actor network:}  outputs a deterministic action given the state.  The deterministic policy function is defined as: \(\displaystyle a_t = \mu(s_t | \theta^\mu) \)   where $s_t$ is the current state, $\mu(s_t)$ is the learned policy, and $\theta^\mu$ are the parameters of the actor network. The actor network is updated using the \ac{DPG}:  
\begin{equation*}
    \nabla_{\theta^\mu} J \approx \mathbb{E} \left[ \nabla_a Q(s, a | \theta^Q) |{a = \mu(s)} \nabla{\theta^\mu} \mu(s | \theta^\mu)\right]
\end{equation*}

This update ensures that the actor selects actions that maximize the critic’s estimated Q-values. 

\noindent\textbf{(ii) Critic network:}  estimates the Q-value (expected return) for state-action pairs. The action-value function is approximated as: 
\begin{equation} Q(s, a | \theta^Q) = \mathbb{E} \left[ r_t + \gamma Q(s_{t+1}, \mu(s_{t+1} | \theta^\mu) | \theta^Q) \right] \end{equation} 

where, $r_t$ is the reward at time $t$,
$\gamma$ is the discount factor, and
$\theta^Q$ are the parameters of the critic network. The loss function for updating the critic network is:  
\begin{equation}
L(\theta^Q) = \mathbb{E} \left[ (y_t - Q(s_t, a_t | \theta^Q))^2 \right] , \hspace{1cm}   
\end{equation}

where \(\displaystyle  y_t = r_t + \gamma Q(s_{t+1}, \mu(s_{t+1} | \theta^\mu) | \theta^Q) \displaystyle\). This minimizes the error between the predicted Q-value and the target Q-value. \textcolor{black}{As an example,  \cite{usmani2021reinforcement}  employed \ac{DDPG} as \ac{RL} method for skin lession segmentation. The method mimics physicians’ delineation of \ac{ROI}, training an agent to refine segmentation through continuous actions using the \ac{DDPG} algorithm.}

\subsubsection{Other RL methods}

 Many other types of \ac{RL} methods have been investigated in the field of cancer detection, including:  

 \begin{itemize}

\item \textbf{\Ac{A3C}:} optimizes policy parameters using the advantage function, which reduces variance in gradient estimation. The policy gradient update is given by:
\begin{equation} \nabla_{\theta} J = \mathbb{E} \left[ \nabla_{\theta} \log \pi_{\theta}(a|s) A(s,a) \right] \end{equation} 
where $A(s,a) = Q(s,a) - V(s)$ is the advantage function, and $\pi_{\theta}(a|s)$ is the stochastic policy parameterized by $\theta$. Unlike standard \ac{AC} methods, \ac{A3C} runs multiple agents asynchronously, reducing training time and preventing correlation in training samples. For example,  \cite{renith2024automated}  addresses the challenge of skin cancer diagnosis by developing an automated system using a novel deep \ac{RL} technique based on asynchronous advantage \ac{A3C}. This assumes multiple independent \ac{CNN} agents that interact. These agents interact with skin images to perform segmentation, guided by policies that maximize rewards and minimize errors. 

\item \textbf{\Ac{RAN}:} integrates attention mechanisms (cf. Section \ref{sec4.5}) into \ac{RL}, where an agent sequentially focuses on informative regions of an observation, and it is formulated as:
\begin{equation} g_t = f_g(s_t, l_t; \theta_g) \end{equation} 
where $g_t$ represents the extracted features at time step $t$, $s_t$ is the state, $l_t$ is the attention location, and $\theta_g$ are the learnable parameters. The agent optimizes its policy to select $l_t$ such that the accumulated reward is maximized. This is particularly effective in image-based RL tasks, reducing computation by selectively processing high-information regions. For example, \cite{pesce2019learning} employed \ac{RAN} to enhance cancer detection by guiding neural networks to focus on lesions in chest radiographs. \ac{RL} penalizes irrelevant areas, while attention improves lesion localization for accurate classification.

\item \textbf{\Ac{RTL}:} 
 Leverages pre-trained policies to accelerate learning in new environments. The knowledge transfer objective is:
 \begin{equation} J_{\text{TL}} = \mathbb{E}{s,a} \left[ \lambda Q_{\text{source}}(s,a) + (1-\lambda) Q_{\text{target}}(s,a) \right] \end{equation} 
where $Q_{\text{source}}(s,a)$ is the action-value function from a pre-trained model, and $Q_{\text{target}}(s,a)$ is the value function for the new task. The parameter $\lambda$ controls the influence of the transferred knowledge. This approach is effective in robotics, where motor skills learned in simulated environments can be transferred to real-world applications.
\textcolor{black}{For instance, Xu et al. \cite{xu2023spatiotemporal} have employed  \ac{RTL} to to improve the quality of \ac{MRI} images and extract significant tumor features. The aim is to   improve the  early detection of brain tumors through the use of advanced \ac{DL} networks for segmentation and classification.  The proposed method employed a specific type of \ac{RL} called \ac{AC}, with  U-Net and ResNet architectures for multi-classification of brain tumors. This approach leverages these models to extract significant features from \ac{MRI} slices, thereby improving diagnostic accuracy and efficiency.}

\item \textbf{\Ac{IRL}:}
Infers a reward function $R(s,a)$ from expert demonstrations. The optimal reward function is estimated by solving:
\begin{equation} \max_{R} \sum_{t} \gamma^t R(s_t, a_t) - \lambda ||\nabla R||^2 \end{equation} 
where $\gamma$ is the discount factor, and $\lambda ||\nabla R||^2$ is a regularization term to prevent overfitting. \ac{IRL} enables learning human-like policies by inferring implicit goals from observed expert behavior, making it a powerful tool in imitation learning applications. In \cite{petousis2019using}, \ac{IRL} enhances cancer screening by deriving reward functions from expert decisions. Utilizing maximum entropy \ac{IRL}, partially observable \ac{MDP} models optimize screening strategies, achieving expert-level recommendations for breast and lung cancer detection. Furthermore, \ac{IRL} improves specificity, reduces false positives, and enhances early cancer detection while maintaining expert-level accuracy.

 \end{itemize}

\begin{center}
\scriptsize
\begin{longtable}{m{0.5cm} m{1.5cm} m{2.5cm} m{5cm} m{1.5cm} m{4.5cm}}
\caption{Summary of \ac{RL} in cancer diagnosis.}
\label{table:6} \\
\hline
\textbf{Ref} & \textbf{Model(s) used} & \textbf{Image dataset} & \textbf{Contribution} & \textbf{Best result (\%)} & \textbf{Limitation} \\
\hline
\endfirsthead

\hline
\textbf{Ref} & \textbf{Model(s) Used} & \textbf{Image dataset} & \textbf{Contribution} & \textbf{Best result (\%)} & \textbf{Limitation} \\
\hline
\endhead

\cite{luo2023lung}& RL-EL& Lung nodule &Combines \ac{RL} and \ac{EL} improve for lung nodule detection.&F1: 89.80 \newline {Pre: 87.70} &Increased computational demands limit scalability.\\ 

\cite{dahdouh2023new} &CNN-DQN &HAM10000  &Combining \ac{CNN} and \ac{DQN} for accurate skin cancer classification.& Acc:  80.00&Reliance on one dataset limits model's coverage.\\ 

\cite{kumar2023attention}& Attention-RL &Skin lesion &\ac{RL} for skin lesion segmentation with attention mechanism.& Acc:  97.10&Computational complexity of attention-guided model.\\ 

\cite{renith2024automated}& AC-CNN& PH2, ISIC &Deep \ac{RL} for skin cancer localization.& Acc: 98.80&Increased computational complexity with \ac{CNN} agents.\\ 

\cite{arora20243d}& RL-DBN& 3D Brain tumor &Automated brain tumor segmentation using \ac{RL} models.& Acc:  97.57 \newline {Pre: 97.15} \newline {AUC: 69.59} &Preprocessed data may miss tumor variations.\\ 

\cite{xu2023spatiotemporal}&SKT-RL & Liver tumor &Liver tumor detection without contrast agents using \ac{SKT}-\ac{RL}.& Acc:  97.35\newline {Rec: 74.60}\newline {Spe: 98.44}&High computational resources required.\\ 

\cite{tao2022seqseg}& DQL-RANet& nasopharyngeal carcinoma  &SeqSeg framework for carcinoma segmentation.&Dice: 80.32 \newline {Rec: 87.57}&High computational complexity of SeqSeg.\\ 

\cite{praneeth2022rl2ndgsnet}& DL-RL& Lymph nodes&\ac{RL} for benign and malignant lymph node diagnosis.& Acc:  98.20 \newline {Sen: 98.03} \newline {AUC: 98.19}&Complexity of \ac{RL} increases demands.\\

\cite{usmani2021reinforcement}& DDPG &ISIC, PH2, HAM10000 & Deep \ac{RL} for skin lesion segmentation.& Acc: 96.33\newline {Spe: 98.60} \newline {Sen: 96.79}&Requires accurate annotations and high computational demands.\\

\cite{maicas2019pre}&RL-Attention  & Breast \acs{DCE}-MRI&Post-hoc breast cancer screening using \ac{RL}.&AUC: 91.00&Challenges with convergence and dataset consistency.\\ 

\cite{huang2022extracting}&Customized reward &Breast ultrasound video &\ac{RL} for keyframe extraction in breast ultrasound videos.& AUC: 84.15 &Requires extensive data for class imbalance.\\

\cite{balajee2023pulmonary}& TL-RL& Lung cancer &Adaptive \ac{RL} for early lung cancer detection.& Acc: 92.00 &Substantial computational resources required.\\ 

\cite{zhang2023detection}& TL-RL& Thyroid nodule &\ac{RL} for thyroid nodule feature extraction and localization.&Acc: 98.32 \newline {Rec: 93.84}&Complexity and computational constraints for implementation.\\ 

\cite{narad2023efficient} & DQN-CNN &Lung cancer& DQN for non-small cell lung cancer prognosis.& Pre: +12.5 & Computational complexity of integrating multiple techniques.\\

\cite{thakur2024reinforcement}& Attention-CNN-RL & Breast cancer &\ac{RL}-based CNN for early breast cancer detection.& Acc:  99.35 \newline {FNR: 0.34}  &Complexity limits practical clinical implementation.\\

\hline
\end{longtable}
\end{center}

\subsection{Federated learning}

\ac{FL} is a distributed \ac{ML} paradigm that allows multiple decentralized devices (clients) to collaboratively train a shared model while retaining their data locally, thereby enhancing privacy preservation. A central server orchestrates the training process by aggregating updates from individual clients without directly accessing their data \cite{himeur2023federated}. \ac{FL} encompasses various approaches, such as \ac{HFL}, \ac{VFL}, and \ac{FSSL}. The overall \ac{FL} process can be formally expressed as:

\begin{equation}
w^{t+1} = \sum_{i=1}^{N} \frac{|D_i|}{\sum_{j=1}^{N} |D_j|} w_i^t
\end{equation}

\begin{figure}[ht]
\begin{center}
\includegraphics[width=1\columnwidth]{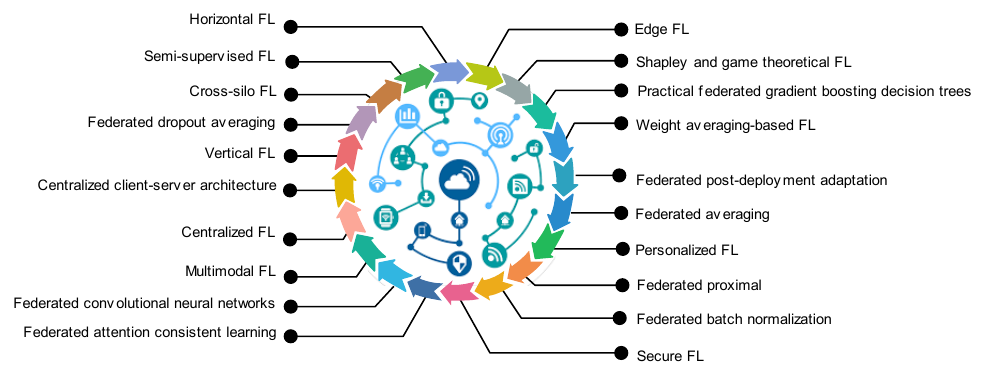}
\end{center}
\caption{\ac{FL} types used in the diagnosis of cancer. Cross-silo \ac{FL} \cite{heidari2023new,  agbley2023federated}; FSSL \cite{bdair2021fedperl};  HFL \cite{lan2022many}; VFL \cite{mohammed2023federated}; Centralized client-server architecture \cite{kumbhare2023federated};  Centralized FL \cite{omran2023detecting}; Federated dropout averaging \cite{gunesli2023federated}; Federated averaging  \cite{alsalman2024federated};  Federated-\ac{CNN} \cite{jindal2023modernizing};   Practical federated gradient boosting decision trees \cite{vibith2023gbdtmo}; Shapley and game theoretical \ac{FL} \cite{supriya2024breast}; Weight averaging-based \ac{FL} \cite{salma2022privacy}; Edge \ac{FL} \cite{kanjula2023edge}; Federated post-deployment adaptation \cite{wagner2023post}; personalized FL \cite{ayekai2023personalized};  Federated proximal \cite{gad2023novel}; Secure \ac{FL} \cite{lessage2024secure}; Federated attention-consistent learning \cite{kong2024federated}.}
\label{fig10}
\end{figure}

Where $w^{t+1}$ denotes the updated global model parameters after the aggregation in the $(t+1)$-th round, $w_i^t$ stands for local model parameters from client $i$ after local training at round $t$ and $|D_i|$ is the number of data samples at client $i$, and this determines the weight of each client's contribution to the global model. Figure \ref{fig10} summarizes the most commonly used \ac{FL} techniques in the context of cancer detection, including both well-established methods from the literature and approaches specific to certain schemes.

\begin{figure}[tbph]
\begin{center}
\includegraphics[width=0.6\columnwidth]{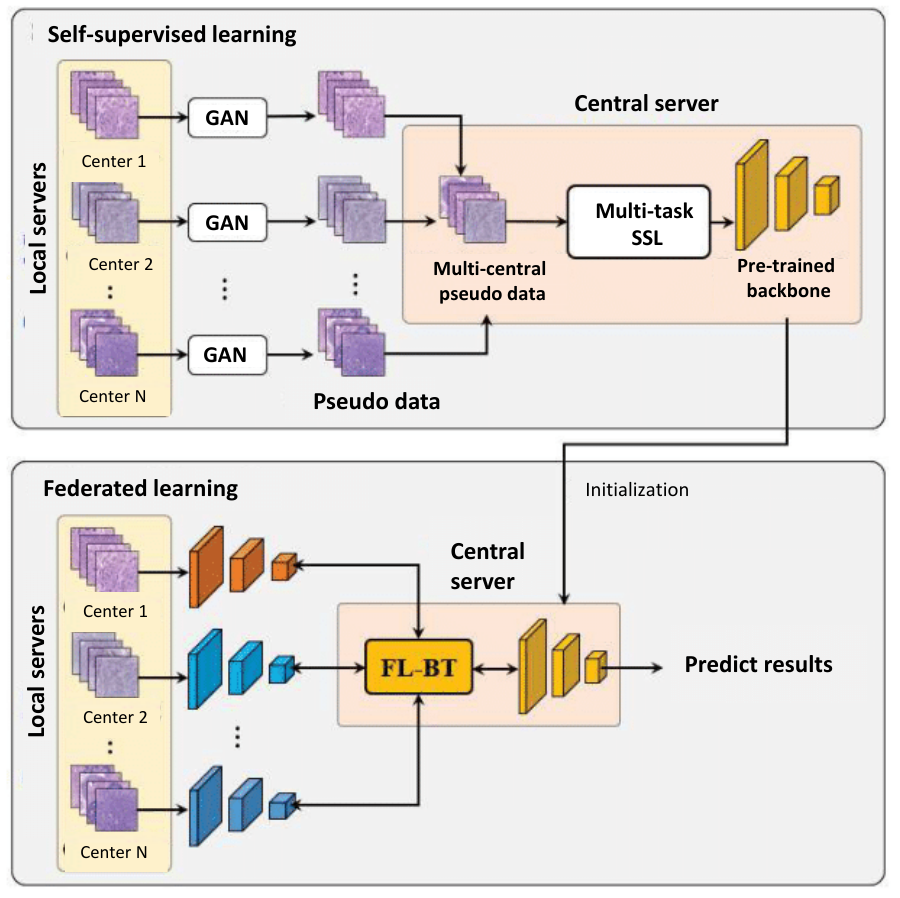}
\end{center}
\caption{{The proposed SSL-FL-BT classification pipeline is designed for histopathological image analysis, which is widely used in cancer diagnosis. The scheme consists of two main stages: the \ac{SSL} stage and the FL stage. During the \ac{SSL} stage, a specially designed multi-task \ac{SSL}  approach is applied to all pseudo images to pre-train the backbone network. In the \ac{FL} stage, the pre-trained backbone serves as the initialization network for the FL-BT model \cite{zhang2023pseudo}.}}
\label{figFL}
\end{figure}

\ac{HFL}, also known as sample-based federated learning, is a type of \ac{FL} where the participants (e.g., clients or institutions) have datasets that share the same features (i.e., the same feature space) but represent different individuals or entities (i.e., different samples). In this scenario, all participants hold data about different people or entities, but the structure of the data (the features) is consistent across them \cite{liu2023federated}. However, \ac{VFL} trains the models on datasets with identical sample spaces but different feature spaces. Participants collaborate by aligning data entities and sharing encrypted model updates, ensuring privacy while leveraging complementary features for improved performance \cite{himeur2023federated}. \ac{FSSL} is a specialized variation of \ac{FL} designed to handle distributed datasets where some clients (devices or organizations) have labeled data while others have unlabeled or partially labeled data. \ac{FSSL} combines the principles of \ac{FL} and \ac{SSL} to collaboratively train models without requiring all participants to have labeled data \cite{qiu2023federated}.  {As shown in Figure \ref{figFL}, the proposed SSL-FL-BT pipeline enhances histopathological image classification by integrating \ac{SSL} with \ac{FL}. This approach improves feature extraction and model generalization, making it well-suited for cancer diagnosis. }

In the context of cancer detection, numerous \ac{FL} techniques have been recently reported. Reference \cite{salmeron2020privacy,iqbal2023privacy,zhu2022knowledge,zubair2023comparative,dagli2021proposed,yan2023pla,mostafa2023detecting,hsiao2024precision,liu2023federated} \ac{HFL} as main technique for cancer detection.  Study \cite{salmeron2020privacy} introduced a distributed learning methodology for particle swarm optimization-based fuzzy cognitive maps that prioritizes data privacy. Applied to cancer detection, this approach demonstrates improved model performance through \ac{FL}, achieving results comparable to those found in existing literature. Reference \cite{iqbal2023privacy} introduced "Skin-net," a \ac{CNN} model for skin cancer detection using progressively private  \ac{HFL}. This method ensures data confidentiality during training, achieving high performance while addressing privacy concerns in medical image analysis. \cite{zhu2022knowledge} proposed a knowledge-sharing model \ac{HFL} using a Unet-based mask generator featuring classification-guided discriminator and an adversarial network for improved detection performance of pulmonary nodules.  The paper \cite{zubair2023comparative}  addressed the challenge of accessing large datasets for disease detection while ensuring data privacy.  \ac{HFL} is proposed as an alternative to conventional methods,  using X-ray images for detecting lung cancer and tuberculosis. Reference  \cite{dagli2021proposed} presented a  \ac{HFL} approach for predicting breast cancer while ensuring data privacy. The study \cite{yan2023pla} introduced the \ac{PLA} method for breast cancer diagnosis within the framework of \ac{IoMT}. This approach combines privacy-preserving techniques, efficiency, and automation. The \ac{PLA} model utilizes a \ac{ViT} backbone optimized for \ac{IoMT} environments, ensuring lightweight classification and effective global information processing of breast cancer data. It also employs  \ac{HFL} to protect patient privacy and incorporates texture analysis as an auxiliary task alongside the primary classification task. The \ac{PLA} framework achieves impressive performance metrics. The study \cite{mostafa2023detecting}, addressed the challenges in lung cancer detection caused by fragmented patient data across various medical institutions and privacy concerns that hinder centralized data analysis.  \ac{HFL} is proposed as a solution to train models on decentralized data and leveraging \ac{TL} to set initial weights. The  approach achieved a high accuracy in detecting lung cancer from medical images. Reference  \cite{hsiao2024precision} tackled the challenge of detecting hepatocellular carcinoma by integrating 2D and 3D \ac{DL} models within  \ac{HFL} framework for accurate liver and tumor segmentation in \ac{CT} scans. Using 131 scans from the liver tumor segmentation challenge, the Hybrid-ResUNet model outperformed ResNet and EfficientNet models, achieving a high dice score and AUC. The horizontal \ac{FL} approach ensures privacy and facilitates large-scale clinical trials, addressing data imbalances and demonstrating robust local model performance. Likewise,  \cite{liu2023federated} proposed a \ac{HFL} framework combined with a ResNet18-based dual path \ac{DL} model for lung nodule detection. This approach addresses privacy issues while training a global model across multiple institutions. The dual path ResNet18 architecture improves feature extraction and detection accuracy. The method demonstrates effective lung nodule detection while preserving patient data privacy.

Reference \cite{mohammed2023federated} presents a new paradigm for cancer detection using a combination of  \ac{VFL}, \ac{AE}, and XGBoost methods within a distributed fog computing environment. This approach addresses challenges in digital healthcare such as security, execution delay, and accuracy. The proposed \ac{MCMOCL} scheme integrates multi-omics data (RNA, miRNA, and methylation) for improved cancer prediction. The method achieved high  accuracy, reduced processing delay, and enhanced security compared to existing models in heterogeneous fog cloud computing environments.

Cross-silo \ac{FL} is another popular method that boost the boundary of cancer detection. For example, Heidari et al \cite{heidari2023new} propose a novel method for lung cancer detection, addressing challenges related to data privacy and inter-hospital collaboration by employing blockchain-based cross-silo \ac{FL}.   The proposed method employs capsule networks for local lung cancer classification and introduces a data normalization technique to handle variability in \ac{CT} data.  Extensive experiments demonstrated the effectiveness of this technique, achieving a high accuracy. Similarly,  \cite{jimenez2023memory}, contributes by proposing a novel memory-aware curriculum cross-silo \ac{FL} approach for breast cancer classification using mammography images. This method addresses the challenge of imbalanced datasets, particularly the scarcity of positive samples in routine screenings, by controlling the order of training samples in the \ac{FL} setting. This method prioritizes forgotten samples to improve local model consistency and overall global model performance. Additionally, the method incorporates unsupervised domain adaptation to handle domain shifts while ensuring privacy.  Agbley et al.  This study demonstrates that the \ac{FL} model performed comparably to a centralized learning model, with only minor differences in F1-Score. Agbley et al \cite{agbley2023federated} proposed a novel approach for breast tumor classification that integrates different magnification factors of histopathological images using a residual network and information fusion in a cross-silo \ac{FL} framework. This method addresses the challenge of limited publicly available medical data by preserving privacy and enabling collaborative model training. The approach was evaluated using the BreakHis dataset, showing superior performance compared to centralized learning models. Additionally, visualizations for explainable \ac{AI} were provided, and the models are intended for deployment in healthcare institutions' \ac{IoMT} systems for timely diagnosis and treatment.

Other proposed methods are based on \ac{FSSL} for identifying cancer. For instance,   \cite{bdair2021fedperl}, suggested the FedPerl scheme , an  \ac{FSSL} method that enhances skin lesion classification using peer learning and ensemble averaging techniques. This approach addresses the challenge of limited annotated data in the medical field by allowing models to learn from each other through community-based peer learning and producing accurate pseudo labels. The peer anonymization technique preserves privacy and reduces communication costs without adding complexity.

Moreover, various aggregation techniques have been proposed and implemented to construct the global model on the central server within the \ac{FL} framework. For example,  \cite{alsalman2024federated} introduces a \ac{FedAvg} aggregation approach for breast cancer detection along with  \ac{CNN}. This method addresses privacy concerns associated with sharing patient data and the challenge of training on limited, localized datasets. The approach is tested on many large-scale datasets, achieving a high detection accuracy. The study demonstrates that FedAvg can enhance detection accuracy while preserving data privacy, offering a robust and scalable solution for breast cancer diagnostics. The work in \cite{gunesli2023federated}, introduced  FedDropoutAvg, an aggregation  approach for tumor detection in histology images that integrates dropout techniques into both client selection and federated averaging processes. This method leverages dropout to mitigate overfitting and improve model generalization. Another aggregation technique proposed in \cite{wagner2023post} employs \ac{FedPDA}, which integrates \ac{FL} to address distribution shifts in medical imaging models. This approach enables remote gradient exchange between the deployed model and source data by maximizing gradient alignment between source and target domains, facilitating more effective learning. \ac{FedPDA} customizes the model for target domains, enhancing performance in cancer metastasis detection and skin lesion classification. Addition, the study in \cite{gad2023novel} addresses the challenge of developing accurate and private \ac{AI} models for breast cancer detection using ultrasound imaging. It proposes leveraging \ac{FL} to manage sensitive medical data while preserving privacy. To enhance model performance on \ac{non-IID} datasets, the study incorporates the \ac{FedProx} aggregation method, combined with a modified U-Net model featuring attention mechanisms. \ac{FedProx} mitigates heterogeneity in \ac{FL} by introducing a proximal term into the local objective function, ensuring stability and improving convergence. This approach resulted in a global model with high accuracy for tumor segmentation, demonstrating the effectiveness of \ac{FedProx} in addressing \ac{non-IID} medical data and improving segmentation performance. Building on this,  \cite{sikandar2024integrating} explored leveraging genomic big data for stomach adenocarcinoma detection using \ac{AI} and \ac{DL}. The study introduces \textit{Fed\_ANN11}, an aggregation technique developed within a \ac{FL} framework, integrating novel feature extraction methods based on Electro-Ion interaction pseudo-potential values and Kidera factors. \textit{Fed\_ANN11} demonstrates superior performance, achieving high testing accuracies in federated environments and showing significant improvements over existing methods, all while maintaining strong data privacy and security.

Researchers have suggested methods that combine both \ac{FL} and \ac{TL} have been reported in \cite{salma2022privacy,tan2024joint}. The study \cite{salma2022privacy} addresses the challenge of prostate cancer detection by proposing a \ac{FL}-based approach that ensures data confidentiality while using weight averaging to detect anomalies. The performance of a customized basic \ac{CNN} model with three Conv2D layers, along with VGG19 and Xception models, was compared in both centralized and decentralized (federated) settings. The results demonstrate that the decentralized method achieves accuracy comparable to state-of-the-art centralized models, effectively balancing data privacy with high detection performance. Similarly,\cite{tan2024joint} report a novel \ac{FL} framework for breast cancer segmentation that addresses data inconsistencies and privacy concerns. It employs random \ac{ROI} and bilinear interpolation to augment data, and a U-Net model with a pretrained VGG backbone. The Gaussian mixture model is applied to enhance segmentation quality by managing diverse data distributions and improving tumor detection. Experiments show that this approach, using \ac{FedAvg} and federated \ac{BN}, outperforms several state-of-the-art methods on five public breast cancer datasets.

Shapley values and game theory were utilized in \cite{supriya2024breast} to develop a unique \ac{FL} approach specifically designed for breast cancer prediction. Leveraging the Wisconsin Diagnostic Breast Cancer dataset, this approach enhances privacy and prediction accuracy by identifying key features and incentivization high-performing clients.

In addition to the previously discussed types of \ac{FL}, other variations are tailored to specific applications, such as \ac{IoT}. For instance, \cite{kanjula2023edge} proposes a lightweight and scalable \ac{IoT} framework for skin cancer detection, leveraging edge \ac{FL} to enhance real-time processing and privacy at the edge. The framework supports integration with other computer vision models and features a mobile application that detects skin cancer.

\begin{table}[ht]
\centering
\scriptsize
\caption{Summary of \ac{FL} in cancer diagnosis. }
\label{table:7}
\begin{tabular}{m{0.5cm} m{2.5cm} m{1.7cm} m{5cm} m{1.6cm} m{4.5cm}}
\hline
\textbf{Ref} & \textbf{Model used} & \textbf{Images  dataset} & \textbf{Contribution} & \textbf{Best perf. (\%)} & \textbf{Limitation} \\
\hline

\cite{heidari2023new} & Blockchain-based \ac{FL} & Chest \ac{CT}  & Proposed \ac{TL}, Blockchain, \ac{FL}, and CapsNets framework enhances lung cancer detection. & Acc: 99.69 & Blockchain  increases computational overhead. \\ 

\cite{mohammed2023federated} & \ac{FL}, \ac{AE}, XGBoost & Omics data & Integrates federated \ac{AE} with XGBoost to detect multi-omics cancer. & Acc: 98.00 & Scalability of the scheme unaddressed.\\

\cite{omran2023detecting} & \ac{FL} & Skin cancer  & Presents a secure \ac{FL} approach for skin cancer detection, mitigating data poisoning attacks. & Acc: 95.20\newline {Pre: 94.80}\newline {Rec: 96.10} & Scalability issues with larger datasets.\\

\cite{alsalman2024federated} & CNN-FL & Breast cancer  & Introduces a \ac{FL} using \ac{CNN}s for breast cancer detection across diverse datasets. & Acc: 98.90\newline {Sen: 95.00}\newline {Spe: 98.00} & Data privacy management across federated networks.\\

\cite{jindal2023modernizing} & CNN-FL & Lung cancer  & Demonstrates \ac{FL} with \ac{CNN}s for scalable, privacy-preserving lung cancer severity diagnosis. & Acc: 95.76 & Privacy and data integration issues.\\

\cite{vibith2023gbdtmo} & GBDTMO-FL & Breast cancer  & \Ac{GBDTMO} with \ac{FL} for breast cancer diagnosis. & Acc: 94.47 \newline{Rec: 98.52}\newline{F1: 97.52} \newline{ROC: 96.32} & Communication overhead, complexity in synchronization.\\

\cite{supriya2024breast} & \ac{FL}, Shapley values & Breast Cancer & Integrates shapley values, game theory, and horizontal \ac{FL} to enhance breast cancer prediction. & Acc: 94.73 \newline{Pre: 95.28}\newline{Rec: 95.48} \newline{ROC: 98.97} & Challenges scaling to larger datasets.\\

\cite{gad2023novel} & \ac{FL} & Breast cancer  & A novel \ac{FL} integrates FedProx and Attention U-Net for precise breast cancer segmentation. & Acc: 96.07 \newline{Spe: 99.19}\newline{F1: 70.76} & Computational overhead unaddressed.\\

\cite{kong2024federated} & \ac{FL}-Attention & Prostate cancer  & Federated attention-consistent learning integrates \ac{FL} and attention consistency to enhance prostate cancer diagnosis and grading. & AUC: 97.18 & \ac{FL} adds communication overhead.\\

\cite{iqbal2023privacy} & Skin-net with \ac{FL} & Skin cancer  & Skin-net \ac{CNN} with privacy-preserving \ac{FL} for skin cancer. & Acc: 98.30\newline{Sen: 98.80}\newline{Spe: 97.90} & Complexity of progressively private \ac{FL} implementation.\\

\cite{zubair2023comparative} & \ac{FL} & X-ray lung  & Compares sequential and \ac{EM}s in \ac{FL} and centralized approaches for disease detection. & Acc: 99.00 & Sequential models add computational complexity.\\

\cite{yan2023pla} & FL-ViT & Breast dataset & Proposes a privacy-preserving, lightweight PLA model using \ac{ViT} for breast cancer diagnosis. & Acc: 95.30\newline{Rec: 99.80} \newline{Pre: 98.80} & Dataset details lacking, scalability concerns.\\

\cite{mostafa2023detecting} & \ac{FL}-\ac{TL} & Lung cancer  & Developed a \ac{FL} framework leveraging \ac{TL} for robust lung cancer detection. & Acc: 91.03\newline{Rec: 89.44}\newline{AUC: 98.55} \newline{Pre: 98.80} & \ac{FL} synchronization adds complexity.\\

\cite{hsiao2024precision} & \ac{DL}-\ac{FL} & Liver \ac{CT}  & Designed a robust Hybrid-ResUNet with \ac{FL} for accurate liver tumor segmentation. & Dice: 94.33\newline{AUC: 99.65} & Communication overhead affects performance.\\

\cite{jimenez2023memory} & Memory-aware curriculum \ac{FL} & Mammography  & Integrated curriculum learning into federated settings for improved consistency, and multi-site breast cancer classification. & AUC: 79.00 & Increased computational complexity.\\

\cite{sikandar2024integrating} & Fed\_ANN11 & Stomach Adenocarcinoma & Integrates generative \ac{AI} and \ac{FL} for sequence-based stomach adenocarcinoma detection. & Acc: 99.00 & Communication overhead and latency issues.\\

\cite{bhadauriya2023detection} & \ac{CNN}-\ac{FL} & Brain tumor  & Proposed \ac{CNN}-based \ac{FL} for brain tumor detection. & Acc: 96.00 & Data heterogeneity and communication overhead.\\

\cite{trivedi2024federated} & \ac{FL} & Breast cancer  & Federated YOLO-ResNet fusion ensures high-accuracy breast cancer detection. & Acc: 98.73\newline{Pre: 98.73}\newline{Rec: 98.73} & \ac{FL} with multiple clients adds complexity.\\

\hline
\end{tabular}
\end{table}

Other \ac{FL} approaches focus on securing the global model (aggregated model). For example,   \cite{lessage2024secure} investigated the integration of fully homomorphic encryption with secure \ac{FL} using mammogram data from Belgian medical records. The research focuses on evaluating memory constraints when applying fully homomorphic encryption to sensitive medical data. Despite notable limitations in memory usage, the results demonstrate that fully homomorphic encryption maintains comparable performance in terms of ROC curves, showcasing its robustness in secure \ac{ML} applications. This approach preserves data confidentiality and enhances the security of exchanges between participants and the central server. The study \cite{jindal2023modernizing} incorporate \ac{CNN} into \ac{FL} to enhance lung cancer diagnosis. Reference \cite{ayekai2023personalized} incorporates client-specific \ac{AE}s and hierarchical clustering to address the challenges posed by non-independent and identically distributed data across clients. This framework aims to enhance the effectiveness of \ac{FL} in heterogeneous medical data environments while preserving data privacy.

Distinct from other methods,  \cite{kong2024federated} introduced a federated attention-consistent learning framework designed to advance \ac{AI} in medical imaging by addressing challenges related to large-scale pathological images and data heterogeneity. Federated attention-consistent learning enhances model generalization by ensuring attention consistency between local clients and the central server model. To further safeguard data, differential privacy is incorporated by adding noise during parameter transfers. The framework was evaluated using a substantial dataset of prostate cancer images from various centers, demonstrating improved performance in cancer diagnosis and Gleason grading (aggressiveness of prostate cancer assessment) compared to existing methods. This approach offers a robust, privacy-preserving solution for training \ac{AI} models in medical imaging.

\subsection{Transfer learning}

\ac{TL} is a \ac{ML} strategy designed to improve the performance of models on a target task by transferring knowledge from a related source task, especially when labeled data for the target task is scarce. This technique capitalizes on pre-trained models that have been developed using large and diverse datasets, which can then be fine-tuned to specific problems in domains with limited data such as medical imaging \cite{habchi2024deep,mazari2023deep}, 3D data representation \cite{sohail2024advancing}, and  \ac{NLP} \cite{kheddar2023deep} tasks. The rationale behind \ac{TL} is that features learned from large-scale tasks, such as detecting edges, shapes, or general patterns in images, can be beneficial for smaller, related tasks without starting the learning process from scratch. The essential mathematical foundation of \ac{TL} lies in the optimization of a loss function that allows the model to learn and adapt to new tasks. The general equation for loss function optimization in \ac{TL} is given by Equation \ref{eq6}:

\begin{equation}
\label{eq6}
L(\theta)=\frac{1}{N} \sum_{n=1}^{N}L(y_i), f(x_i;\theta) \\
\end{equation}

\noindent Where N denotes the number of samples, $\theta$ represents the model parameters, $y_i$ is the true label, and $f(x_i;\theta)$  is the model's prediction. This loss function is minimized using optimization algorithms like \ac{SGD}, adjusting the model's weights for better generalization on the target task. \ac{TL} encompasses various approaches, including inductive, transductive, and unsupervised \ac{TL}, allowing models to generalize better to new tasks while improving training efficiency. Regularization techniques like dropout help mitigate overfitting during fine-tuning, and feature extraction methods utilize pre-trained models as fixed feature extractors for simpler models. Knowledge distillation further enhances efficiency by training smaller models to replicate the performance of larger ones \cite{kheddar2023deep,himeur2023video, kheddar2023deepa, sohail2024advancing}. Table \ref{Tab:TL} provides a summary of the types of \ac{TL}, including inductive, transductive, and unsupervised \ac{TL}, highlighting their distinct characteristics and key differences. Table \ref{table:8} summarizes several state-of-the-art studies in terms of the \ac{AI} models used, datasets, contributions, limitations, and best performances achieved. Additionally, popular pre-trained models are comprehensively reviewed in \cite{kheddar2023deep,himeur2023video,kheddar2023deepa}. { Figure \ref{fig:TLEX} illustrates an example of the proposed approach, which involves unsupervised pre-training of a \ac{CAE} to extract features from CT images, combined with the \ac{TL} technique to enhance lung cancer prediction.}

\begin{table}[ht!]
\centering
\caption{Comparison of various \ac{TL} techniques. } \label{Tab:TL}

\begin{tabular}{m{2cm}m{1.5cm}m{1.5cm}m{2cm}m{8cm}}
\hline
\textbf{\ac{TL} type} & \textbf{\(D_S\)} & \textbf{\(D_T\)} & \textbf{Labels in \(D_T\)} & \textbf{Objective function} \\ \hline

Inductive   & Same/Diff & Same/Diff & Labeled & \(\hat{\theta}_T = \arg\min_{\theta} \mathcal{L}_T(f_T(\theta, D_T), T_T)\) \\ \hline

Transductive & Diff & Diff & Unlabeled & \(\hat{\theta}_T = \arg\min_{\theta} \mathcal{L}_S(f_S(\theta, D_S), T_S) + \lambda\cdot \mathcal{L}_T(f_T(\theta, D_T), T_T)\) \\ \hline

Unsupervised & Diff & Diff & Unlabeled & \(\hat{\theta}_T = \arg\min_{\theta} \mathcal{L}_S(f_S(\theta, D_S), T_S) + \lambda\cdot \mathcal{L}_U(f_U(\theta, D_T))\) \\ \hline
\end{tabular}
\end{table}

\begin{figure}[ht!]
\begin{center}
\includegraphics[width=0.5\columnwidth]{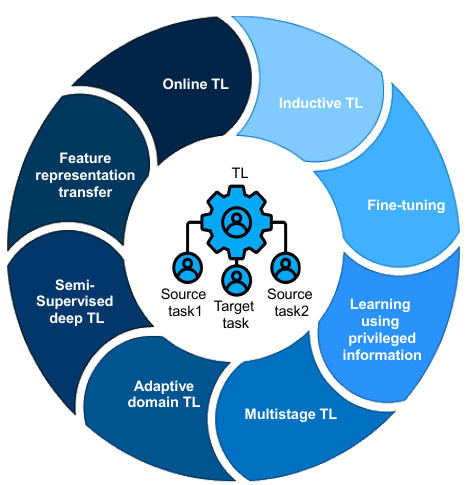}
\end{center}
\caption{\ac{TL} types used in the diagnosis of cancer and their respective related work. Inductive \ac{TL} \cite{aljuaid2022computer}; Fine-tuning \cite{kalbhor2023cervical, bansal2022deep,dadgar2022comparative}; Learning using privileged information \cite{shaikh2020transfer};  Multistage \ac{TL} \cite{ayana2022novel}; Adaptive domain \ac{TL} \cite{rong2021diagnostic}; Semi-supervised deep \ac{TL} \cite{shi2021semi}; Feature representation \ac{TL} \cite{mehmood2022malignancy};  Online \ac{TL} \cite{zhou2020online}.}
\label{figTL}
\end{figure}

Recently, numerous \ac{AI} methods based on \ac{TL} have been proposed to enhance cancer diagnosis and detection as summarized in Figure \ref{figTL}, with most relying on inductive learning and fine-tuning techniques. For instance  in \cite{aljuaid2022computer}, the authors addressed the critical issue of breast cancer detection and classification by proposing a novel \ac{CAD} method utilizing ResNet18, ShuffleNet, and Inception-V3Net in an inductive \ac{TL} mode. The method is tested on the BreakHis dataset, with final image dimensions of 224$\times$224 for ResNet18 and ShuffleNet, and 299$\times$299 for Inception-V3Net. This approach demonstrates high accuracy in both binary and multi-class classification. Similarly, \cite{khan2019novel} proposed a novel \ac{DL} framework for detecting and classifying breast cancer in cytology images using inductive \ac{TL}. Pre-trained \ac{CNN} architectures, including GoogLeNet, VGGNet, and ResNet, are employed for feature extraction, followed by classification of malignant and benign cells. Saber et al. \cite{saber2021novel} contributed by developing a \ac{DL} model based on inductive \ac{TL} to assist in the automatic detection and diagnosis of breast cancer. This study utilizes pre-trained \ac{CNN} architectures such as Inception V3, ResNet50, VGG-19, VGG-16, and Inception-V2 ResNet and evaluates the model on the MIAS dataset. The findings demonstrate that VGG16 is the most effective architecture for breast cancer diagnosis using mammographic images. Reference \cite{anand2022enhanced} introduced a \ac{DL} model for detecting skin cancer in its benign and malignant stages using inductive \ac{TL}. The model builds upon the pre-trained VGG16 architecture by adding a flattened layer, two dense layers with LeakyReLU activation, and a final dense layer with sigmoid activation to improve accuracy.  This model aims to assist dermatologists in early skin cancer diagnosis. Continuing diagnois scheme cancer,   \cite{balaha2023skin}, proposed an automatic  detection system using the sparrow search algorithm (SpaSA) for hyperparameter optimization. This study employs five U-Net models (U-Net, U-Net++, Attention U-Net, V-net, Swin U-Net) for segmentation and eight pre-trained \ac{CNN} models (VGG16, VGG19, MobileNet variants, NASNet variants) for classification.  For segmentation, U-Net++ with DenseNet201 achieved the best results on “skin cancer segmentation and classification” dataset, while Attention U-Net with DenseNet201 performed best on the “PH2” dataset. MobileNet pre-trained models achieved the highest accuracy on the “ISIC 2019 and 2020 Melanoma” and “HAM10K” datasets, and MobileNetV2 excelled with the “skin diseases image” dataset. The proposed method is compared with 13 related studies, demonstrating its effectiveness in skin cancer diagnosis. A breast cancer diagnostic system that integrates deep \ac{TL} with \ac{IoT} and fog computing was proposed in \cite{pati2023breast}. Using mammography images from the Cancer Imaging Archive, the system employed ResNet50, InceptionV3, AlexNet, VGG16, and VGG19 architectures alongside a \ac{SVM} classifier. Fog computing played a crucial role in enhancing privacy, reducing server load, and improving overall system efficiency.

\begin{figure}
    \centering
    \includegraphics[width=0.9\linewidth]{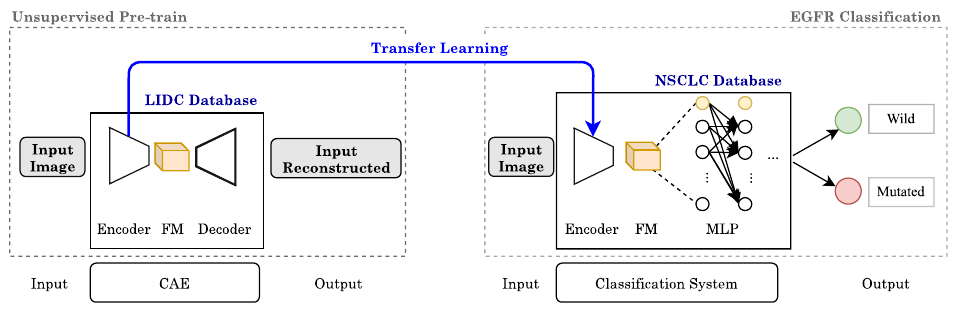}
    \caption{{Example of a proposed approach, which involves unsupervised pre-training of a \ac{CAE} to extract features from CT images, followed by an end-to-end classifier for predicting the mutation status of the \ac{EGFR}.This latter is a transmembrane protein involved in cell signaling pathways that regulate cell proliferation, survival, and differentiation; mutations in this receptor are commonly associated with non-small cell lung cancer. \ac{TL} enables the reuse of the encoder, pre-trained on unlabeled data from the \textit{LIDC-IDRI} database, as a feature extractor for EGFR mutation classification in the \textit{NSCLC-Radiogenomics} database \cite{silva2021egfr}.}}
    \label{fig:TLEX}
\end{figure}

Reference \cite{bansal2022deep} applies deep \ac{TL} fine-tuning of a pre-trained model  combined with hybrid optimization, using \ac{adam}, RMSprop, and SGD optimizers, to improve detection and diagnosis of oral cancer. This study utilizes models such as ResNet50, MobileNetV2, VGG19, VGG16, and DenseNet on both real-time and histopathologic datasets. Image preprocessing techniques, including Gaussian blur and morphological operations, are used to prepare the data. Likewise,  \cite{dadgar2022comparative}  presented a fine-tuned model to classify three widespread lung cancer types: Squamous cell carcinoma, large cell carcinoma, and Adenocarcinoma. The models used include VGG16, ResNet152V2, MobileNetV3 (small and large), InceptionResNetV2, and EfficientNetV2.  InceptionResNetV2 emerged as the best model with the highest performance metrics (accuracy, precision, AUC, and F1-score) in classifying the lung cancers from normal samples. Study \cite{faghihi2024diagnosis} presented a novel approach to skin lesion classification by combining VGG16 and VGG19 architectures into a modified AlexNet network. This combined model was fine-tuned on a dermatology dataset of 2,541 images without relying on data augmentation techniques. Dropout was used to address overfitting, and the model was evaluated using K-fold cross-validation. The proposed method achieved a significant improvement in classification accuracy. In a different study \cite{rai2021real} contributed by proposing a real-time data augmentation-based fine-tuning \ac{TL} model for breast cancer diagnosis using histopathological images. It compares two popular models, InceptionV3 and Xception, trained on the BreakHis dataset. The findings show that the Xception model, when trained using fine-tuning \ac{TL} from ImageNet weights, achieved high accuracy, outperforming models trained from scratch and surpassing previous state-of-the-art results on the BreakHis dataset.  Deepak et al. \cite{deepak2019brain} suggested a brain tumor classification system using fine-tuning \ac{TL} by employing a pre-trained GoogLeNet to extract features from \ac{MRI} images of glioma, meningioma, and pituitary tumors, and integrates proven classifiers for accurate classification.

\begin{center}
\scriptsize
\begin{longtable}{m{0.5cm} m{1.5cm} m{2cm} m{5cm} m{1.5cm} m{4.5cm}}
\caption{Summary of \ac{TL} in cancer diagnosis.}
\label{table:8} \\
\hline
\textbf{Ref} & \textbf{Model(s) used} & \textbf{Image dataset} & \textbf{Contribution} & \textbf{Best result (\%)} & \textbf{Limitation} \\
\hline
\endfirsthead

\hline
\textbf{Ref} & \textbf{Model(s) used} & \textbf{Image dataset} & \textbf{Contribution} & \textbf{Best result (\%)} & \textbf{Limitation} \\
\hline
\endhead
\hline
\multicolumn{6}{r}{} \\
\endfoot

\cite{aljuaid2022computer} &DNN-TL & BreakHis & Fine-tuning multiple \ac{DNN} models with \ac{TL} for accurate breast cancer classification. & Acc: 97.81\newline{Pre: 97.65}\newline{Sen: 97.65} & Single dataset may cause overfitting. \\

\cite{kalbhor2023cervical} &  ResNet-50 & Herlev, UIC& \ac{TL} enhances cervical cancer detection using Pap smear images. & Acc: 92.03 & Requires significant computational power. \\

\cite{bansal2022deep} & DL-TL & Oral cancerous, histopathologic & Efficient oral cancer diagnosis using DenseNet with fine-tuning and hybrid optimization. & Acc: 95.41\newline {Loss:  0.70} & Hybrid optimization adds computational complexity. \\

\cite{ayana2022novel} & EfficientNetB2 & Ultrasound breast cancer & Proposed multistage \ac{TL} method significantly improves breast cancer ultrasound classification. & Acc: 99.00\newline{F1: 98.90} & Extensive pre-processing may introduce biases. \\

\cite{mehmood2022malignancy} &  AlexNet & Histopathology  & \ac{TL} for lung and colon histopathology image classification. & Acc: 98.40 & Contrast enhancement method limits class generalizability. \\

\cite{zhou2020online} & Online \ac{TL} & Thyroid cancer & Online \ac{TL} distinguishes thyroid nodules using ultrasound. & AUC: 98.00\newline{Sen: 98.70}\newline{Spe: 98.80} & Relies on specific datasets. \\

\cite{khan2019novel} & CNN-TL & Breast cancer & \ac{DL} framework for breast cancer classification using \ac{TL}. & Acc: 97.67 & Requires substantial computational resources. \\

\cite{saber2021novel} & DL-CNN & Mammographic  & \ac{TL} for breast cancer detection using pre-trained \ac{CNN}s. & Acc: 98.96\newline{Sen: 97.83}\newline{AUC: 99.50} & Relies on single dataset. \\

\cite{anand2022enhanced} & DL-TL & Skin cancer & Fine-tuned VGG16 with added layers, data augmentation, and hyperparameter optimization for skin cancer diagnosis. & Acc: 89.09 & Model complexity increases training time. \\

\cite{balaha2023skin} & U-Net  & PH2, ISIC, HAM10K &  Fine-tuning \ac{CNN}s with SpaSA optimizer improves skin cancer segmentation, classification, and detection efficiency. & Acc: 98.83\newline{AUC: 99.45} & Model complexity adds challenges. \\

\cite{pati2023breast} & CNN-TL & Mammography  & Integrates \ac{TL} with fog computing for breast cancer diagnosis. & Acc: 97.99\newline{Pre: 99.51}\newline{Spe: 80.08} & High computational power needed. \\

\cite{faghihi2024diagnosis} & VGG16, VGG19, \ac{TL} & Dermatology dataset & Introduces a fine-tuned ac{TL} approach combining VGG16/VGG19 architectures for superior melanoma detection. & Acc: 98.18 & Dataset size limits generalization. \\

\cite{rai2021real} & InceptionV3, Xception & BreakHis & Proposed a \ac{TL}-based fine-tuning and real-time augmentation model for breast cancer classification. & Acc: 90.86\newline{Sen: 92.51}\newline{Spe: 80.85} & Limited data increases overfitting risk. \\

\cite{deepak2019brain} & TL-CNN & \ac{MRI} & \ac{TL} for brain tumor classification using \ac{MRI}. & Acc: 98.00 & Dataset may not cover all tumor types. \\

\cite{rao2024intelligent} & TL-MLP-SVM & Breast cancer & Intelligent breast cancer diagnosis with ensemble \ac{TL} models. & AUC: 94.70\newline{Rec: 85.80} & Data quality limits real-time use. \\

\cite{celik2020automated} & ResNet-50, DenseNet & Breast cancer & Automated breast cancer detection using \ac{TL}. & {Acc: 90.96}\newline F1: 94.11 & Larger dataset required for validation. \\

\cite{chen2020deep} & CNN-TL & Cervical histopathological & Proposed a fine-tuned deep \ac{TL} framework for interpretable and automated cervical histopathological image diagnosis. & Acc: 97.42\newline{Spe: 98.93}\newline{Sen: 95.88} & Requires substantial computational resources. \\

\cite{samee2022hybrid} & \ac{TL} & Mammogram  & \ac{TL} for breast lesion diagnosis with hybrid \ac{DL}. & Acc: 98.80 & Computational complexity due to the multi-step pipeline. \\

\cite{samee2022classification} & Hybrid \ac{TL} & CE-MRI & \ac{TL} model for brain tumor classification using \ac{CE}-\ac{MRI}. & Acc:  99.51\newline{Sen: 98.90} & Relies on specific \ac{CE}-\ac{MRI} dataset. \\

\cite{aziz2023ivnet} & VGG16 & JUMC & IVNet uses \ac{TL} for breast cancer gradingt. & Acc: 97.00 & Despite employing \ac{TL}, the relatively small dataset size could still cause overfitting. \\

\hline
\end{longtable}
\end{center}

Another \ac{TL} technique consist  of multistage \ac{TL} as suggested in  \cite{ayana2022novel} dedicated  for ultrasound breast cancer image classification. It employs three pre-trained models EfficientNetB2, InceptionV3, and ResNet50—along with \ac{adam}, Adagrad, and SGD optimizers. Study \cite{rong2021diagnostic} employed domain adaptation \ac{TL} techniques for early lung cancer diagnosis using multi-omics data. The approach combines \ac{CNN}s and convolutional \ac{AE}s to handle the challenges of high-dimensional, low-sample-size, and noisy omics data. The \ac{AE}s reduce dimensionality to enhance migration learning, while the \ac{CNN} model is trained on both the original and labeled datasets. The proposed method outperforms five other \ac{ML} models, demonstrating superior performance in classifying and predicting lung cancer from gene datasets. Similarly in  \cite{alzubaidi2021novel}, where the domain adaptation approach have been used  to address the issue of limited labeled data in medical imaging. Traditional \ac{TL}  methods, which often use pretrained models from datasets like ImageNet, may be ineffective due to feature mismatches between natural and medical images. To overcome this, the authors propose training \ac{DL} models on large unlabeled medical image datasets before fine-tuning them on smaller labeled datasets. They also introduce a new deep \ac{CNN} model that incorporates recent advancements in the field. The proposed method has been empirically validated through experiments on skin and breast cancer classification tasks, showing significant performance improvements.

The scheme proposed in  \cite{shi2021semi}  is a semi-supervised  \ac{TL}  framework for diagnosing benign and malignant pulmonary nodules in chest CT images. This approach combines \ac{TL} with a pre-trained classification network and an iterated feature-matching-based semi-supervised method to handle the challenge of limited and imbalanced pathological datasets. Unlabeled dataset  contains 14,735 nodules from 4,391 subjects and are used by the semi-\ac{SL} framework. However,  the work in \cite{samee2022hybrid} proposed a \ac{TL}-based feature extraction approach for breast lesion diagnosis using mammograms. It addresses two key challenges: enhancing input data by generating pseudo-colored images using CLAHE and pixel-wise intensity adjustment, and mitigating multicollinearity in high-level features through a novel LR-PCA method. The proposed system achieves high performance with accuracies demonstrating its effectiveness for breast cancer diagnosis.

\subsection{Transformer-based  learning}
\label{sec4.5}

Transformers have revolutionized computer vision by modeling long-range dependencies through self-attention mechanisms \cite{kheddar2024Transformers}, unlike \acp{CNN}, which rely on local receptive fields. Since cancer diagnosis is image-based, the Transformer architecture employed in this domain is \ac{ViT}, which enables more effective feature extraction for tumor detection, segmentation, and classification. Various \ac{ViT} architectures have been designed to address challenges such as computational complexity, spatial information loss, and multi-scale feature representation. \Ac{ViT} replaces traditional \ac{CNN}s by dividing images into patches and processing them as sequences, similar to how words are handled in \ac{NLP}  \cite{kheddar2024automatic,habchi2024machine}. 

In general, \Ac{ViT} divides an image of size \( H \times W \times C \) into \( N = \frac{H \times W}{P^2} \) patches, where each patch is flattened into a vector and linearly transformed using a learnable embedding matrix \( E \in \mathbb{R}^{(P^2 \cdot C) \times D} \), resulting in patch embeddings:
\begin{equation}
z_0^i = x^i \cdot E
\end{equation}
for each \( i \)-th patch. To retain spatial information, positional encodings \( p_i \) are added:
\begin{equation}
z_0^i = z_0^i + p_i.
\end{equation}
These embeddings pass through the Transformer encoder, which applies the \ac{MHSA} mechanism:

\begin{equation}
\text{Attention}(Q, K, V) = \text{softmax} \left( \frac{QK^T}{\sqrt{d_k}} \right) V
\end{equation}
where \( Q, K, V \) represent the query, key, and value matrices, and \( d_k \) is the dimension of the keys. The \ac{MHSA} operation aggregates multiple attention heads, such that:  \(\displaystyle\text{MHSA}(X) = \text{Concat}(\text{head}_1, \dots, \text{head}_h)W^O \). A \ac{FFN} is applied to each token:
\begin{equation}
\text{FFN}(x) = \max(0, x.W_1 + b_1).W_2 + b_2,
\end{equation}
where \( W_1, W_2 \) are learnable weight matrices and \( b_1, b_2 \) are biases. A special classification token (CLS token) is added to the patch sequence, and after the final Transformer layer, the CLS token is passed to a multi-layer perceptron (MLP) classifier:
\begin{equation}
y = \text{MLP}(z_L^{\text{CLS}}),
\end{equation}
where \( z_L^{\text{CLS}} \) is the processed CLS token.

While the standard \ac{ViT} structure has shown competitive performance, several variants have been developed to enhance efficiency, improve local and global feature extraction, and incorporate convolutional priors. Table \ref{tab:vit_variants} summarizes key \ac{ViT} variants, their advantages, applications in cancer diagnosis, and limitations. These variants address some of the computational and structural challenges of \ac{ViT}, making them more suitable for tasks such as medical image analysis, where feature locality and efficiency are critical. A comprehensive overview of these architectures can be found in \cite{habchi2024machine,kheddar2024Transformers}.

\begin{table}[h]
    \centering
    \scriptsize
    \caption{{Comparison of \ac{ViT} variants in cancer diagnosis.}}
    \label{tab:vit_variants}
    \renewcommand{\arraystretch}{1.3}
    \begin{tabular}{m{2.5cm}m{5cm}m{4cm}m{5cm}}
        \hline
        \textbf{{ViT variant}} & \textbf{{Advantages}} & \textbf{{Use case} } & \textbf{{Limitations}} \\
        \hline
        {Self-Supervised ViT} & {Learns representations without labeled data, improves generalization} & {Detects rare cancer types with limited labeled data }& {Requires extensive computational resources for pretraining} \\
        \hline
        {Multi-Axis ViT} & {Processes spatial axes separately for better feature extraction} & {Enhances tumor boundary delineation in histopathology images} & {Increased model complexity affects real-time processing} \\
        \hline
        {Swin Transformer} & {Uses shifted windows to reduce computation and improve efficiency} & {Efficient large-scale whole-slide image analysis} & {Limited ability to model global dependencies} \\
        \hline
        {Tokens-to-Token ViT} & {Preserves spatial relationships via progressive tokenization} & {Improves segmentation of tumor regions in MRI scans} & {Higher memory requirements due to hierarchical tokenization} \\
        \hline
        {Convolutional ViT (CvT)} & {Incorporates convolutional layers for local feature extraction} & {Better texture recognition for skin and breast cancer detection} & {Partially loses global receptive field advantage of ViTs} \\
        \hline
        {CrossFormer \& Cross ViT} & {Multi-scale feature fusion improves robustness} & { Enhances multi-modal imaging (e.g., PET-CT fusion for cancer staging)} & {Computationally intensive due to cross-scale processing} \\
        \hline
        {Separable ViT} & {Reduces computational complexity by factorizing attention} & {Speeds up inference for real-time cancer detection systems} & {Trade-off in accuracy due to reduced interaction between features} \\
        \hline
        {Multi-View ViT} & {Aggregates different perspectives for robust representations} & {Assists in multi-angle breast cancer mammography analysis} & {High data requirements for multi-view learning} \\
        \hline
        {Boosted ViT} & {Applies ensemble principles to enhance prediction accuracy} & {Improves classification of rare cancer subtypes} & {Training complexity increases significantly} \\
        \hline
        {Local-Global ViT} & {Balances fine-grained and high-level feature extraction} & {Helps identify tumor microenvironments in histopathology} & {Computationally expensive for large images} \\
        \hline
        {Pooling-Based ViT} & {Reduces dimensionality using pooling mechanisms} & {Speeds up processing in large-scale cancer imaging datasets} &{ May lose fine-grained spatial information} \\
        \hline
        {Nested Hierarchical Transformer} & {Multi-level feature extraction enhances contextual understanding} & {Improves differentiation of overlapping cell structures in pathology} & {Increased model depth leads to high training time} \\
        \hline
    \end{tabular}
\end{table}

\begin{figure}[tbph]
\begin{center}
\includegraphics[width=0.6\columnwidth]{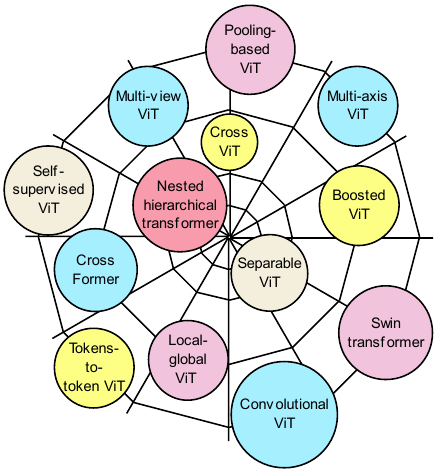}
\end{center}
\caption{\ac{ViT} types used in the diagnosis of cancer. Separable ViT \cite{abbas2023assist}; Cross-validation \ac{ViT} \cite{gulsoy2023diagnosis}; Multi-Axis \ac{ViT} \cite{pacal2024maxcervixt};  Boosted \ac{ViT} \cite{ayana2024pathological}; Swin Transformer \cite{zhou2021swin}; Tokens-to-token \ac{ViT} \cite{zhao2022improving};  Local-global \ac{ViT} \cite{wang2023lgvit}; Self-supervised \ac{ViT} \cite{chen2022self}; \ac{CNN}-\ac{ViT} \cite{tabatabaei2023attention}; CrossFormer \cite{chen2023crossformer}; CrossViT \cite{sriwastawa2024vision}; and nested hierarchical Transformer \cite{sriwastawa2024vision}. }
\label{figVIT}
\end{figure}

Among the types of \ac{ViT}, there is a swin transformer, which is a type of \ac{ViT} designed to efficiently handle high-resolution visual data and complex vision tasks. Unlike the original \ac{ViT}, which treats an entire image as a sequence of patches, the swin Transformer introduces a hierarchical structure and operates on non-overlapping windows (patches). This structure progressively increases the model's ability to capture both local and global contextual information \cite{pina2024unsupervised}.  The compact \ac{ViT} is another variant of the \ac{ViT} designed to address some limitations of the original \ac{ViT}, particularly its need for large amounts of data and lack of inductive biases commonly found in \ac{CNN}s.  The compact \ac{ViT} introduces mechanisms like sequence pooling and multi-scale features to improve the model's efficiency and performance, especially when working with smaller datasets or when computational resources are limited \cite{mali2022detection}. Figure \ref{figVIT} provides the taxonomy of different types of \ac{ViT}, including some commonly used models and others proposed in specific studies.

Many suggested strategies for cancer detection are based on Transformers. In particular, the study in \cite{sriwastawa2024vision} compares different \ac{ViT} models including \ac{PiT}, \ac{CvT}, CrossFormer, CrossViT, NesT, MaxViT, and \ac{SepViT} for classifying breast cancer in digital histopathology images. MaxViT was identified as the best-performing model.   \cite{ikromjanov2022whole} proposed  a \ac{ViT} model designed for the detection and classification of prostate cancer  by first extracting patches from \ac{ROI} in the whole slide images and then applying the \ac{ViT} model for classification. The classified patches are subsequently scored and graded according to the Gleason system.  Mali et al \cite{mali2022detection} investigate the use of \ac{ViT} focusing on data preprocessing and the Transformer models are utilized for detecting colorectal cancer. This study implements three transformer-based models \ac{ViT}, \ac{CvT}, and \ac{CCT}  to classify histopathological images, exploring the impact of different patch sizes and model configurations, demonstrating that \ac{ViT}s outperform traditional methods in accuracy. It provides insights into optimization techniques specific to colorectal cancer detection. Similarly, the research in  \cite{tummala2022classification}, explored the use of an ensemble of \ac{ViT} for classifying brain tumors from \ac{MRI} scans. Pretraining and fine-tuned four ViT models is performed on ImageNet. The dataset comprised T1-weighted contrast-enhanced MRI slices with meningiomas, gliomas, and pituitary tumors. It shows that combining multiple \ac{ViT} models enhances classification accuracy and reliability compared to single \ac{ViT} models and traditional methods. Pachetti et al  \cite{pachetti2022effectiveness} evaluate the effectiveness of 3D \ac{ViT} in predicting prostate cancer aggressiveness. It demonstrates that 3D \ac{ViT}s can effectively capture and analyze spatial information in 3D medical images, potentially outperforming traditional 2D approaches and 3D \ac{CNN}. This study optimized 3D Vision Transformer models for prostate cancer aggressiveness prediction by selecting five key MRI slices per lesion, harmonizing pixel dynamics, and rescaling and center-cropping images to focus on the prostate gland. Ayana et al. \cite{ayana2024pathological} investigated the use of a novel type of Transformer called ViTCol, a boosted vision Transformer model specifically designed for classifying endoscopic pathological findings. Additionally, they proposed PUTS, a Swin-Unet transformer-based model for polyp segmentation. Both models represent significant advancements in classification and segmentation tasks, which are critical for early CRC diagnosis. Their findings demonstrate that these enhanced \ac{ViT}s can significantly improve detection performance and provide valuable insights in pathology. Sun et al. \cite{sun2023classification} propose thyroid cancer-\ac{ViT}, which integrates contrastive learning to enhance the classification accuracy of thyroid nodules in ultrasound images. The main issue addressed is the lack of distinguishing features between benign nodules, particularly those classified as TI-RADS level 3 and malignant nodules, which can lead to diagnostic inconsistencies, overdiagnosis, and unnecessary biopsies. The thyroid cancer-\ac{ViT} model leverages \ac{ViT}'s capability to capture global features and contrastive learning to reduce the representation distance between nodules of the same category, improving the consistency of global and local feature representations. Similarly, Chen et al. \cite{chen2022Transformers} demonstrated that transformer-based models improve breast cancer diagnosis by utilizing unregistered multi-view mammograms (CC and MLO) from both sides (right and left breasts). The outputs are concatenated and processed through global Transformer blocks to jointly learn patch relationships across the four images. The model was trained and evaluated on a balanced dataset of mammogram sets, comprising malignant and normal cases, and showed that this approach significantly outperforms traditional methods in handling unregistered images.

Several studies \cite{yang2023novel,ayana2024vision,nejad2023hvit4lung,hossain2023vision,yang2024enhancing} have combined \ac{ViT} with \ac{TL} to enhance cancer detection. Yang et al. \cite{yang2023novel} developed a model for skin cancer classification, where the network was pretrained on the ImageNet dataset and fine-tuned on the HAM10000 dataset. Ayana et al. \cite{ayana2024vision} focused on breast cancer mass classification, introducing three \ac{ViT}-based \ac{TL} architectures pretrained on ImageNet and evaluating their performance using ultrasound and mammogram datasets. The comparative analysis demonstrated that \ac{ViT}-based \ac{TL} achieved superior performance. Similarly, Nejad et al. \cite{nejad2023hvit4lung} proposed a hybrid \ac{DL} framework combining \ac{ViT}s and \ac{CNN}s with \ac{TL} for improved lung cancer detection. Their hybrid \ac{ViT} model effectively addressed the challenges of lung nodule detection by extracting features from chest \ac{CT} images to classify nodules as normal, benign, or malignant. Tested on a dataset of \ac{CT} images, the model achieved high accuracies in training, validation, and testing, demonstrating superior performance compared to existing methods. Hossain et al. \cite{hossain2023vision} explored advanced methods for brain tumor detection and classification using \ac{MRI} images. The study addressed the challenges of inconsistent diagnostic results among specialists and emphasized the need for reliable multi-class tumor classification. It evaluated several \ac{DL} architectures, including VGG16, InceptionV3, VGG19, ResNet50, InceptionResNetV2, and Xception. A new \ac{TL}-based model, IVX16, was proposed by combining features from the best-performing architectures. Tested on a dataset of 3264 images, the model demonstrated improved classification performance. Additionally, \ac{XAI} was employed to assess model performance, while \ac{ViT} models were compared with other \ac{TL} methods and \ac{EM}s, highlighting their effectiveness.

Several frameworks combined \ac{ViT} with \ac{CNN} as an efficient solution for detecting various types of cancer. Zhou et al. \cite{zhou2023rfia} introduced a hybrid architecture, RFIA-Net, for classifying stage I multi-modality oesophageal cancer images. RFIA-Net leveraged the local modeling capabilities of \ac{CNN}s and the global information extraction of \ac{ViT}s, enhanced by a structural reparameterization strategy. The architecture included a feature extraction module and a feature enhancement module, which exchanged information between branches to improve performance. An asymmetric fusion module further strengthened feature relationships through spatial translation and channel swapping. Tested on the XJMU-XJU dataset, RFIA-Net achieved high performance. Xin et al. \cite{xin2022improved} proposed an enhanced \ac{ViT} model, SkinTrans, for classifying skin cancer from dermoscopic images. Traditional \ac{CNN}-based methods, while effective, struggled to extract features from large, critical regions in images. To address this, SkinTrans utilized self-attention mechanisms to focus on significant features while suppressing noise. The model incorporated multi-scale patch embedding, overlapping sliding windows, and contrastive learning to enhance classification accuracy. Similarly, Zhang \cite{zhang2023msht} developed \ac{MSHT}, a hybrid model that combined the local spatial feature extraction capabilities of \ac{CNN}s with the global feature capture and long-range dependency handling of Transformers. The \ac{MSHT} architecture integrated a \ac{CNN} backbone to extract multi-scale local features, guiding the Transformer in capturing global features. This approach enabled the model to effectively leverage both local and global information. Lastly, Tabatabaei et al. \cite{tabatabaei2023attention} explored an advanced framework for classifying brain tumors in \ac{MRI} images using a hybrid architecture combining self-attention units and \ac{CNN}s. Their proposed model enhanced classification accuracy by combining local features extracted by \ac{CNN}s with global features captured by Transformers through a cross-fusion strategy. Additionally, the study introduced an improved \ac{CNN} architecture, iResNet, optimized for distinguishing tumor features in \ac{MRI} images.

\Ac{T2T-ViT} also attracted significant attention from researchers in the field of cancer diagnosis. For instance, Yin et al. \cite{yin2022pyramid} introduced a novel lightweight architecture, Pyramid \ac{T2T-ViT}, to address challenges in classifying histopathological thyroid images. Traditional methods using \ac{CNN}s struggled with high-resolution whole slide images due to increased model parameters and computational complexity. The Pyramid \ac{T2T-ViT} combined \ac{T2T-ViT} with an image pyramid approach to balance local and global features effectively while reducing computational demands. The model incorporated a feature extractor to minimize parameters and employed multiple receptive fields to enhance classification accuracy. In the same manner, Zhao et al. \cite{zhao2022improving} tackled the challenges of cervical cancer screening, such as limited public datasets, imbalanced class distribution, and varying image quality. They proposed a cervical cell image generation model, CCG-taming Transformers, to create high-quality, balanced datasets. The researchers enhanced the model's encoder with SE-blocks and multi-Res-blocks for improved feature extraction and introduced a normalization layer to optimize data processing. Additionally, SMOTE-Tomek links were used to balance the dataset, and \ac{T2T-ViT} combined with \ac{TL} was employed for classification. The model was validated on three public datasets, demonstrating its effectiveness.


\begin{center}
\scriptsize
\begin{longtable}{m{0.5cm} m{1.5cm} m{2cm} m{5.5cm} m{1.5cm} m{4.5cm}}
\caption{Summary of Transformer-based  cancer diagnosis.} 
\label{table:10} \\
\hline
\textbf{Ref} & \textbf{Model(s) used} & \textbf{Image dataset} & \textbf{Contribution} & \textbf{Best result (\%)} & \textbf{Limitation} \\
\hline
\endfirsthead

\hline
\textbf{Ref} & \textbf{Model(s) used} & \textbf{Image dataset} & \textbf{Contribution} & \textbf{Best result (\%)} & \textbf{Limitation} \\
\hline
\endhead

\hline
\multicolumn{6}{r}{\textit{Continued on next page}} \\
\endfoot

\hline
\endlastfoot

\cite{ayana2024pathological} & \ac{ViT} & Colorectal pathology  & Enhanced \ac{ViT} improves colorectal cancer early detection. & AUC: 99.99\newline{MCC: 90.48} & Model complexity increases computational demands. \\

\cite{zhao2022improving} & T2T-ViT & Cervical cancer  & Hybrid \ac{T2T-ViT} model improves cervical cancer classification. & Acc:  99.98\newline{Sen: 98.34} & High computational complexity. \\

\cite{tabatabaei2023attention} & ATM-CNN & Brain tumor \ac{MRI} & Combines attention Transformers and \ac{CNN} for brain tumor classification. & Acc:  99.30 & High computational resource requirements. \\

\cite{sriwastawa2024vision} & \ac{ViT} & Breast cancer histopathology & \ac{ViT} models compared for breast cancer classification. & Acc:  92.12 & Study only compares \ac{ViT} models. \\

\cite{ikromjanov2022whole} & \ac{ViT} & Prostate tissue slides & \ac{ViT} improves prostate cancer detection from slide images. & Pre: 99.00\newline{F1: 94.00} & High computational resources needed. \\

\cite{tummala2022classification} & \ac{ViT} & Brain tumor \ac{MRI} scans & Ensemble \ac{ViT}s improve brain tumor classification accuracy. & Acc:  98.70\newline{Sen: 97.78} & Ensemble models increase computational complexity. \\

\cite{pachetti2022effectiveness} & 3D \ac{ViT} & 3D \ac{MRI} prostate scans & 3D \ac{ViT} improves prostate cancer aggressiveness prediction. & AUC: 77.50\newline{Spe: 75.00} & High computational cost for training. \\

\cite{sun2023classification} & ViT-CL  & Thyroid nodule ultrasound & Combines \ac{ViT} with \ac{CL} for thyroid classification. & Acc:  86.90 & High computational demands for \ac{CL}. \\

\cite{chen2022Transformers} & Multi-view \ac{ViT} & Breast cancer  & Multi-view \ac{ViT} improves feature management in cancer diagnosis. & AUC:  81.40\newline{Pre: 79.70} & Depends on availability of multi-view data. \\

\cite{yang2023novel} &ViT-TL & Skin cancer  & ViT-TL enhances skin cancer classification. & Acc:  80.5 & Requires significant computational resources. \\

\cite{ayana2024vision} & ViT-TL & Breast mass  & ViT-TL improves breast mass classification. & AUC: 100 & Depends on diverse diagnostic modalities. \\

\cite{nejad2023hvit4lung} & ViT-TL & Lung cancer  & Hybrid ViT-TL improves lung cancer diagnosis accuracy. & Acc:  99.09 & Hybrid model increases computational complexity. \\

\cite{hossain2023vision} &ViT-EM-TL & Brain tumor & Ensemble \ac{ViT}s improve brain tumor classification accuracy. & Acc:  96.94 & Requires diverse imaging data. \\

\cite{zhou2023rfia} & ViT-CNN & Oesophageal cancer & Enhances oesophageal cancer diagnosis using \ac{ViT} + \ac{CNN}. & Acc: 99.02\newline AUC: 99.73  & Complex model increases computational demands. \\

\cite{xin2022improved} & SkinTrans & Skin cancer  & Improved Transformer network enhances skin cancer classification. & Acc:  94.30 & High computational demands. \\

\cite{zhang2023msht} & MSHT & Pancreatic cancer  & \ac{MSHT} improves pancreatic cancer diagnosis using Transformers. & Acc:  95.68\newline{Spe: 96.95}\newline{NPV: 96.35} & Increased computational complexity. \\

\cite{wagner2023transformer} & \ac{ViT} & Colorectal histology  & \ac{ViT} predicts biomarkers from colorectal histology images. & Sen: 99.00 & Limited applicability to other data types. \\

\cite{alotaibi2023vit} & ViT-DeiT & Breast histopathology & \ac{ViT} + \ac{DeiT} improve breast cancer classification. & Acc:  98.17\newline{Pre: 98.18}\newline{F1: 98.12} & High computational resources required. \\

\cite{gulzar2022skin} & ViT-CNN & Skin lesion  & Compares \ac{ViT} and \ac{CNN} for skin lesion segmentation. & Acc:  92.11\newline{Dice: 89.84} & High computational demands for both methods. \\

\cite{boudouh2024advancing} & ViT-CNN & Mammography  & Integrating \ac{ViT} with \ac{CNN} improves breast cancer detection. & Acc:  99.22 & Increased computational complexity. \\

\cite{ferdous2023lcdeit} & LCDEiT & Brain tumor \ac{MRI} & \ac{LCDEiT} reduces Transformer model complexity in brain classification. & Acc:  98.11\newline{F1: 93.69} & Requires extensive tuning for optimal results. \\

\cite{hashemi2024realism} & YOLOv8-DeiT & Brain tumor  & Combines \ac{YOLOv8} and \ac{DeiT} for brain tumor detection. & Acc:  100\newline{F1: 100} & High computational demands. \\

\cite{ayas2023multiclass} & Swin  & Skin lesion  & Swin Transformer improves multiclass skin lesion classification accuracy. & Acc:  97.20\newline{Spe: 98.00} & Requires extensive computational resources. \\

\cite{pacal2024novel} & Swin -MLP & Brain tumor \ac{MRI} & Hybrid swin Transformer and \ac{MLP} improves brain tumor diagnosis. & Acc:  99.92\newline{Rec: 99.92} & Significant computational resource requirements. \\

\cite{tummala2022breast} & Ensemble swins & Breast histopathology & BreaST-Net uses swin Transformers for breast cancer classification. & Acc:  99.60\newline{MCC:  98.90} & Single dataset limits generalization. \\

\cite{chaudhury2024transforming} & ViT-LSTM & Breast histopathology & \ac{ViT} and \ac{LSTM} improves breast cancer classification accuracy. & Acc:  99.20 & Complexity increases computational demands. \\

\cite{gai2022rmtf} & RMTF-Net & Brain tumor \ac{MRI} & \ac{RMTF-Net} improves brain tumor segmentation accuracy. & Dice: 93.50& Limited to 2D imaging scenarios. \\

\cite{basu2023radformer} & RadFormer & Gallbladder cancer & RadFormer improves gallbladder cancer detection using global attention. & Acc:  92.10\newline{Spe: 96.10} & The complexity of the attention mechanisms. \\

\hline
\end{longtable}
\end{center}

The study \cite{zhou2021swin}, introduces a novel approach for segmenting cholangiocarcinoma histopathological images using hyperspectral imaging. Hyperspectral imaging offers richer spectral information compared to RGB imaging, which can enhance segmentation performance. The study proposes a swin-spectral Transformer network, which integrates spectral and spatial features more effectively. The network employs a spectral multi-head self-attention mechanism to handle the spectral dimension as a sequence, rather than treating it as an additional spatial dimension. The swin-spectral Transformer combines spectral-multi-head self-attention with shifted window-based multi-head self-attention to capture both spectral and spatial features. Additionally, a spectral aggregation token is introduced for effective dimensional reduction, producing a 2D segmentation result. The experimental results demonstrate that the proposed method significantly outperforms existing techniques. 

\begin{figure*}[ht!]
\begin{center}
\includegraphics[width=0.8\columnwidth]{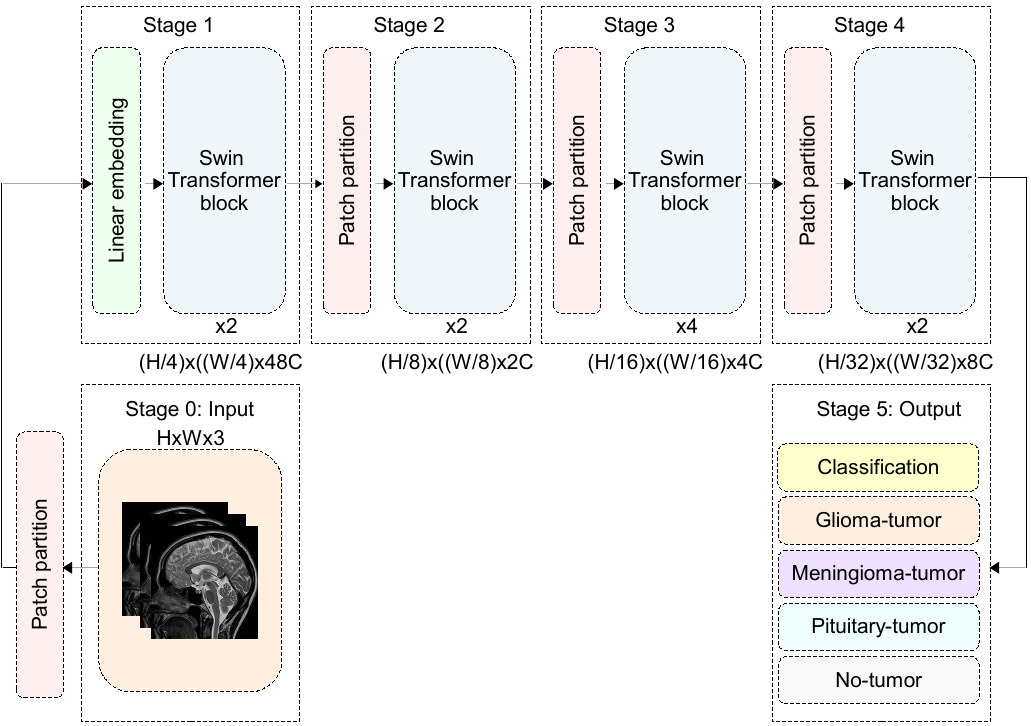}
\end{center}
\caption{An example of the structure of Swin Transformer architecture for brain tumor diagnosis \cite{pacal2024novel}. The Swin Transformer architecture comprises essential blocks for brain tumor diagnosis. The input module, preprocesses brain scans by normalizing and resizing them. The Patch partitioning block, divides images into fixed-size patches, embedding them into tokens. The Swin Transformer blocks, employ hierarchical learning with shifted window attention, ensuring efficient local-global feature extraction while multi-head self-attention identifies critical regions. Feature Aggregation, progressively reduces spatial dimensions, consolidating key features. The classification head, maps aggregated features to tumor classes using fully connected layers and softmax. Finally, the output delivers diagnostic results, excelling in capturing spatial hierarchies for accurate brain tumor detection.}
\label{fig19}
\end{figure*}

Swin Transformer uses a hierarchical feature maps and shifted windows for efficient image modeling and scalability. In particular,  \cite{ayas2023multiclass} proposed a Swin Transformer model for multi-class skin lesion classification, leveraging the strengths of both Transformers and \ac{CNN}s. This model benefits from end-to-end learning capabilities and does not require prior knowledge. To address class imbalance, this approach uses a weighted cross-entropy loss, enhancing the model's performance in dealing with imbalanced datasets. The proposed method was evaluated on the skin lesion imaging. This study demonstrates that the Swin Transformer model outperforms many existing state-of-the-art methods, showing superior balanced accuracy and effectiveness in multiclass skin lesion classification. In contrast \cite{pacal2024novel} addressed the critical need for accurate and timely brain tumor diagnosis. This study introduces an advanced \ac{DL} approach using the Swin Transformer model as depicted in Figure \ref{fig19}. Key innovations include hybrid shifted windows, multi-head self-attention modules and a residual-based \ac{MLP} integrated into the swin Transformer. The Res-\ac{MLP} replaces the traditional \ac{MLP} to further improve accuracy, training speed, and parameter efficiency. The proposed-swin model was evaluated on the publicly available brain \ac{MRI} dataset. The model benefits from \ac{TL} and data augmentation techniques. This approach offers a novel and robust diagnostic tool, supporting radiologists in achieving timely and accurate brain tumor diagnoses. The study \cite{cai2023mist} addresses the challenge of distinguishing benign from malignant colorectal adenomas. The proposed method, \ac{MIST}, leverages the swin Transformer as its backbone for feature extraction. This approach utilizes self-supervised contrastive learning and integrates a dual-stream multiple instance learning network to classify whole slide images based solely on slide-level labels, eliminating the need for labor-intensive manual annotation. \ac{MIST} was trained and validated on a dataset of 666 whole slide images from 480 patients, encompassing six common types of colorectal adenomas. Reference \cite{tummala2022breast}, explored the use of swin Transformers for classifying breast cancer subtypes from histopathological images.

Other proposed Transformers included the multi-axis \ac{ViT} introduced by Pacal et al. \cite{pacal2024maxcervixt}, an advanced architectural framework designed to enhance the early detection of cervical cancer using Pap smear images. The multi-axis \ac{ViT} was specifically adapted for Pap smear data, featuring a lightweight structure that improved both accuracy and inference speed. \cite{tu2022maxvit} replaced MBConv blocks in the MaxViT architecture with ConvNeXtv2 blocks and \ac{MLP} blocks with gated residual network \ac{GRN}-based \ac{MLP}s. These modifications significantly reduced the number of parameters and enhanced the model's generalization capabilities. The proposed method was rigorously evaluated on the SIPaKMeD and Mendeley LBC Pap smear datasets, utilizing 106 \ac{DL} models—53 \ac{CNN}s and 53 \ac{ViT} models for each dataset. Barzekar et al. \cite{barzekar2023multinet} presented a novel \ac{DNN} architecture, MultiNet, designed to improve the multiclass classification of medical images with a focus on cancer diagnosis. This study integrated Transformers within a multiclass framework to enhance data representation and improve classification accuracy. The proposed MultiNet \ac{ViT} model was evaluated on publicly available datasets, employing various assessment metrics to ensure reliability. Results demonstrated that MultiNet significantly improved cancer diagnosis accuracy through image classification, potentially reducing the reliance on costly and time-consuming manual analyses by radiologists and pathologists. This approach offered a scalable solution for large-scale biological image classification, paving the way for more efficient and effective cancer detection. Chaudhury et al. \cite{chaudhury2024transforming} explored the integration of \ac{ViT} and \ac{LSTM} networks for breast cancer diagnosis, comparing ultrasonography and histology as diagnostic modalities. While traditional \ac{CNN}s effectively extracted visual features, this study leveraged bidirectional encoder representations from Transformers, optimized for image processing. The integration of \ac{ViT} and \ac{LSTM} demonstrated significant potential in improving breast cancer diagnosis, providing a powerful tool for more precise and reliable healthcare decisions.

\subsection{Large language models}

\begin{figure}[ht!]
\begin{center}
\includegraphics[width=0.4\columnwidth]{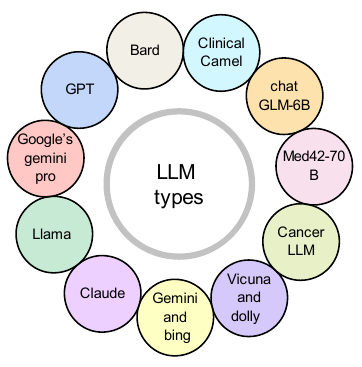}
\end{center}
\caption{Type of \ac{LLM} for cancer diagnosis. CancerLL M \cite{li2024cancerllm}; Vicuna and dolly \cite{sun2024large}; GeneLLM \cite{deng2024genellm}; Llama \cite{chang2024classifying}; GPT \cite{cao2024large}; Gemini and bing \cite{cao2024large}; Claude and Bard \cite{zhou2024performance}; Privacy-preserving \ac{LLM} \cite{lee2023development}; Off-the-shelf 
LLMs \cite{manjunath2024towards};  Google’s Gemini-pro \cite{lammert2024large}; ClinicalCamel \cite{chang2024classifying}; ChatGLM-6B  \cite{wu2024diagnosis}.}
\label{figLLM}
\end{figure}

\ac{LLM}s, such as \ac{GPT}, \ac{BERT}, and their variants, are neural network models designed for processing and understanding human language at scale. {\acp{LLM}, for example, have also been explored in many medical domains, such as diabetes diagnosis, where \ac{LSTM} networks and chatbot-based medical support, such as the WizardLM-based DiabeTalk, have been investigated to improve classification accuracy and patient interaction \cite{rossi2024comparative}}.   \acp{LLM} are typically based on Transformer architectures, which rely on self-attention mechanisms to capture the relationships between words in a sequence. The self-attention computes a weighted sum of the values, with weights determined by the similarity between queries and keys. \ac{LLM}s are trained using large corpora through masked language modeling for \ac{BERT} or autoregressive prediction for \ac{GPT}, optimizing the following cross-entropy loss:

\begin{equation}
L = -\sum_{i} \log p(w_i | w_{<i})
\end{equation}
where $p(w_i \mid w_{<i})$ is the probability of word $w_i$ \cite{kheddar2024Transformers, nazi2024large, li2024cancerllm}. Among the types of advanced methods, GPT-4 Vision (GPT-4V) is a multi-modal \ac{LLM} developed by OpenAI, extending the capabilities of GPT-4 by enabling it to process and generate both text and visual data. This integration of vision and language allows GPT-4V to perform a wide range of tasks involving the understanding and reasoning of images in combination with text \cite{ferber2024context,naik2024applying}. Another example is the \ac{GLaM}, a sparse mixture-of-experts model designed by Google to provide efficient and scalable language modeling with significantly reduced computational cost compared to dense models like GPT-3. \ac{GLaM} achieves this by activating only a subset of its parameters during each forward pass, making it more efficient in terms of memory and processing power while maintaining high performance \cite{wu2024diagnosis}. Additionally, Llama-2-70b-chat is a version of the Llama-2 model, an open-source model developed by Meta (formerly Facebook) as part of their LLaMA series. Specifically, Llama-2-70b-chat is a dialogue-optimized model with 70 billion parameters, designed to generate natural and coherent text for conversation-like tasks. It builds on the success of the original LLaMA models and is optimized for chatbot and conversational applications \cite{change2024classifying, sivarajkumar2024automating}. Figure \ref{figLLM} provides an overview of the key \acp{LLM} explored and analyzed in the context of cancer detection.  Additionally, Table \ref{tableLLM} provides a summary of several state-of-the-art studies that employed \ac{LLM}s as the primary approach for cancer diagnosis.

\begin{figure}[tbph]
\begin{center}
\includegraphics[width=0.4\columnwidth]{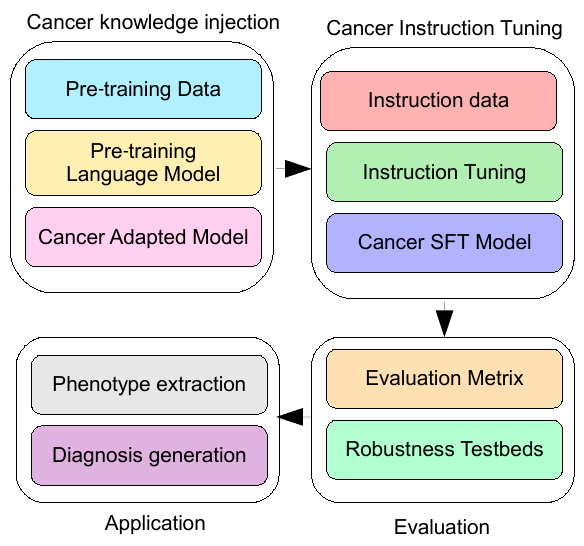}
\end{center}
\caption{An example of a \ac{LLM} in cancer domain \cite{li2024cancerllm}. The process of adapting an \ac{LLM} for the cancer domain begins with pre-training a general-purpose language model on diverse datasets. Cancer-specific knowledge is then injected using specialized datasets, publications, and terminologies. Instruction data is created for domain-specific tasks like phenotype extraction and diagnosis generation. The model undergoes instruction tuning and supervised fine-tuning (SFT) to specialize in cancer-related applications. The resulting Cancer-SFT model is evaluated using metrics and robustness tests to ensure reliability. Finally, the model is applied in real-world scenarios, aiding in phenotype extraction, diagnostic insights, and treatment recommendations, enhancing precision and efficiency in cancer research and clinical workflows.}
\label{fig20}
\end{figure}

The paper \cite{li2024cancerllm} introduced \textbf{CancerLLM}, a specialized \ac{LLM} designed for the cancer domain to address the need for more focused and efficient models in healthcare (see Figure \ref{fig20}). The authors highlighted that while existing \ac{LLM}s such as ClinicalCamel 70B and Llama3-OpenBioLLM 70B are large and computationally expensive, CancerLLM provided a more efficient alternative with 7 billion parameters and a Mistral-style architecture. They provided three datasets specifically designed for cancer phenotype extraction, cancer diagnosis generation, and cancer treatment plan generation. The model was pre-trained on over 2.6 million clinical notes and 500,000 pathology reports covering 17 cancer types and fine-tuned on these tasks. The findings demonstrated that CancerLLM significantly enhanced clinical AI systems, contributing to improved research and healthcare delivery in oncology. Wu et al. \cite{wu2024diagnosis} proposed a liver cancer diagnosis assistant that combined large and small models to improve diagnostic accuracy, particularly for less experienced doctors in primary healthcare settings. The framework addressed limitations such as inadequate understanding of medical images, insufficient consideration of liver blood vessels, and inaccurate medical information extraction. Small models were optimized for precise perception, with specialized methods for liver tumor and vessel segmentation to enhance information extraction. The large model employed Chain-of-Thought technology to mimic the reasoning process of experienced doctors and utilized retrieval-augmented generation for responses based on reliable domain knowledge. Results showed improved segmentation performance and higher evaluation scores from doctors for the assistant's responses compared to control methods, making it a valuable tool for liver cancer diagnosis. Sun \cite{sun2024large} focused on developing five-year survival prediction models for bladder cancer patients undergoing neoadjuvant chemotherapy and radical cystectomy. The study explored the feasibility of using \ac{LLM}s, such as Vicuna and Dolly, to extract clinical descriptors from reports, which were then incorporated into a nomogram model. It also examined the impact of combining these descriptors with radiomics and \ac{DL} features derived from CTU images, leveraging \acp{BPNN}. The models, developed and validated using data from 163 patients, included variations based on clinical descriptors (C), radiomics descriptors (R), \ac{DL} descriptors (D), and their combinations. The performance of models using \ac{LLM}-extracted descriptors was comparable to those using manually extracted descriptors, demonstrating the potential of \ac{LLM}s in clinical information extraction for survival prediction. Deng et al. \cite{deng2024genellm} proposed \textbf{GeneLLM}, a novel \ac{LLM} designed to directly interpret  \ac{cfRNA} sequences for cancer screening, bypassing traditional genome annotations. Unlike conventional methods reliant on bioinformatics tools to count known genes, GeneLLM identified \ac{cfRNA} from previously unknown genes, termed "dark matters," which serve as pseudo-biomarkers for cancer detection. This approach not only improved detection accuracy but also made the process more accessible and cost-effective. The findings highlighted GeneLLM's potential to revolutionize biomarker discovery and enhance understanding of intercellular communication through novel RNA molecules. Chang et al. \cite{chang2024classifying} investigated the use of open-source clinical \ac{LLM}s (Llama-2-70b-chat, ClinicalCamel-70B, and Med42-70B) for classifying cancer stages, specifically extracting pathologic tumor-node-metastasis (TNM) staging information from unstructured clinical reports. Unlike traditional \ac{NLP} approaches that require labor-intensive labeled datasets, this study demonstrated the feasibility of using \ac{LLM}s without labeled training data. The experiments compared \ac{LLM}s with a \ac{BERT}-based model fine-tuned on labeled data. Results showed that while \ac{LLM}s performed sub-optimally in Tumor (T) classification, they achieved comparable results in Metastasis (M) classification and outperformed in Node (N) classification when appropriate prompting strategies were applied. This demonstrated the potential of \ac{LLM}s for efficiently extracting staging information for oncology patients from real-world pathology reports. Google’s \textbf{Gemini} was also employed in the context of cancer detection by Lammert et al. \cite{lammert2024large}, who developed an advanced system named \textbf{MEREDITH} to address the limitations of \ac{LLM}s in precision oncology. MEREDITH integrated PubMed clinical studies, trial databases, and oncology guidelines with \ac{LLM}s to enhance the accuracy and relevance of treatment recommendations. The study evaluated the system using 10 fictional patient cases involving 7 tumor types and 59 molecular alterations, assessed by the \ac{MTB} at the Center of Personalized Medicine (ZPMTUM). MEREDITH employed a retrieval-augmented generation system enhanced by Google’s Gemini Pro, along with chain-of-thought prompting. Iterative improvements were made based on expert feedback from \ac{MTB}, comparing \ac{LLM}-generated recommendations to those annotated by clinical experts. The study concluded that incorporating expert feedback into \ac{LLM} training significantly enhanced the model's alignment with clinical reasoning, presenting a promising tool for clinical decision support in precision oncology. Lee et al. \cite{lee2023development} evaluated a privacy-preserving \textbf{FastChat-T5} for automating question answering from thyroid cancer surgical pathology reports. Eighty-four reports were analyzed by the \ac{LLM} and two independent reviewers across 12 medical questions related to staging and recurrence risk. FastChat-T5 significantly reduced task completion time while maintaining accuracy comparable to human reviewers.

Sivarajkumar et al. \cite{sivarajkumar2024automating} investigated the use of \textbf{LLAMA-2} to automate the extraction of treatment response information, particularly disease progression, from \ac{EHR}s. Their study involved 1,953 primary lung cancer patients from the UPMC Hillman Cancer Center, using approximately 113,000 clinical notes. Disease progression was defined based on RECIST guidelines, with a gold standard dataset created from ~50 manually annotated notes. The researchers fine-tuned LLAMA-2 and compared its performance to a traditional rule-based \ac{NLP} system. The results highlighted the potential of \ac{LLM}s to improve accuracy and scalability in treatment response assessments. Rajaganapathy et al. \cite{rajaganapathy2024synoptic} applied \textbf{LLAMA-2} to automatically generate synoptic reports from 7,774 cancer-related pathology reports annotated with reference synoptic reports from Mayo Clinic \ac{EHR}s. LLAMA-2 was fine-tuned to produce reports containing 22 unique data elements. The evaluation showed that fine-tuned \ac{LLM}s effectively automated synoptic report generation, demonstrating their potential to enhance efficiency and accuracy in clinical documentation.

Multiple versions of ChatGPT have been utilized in \ac{LLM}-based cancer detection and treatment. As examples,  Cao et al. \cite{cao2024large} evaluated the performance of \textbf{ChatGPT-3.5} (OpenAI), \textbf{Gemini} (Google), and \textbf{Bing} (Microsoft) in answering questions related to hepatocellular carcinoma (liver cancer). Their work assessed the accuracy, reliability, and readability of responses to 20 liver cancer-related questions concerning diagnosis and management. Six fellowship-trained physicians from three academic liver transplant centers evaluated the responses, categorizing them as accurate (all information true and relevant), inadequate (true but incomplete or irrelevant), or inaccurate (false information). Mean scores with standard deviations were recorded to determine overall accuracy and reliability, while readability was assessed using the Flesch Reading Ease Score and Flesch-Kincaid Grade Level. Choi et al. \cite{choi2023developing} explored the use of \textbf{GPT-4} for extracting clinical factors from breast cancer pathology and ultrasound reports. Using 2,931 patient records, they developed prompts for \ac{GPT} to extract data more time- and cost-efficiently than manual methods. The work demonstrated that \ac{LLM}s could significantly improve the efficiency of clinical data extraction. Matsuo et al. \cite{matsuo2024exploring} evaluated \textbf{GPT-3.5 Turbo} for \ac{TNM} classification in lung cancer radiology reports, focusing on multilingual capabilities in Japanese and English. The \ac{LLM} achieved its highest accuracy when provided full \ac{TNM} definitions in English, showcasing the relevance of \ac{LLM}s in radiological applications. Deng et al. \cite{deng2024evaluation} assessed the performance of \textbf{ChatGPT-3.5}, ChatGPT-4.0, and Claude 2 in breast cancer clinical scenarios. ChatGPT-4.0 provided the most accurate and relevant responses, particularly in psychosocial support and treatment decision-making, outperforming other models and highlighting its potential in clinical oncology. Ferber et al. \cite{ferber2024context} tested \textbf{GPT-4V} for cancer histopathology tasks using in-context learning. The \ac{LLM} matched or surpassed specialized neural networks in colorectal tissue classification, colon polyp subtyping, and breast tumor detection, demonstrating its potential for medical image analysis without domain-specific training. Sushil et al. \cite{sushil2024comparative} investigated zero-shot breast cancer pathology classification using \textbf{ChatGPT-4} and ChatGPT-3.5. ChatGPT-4 outperformed traditional models, particularly in tasks with label imbalance, highlighting \ac{LLM}s as a viable alternative for reducing data annotation burdens. Shiraishi et al. \cite{shiraishi2023preliminary} explored \textbf{\ac{LLM}s} (GPT-4, Bard, and BingAI) for diagnosing skin lesions. Using prompts with lesion images, they evaluated \ac{LLM}s in determining malignancy and specific diagnoses. The findings underscored their potential for assisting dermatology patients in seeking professional consultations. Similarly, Zhou et al. \cite{zhou2024performance} compared the performance of \ac{LLM}-powered chatbots in addressing colorectal cancer queries with oncology physicians. Eight chatbots, including \textbf{Claude 2.1}, ChatGPT-3.5, ChatGPT-4, and Doctor \ac{GPT}, were tested on 150 questions, alongside responses from nine oncology physicians. Each response was scored for consistency with clinical guidelines. Claude 2.1 outperformed residents, fellows, and attendings in accuracy, while Doctor \ac{GPT} also surpassed residents and fellows. Results indicated that \ac{LLM}s can provide more accurate information on colorectal cancer than human physicians in certain cases.

Additionally, BERT has also been explored in the same context. Karim et al. \cite{karim2023large} addressed challenges in utilizing biomedical data for cancer diagnosis and treatment. They introduced a domain-specific \ac{KG}, supported by the OncoNet Ontology (ONO), which integrated information from structured and unstructured sources. The \ac{KG}, enriched with \textbf{BioBERT} and \textbf{SciBERT}, enabled semantic reasoning for cancer biomarker discovery and interactive question answering, fine-tuned using \ac{LLM}s to incorporate the latest research.

\begin{center}
\scriptsize
\begin{longtable}{m{0.4cm} m{1.6cm} m{2.cm} m{6cm} m{1.5cm} m{4.2cm}}
\caption{Summary of \acp{LLM}-based approaches in cancer diagnosis.}
\label{tableLLM} \\
\hline
\textbf{Ref} & \textbf{Model(s) used} & \textbf{Dataset} & \textbf{Contribution} & \textbf{Best result (\%)} & \textbf{Limitation} \\
\hline
\endfirsthead

\hline
\textbf{Ref} & \textbf{Model(s) used} & \textbf{Dataset} & \textbf{Contribution} & \textbf{Best result (\%)} & \textbf{Limitation} \\
\hline
\endhead

\hline
\multicolumn{6}{r}{\textit{Continued on next page}} \\
\endfoot

\hline
\endlastfoot

\cite{li2024cancerllm} & CancerLLM & Cancer  & \ac{LLM} for cancer diagnosis and treatment recommendations. & F1: 95.52\newline{Rec: 95.52}& Depends on dataset quality and diversity. \\

\cite{wu2024diagnosis} & ChatGLM-6B & Liver cancer & \ac{LLM} for liver cancer diagnosis with domain-specific knowledge. & Acc: 99.50\newline{Rec: 92.90} & Constrained by integrated knowledge sources. \\

\cite{sivarajkumar2024automating} & LLAMA-2 & Lung cancer & \ac{LLM} for detecting lung cancer treatment progression. & F1: 92.00\newline{Spe: 91.00} & Variability in patient records. \\

\cite{sun2024large} & Vicuna and Dolly & Bladder cancer & \ac{LLM} for survival predictions from clinical reports. & AUC: 89.00 & Relies on clinical report quality. \\

\cite{chang2024classifying} & Llama-2-70b-chat & Clinical text & \ac{LLM} to classify cancer stages from clinical text. & Pre: 98.00\newline{Pre: 94.00} & Dependent on text data consistency. \\

\cite{deng2024genellm} & GeneLLM & cfRNA sequencing & \ac{LLM} for processing cfRNA reads for cancer screening. & Acc > 78.00\newline{AUC>90.00} & Requires extensive computational resources. \\

\cite{zhou2024performance} & Claude 2.1 & Colorectal cancer &  \ac{LLM} chatbots for responding to cancer queries. & Acc: 82.67 & Limited by general knowledge base. \\

\cite{rajaganapathy2024synoptic} & LLAMA-2 & Pathology reports & \ac{LLM} for summarizing cancer pathology reports. & F1: 94.00\newline{Acc: 81.00}  & Summaries may miss clinical nuances. \\

\cite{choi2023developing} & GPT & Breast cancer & \ac{LLM} for clinical information extraction from pathology reports. & Acc: 98.20 & Report formatting variability. \\

\cite{matsuo2024exploring} & GPT-3.5-turbo & Lung cancer & Multilingual \ac{LLM} for lung cancer TNM staging classification. & Acc: 74.00 & Variability in report language. \\
\cite{deng2024evaluation} & ChatGPT, and Claude2 & Breast cancer & Comparison of ChatGPT and Claude2 for breast cancer scenarios. & Fleiss’ kappa: 0.345 & Requires further validation. \\
\cite{karim2023large} & BioBERT, and SciBERT & Biomedical literature & \ac{LLM} and \ac{KG}s for cancer biomarker discovery. & Pre: 91.36\newline{Rec: 90.75} & Complex integration of \ac{KG}s. \\

\cite{putz2024exploring} & GPT-3.5 and GPT-4 & Radiation oncology & \ac{LLM} for decision support in radiation oncology. & Acc: 74.57 & Needs real-world clinical validation. \\

\hline
\end{longtable}
\end{center}

\section{Computational approaches}
\label{sec5}

Advanced \ac{DL} methods, including \ac{FL}, \ac{RL}, \ac{TL}, Transformers, and \acp{LLM}, have become essential for addressing computationally intensive tasks, enabled by advancements in algorithms, computational power, and access to vast datasets. These methods tackle challenges such as scalability, privacy, and efficiency, unlocking applications previously unattainable. Modern \ac{DL} architectures vary in layer connectivity and depth, with networks like ResNet evolving toward unprecedented scales, potentially reaching 1000 layers. Optimization strategies, such as \ac{SGD}, fine-tune parameters to boost accuracy, while \ac{FL} ensures privacy-preserving training across distributed datasets. Training large models like ResNet on datasets such as ImageNet, with over 14 million images, often requires tens of thousands of iterations and immense computational power, exceeding $10^{20}$ FLOPS. \ac{RL} optimizes dynamic decision-making tasks, \ac{TL} reduces labeled data requirements by leveraging pretrained models, and Transformers/\acp{LLM} excel in tasks requiring contextual and multi-modal analysis. Despite improvements in hardware like GPUs and memory bandwidth, rising \ac{DL} complexity amplifies computational demands, particularly in networks with varying computation-to-bandwidth ratios. These advanced \ac{DL} techniques are reshaping the landscape of \ac{AI}, delivering scalable, efficient, and privacy-conscious solutions across diverse fields \cite{darwish2020survey, cui2020discovering}.

\subsection{CPU}
Advanced deep learning (\ac{DL}) methods, including \ac{FL}, \ac{RL}, \ac{TL}, Transformers, and \acp{LLM}, often require a balance between computational power, memory, and adaptability, making the choice of hardware critical. \Ac{CPU} nodes are ideal for tasks that prioritize robust network connections, extensive memory, and storage capabilities, offering flexibility and ease of integration into diverse systems. While \acp{CPU} lack the raw computational throughput of specialized hardware like \acp{FPGA} and GPUs, they are well-suited for \ac{DL} techniques like \ac{FL}, where distributed training across nodes necessitates strong networking and memory resources. Similarly, CPUs can support \ac{RL} for real-time decision-making and \ac{TL} for adapting pretrained models to specific tasks without requiring immense computational loads. However, computational-intensive tasks such as fine-tuning Transformers and \acp{LLM} on large datasets often benefit from the parallel processing capabilities of GPUs or the customizability of \acp{FPGA}. Despite this, the adaptability and versatility of CPUs make them indispensable for many \ac{DL} applications where network and memory demands outweigh the need for pure processing speed \cite{biswas2019state, paszke2019pytorch}. 

\subsection{GPU}
Advanced \ac{DL} methods, including \ac{FL}, \ac{RL}, \ac{TL}, Transformers, and \acp{LLM}, often rely on GPUs for their ability to handle computationally intensive tasks with exceptional efficiency. GPUs excel in executing fundamental \ac{DL} operations such as \acp{AF}, matrix multiplication, and convolutions due to their highly parallel computing capabilities. Modern GPUs, equipped with integrated \ac{HBM} stacked memory, significantly boost bandwidth and optimize resource utilization, enabling dense linear algebra operations to outperform CPUs by 10–20 times. This makes GPUs indispensable for tasks like training \ac{DNN}s, fine-tuning Transformers, and optimizing \acp{LLM} for large-scale datasets. For \ac{FL}, their parallel processing power supports the aggregation of models from distributed nodes, while in \ac{RL}, GPUs enable real-time learning and decision-making. \ac{TL} also benefits from GPUs' speed in adapting pretrained models to new tasks, reducing training time. With up to sixty-four computational units, each featuring multiple \ac{SIMD} engines, GPUs achieve peak performances of 25 TFLOPS (fp16) and 10 TFLOPS (fp32), ensuring efficiency in both training and inference. The ability to combine addition and multiplication functions for vector operations and utilize inner product instructions further enhances GPU performance, making them a versatile and essential tool for advanced \ac{DL} applications \cite{gianniti2018performance, hossain2019deep}.

\subsection{\ac{FPGA}}
Advanced \ac{DL} methods, including \ac{FL}, \ac{RL}, \ac{TL}, Transformers, and \acp{LLM}, benefit significantly from \ac{FPGA} technology, particularly in inference acceleration tasks where efficiency and customization are paramount. Unlike GPUs, which offer high floating-point performance, \ac{FPGA}s excel in minimizing unnecessary functions and overhead, delivering low-latency and energy-efficient solutions. Their ability to dynamically reconfigure array characteristics in real-time and support custom designs makes them highly adaptable for specialized \ac{DL} operations. \ac{FPGA}s achieve superior performance per watt through strategies like implementing custom high-performance hardware, pruned networks, and reduced arithmetic precision, often outperforming GPUs and CPUs in these areas. For tasks such as \ac{CNN} inference, \ac{FPGA}s deliver over 15 TOPs in peak performance, reaching more than 80\% efficiency with 8-bit accuracy. Additionally, pruning techniques, particularly in \ac{LSTM} models, enable significant model size reductions up to 20 times facilitating optimal deployment in resource-constrained environments. Recent advancements in lowering arithmetic precision to 8-bit fixed-point or custom floating-point formats further enhance \ac{FPGA} performance, making them ideal for applications requiring highly efficient and tailored solutions, such as in federated learning or \acp{LLM} deployed at edge devices \cite{wang2019overview, namburu2022fpga}.

\section{Research challenges and future directions}
\label{sec6}

\subsection{Challenges}
\ac{DL} is a widely used and highly effective method for training \ac{AI} systems with large datasets, but it also presents significant challenges that require innovative solutions. These challenges include issues like data privacy, computational costs, limited labeled data, and model interpretability. Advanced \ac{DL} techniques offer promising alternatives to address these difficulties. \ac{FL} tackles privacy concerns by enabling collaborative model training across decentralized datasets without sharing sensitive data. \Ac{RL} optimizes decision-making in complex, dynamic environments, making it suitable for tasks like personalized treatment planning. \ac{TL} mitigates the need for extensive labeled data by leveraging pretrained models to adapt to specific tasks with smaller datasets. Additionally, Transformers and \acp{LLM} have revolutionized \ac{AI} by improving the scalability, accuracy, and interpretability of models in tasks like \Ac{NLP} understanding and medical data analysis. Adopting these advanced approaches not only addresses the limitations of traditional \ac{DL} methods but also enhances the effectiveness and deployment of \ac{AI} systems across diverse applications.

\setlength{\leftmargini}{0.4cm}
\begin{enumerate}[label={(\alph*)}]
\item\textbf{Training data: } 
In \ac{DL}, training data is critical for robust model performance, but data insufficiency remains a challenge. \ac{FL} addresses this by training models across decentralized data without sharing sensitive information; however, ensuring data privacy, managing communication overhead, and addressing heterogeneity across devices remain key challenges. \ac{TL} leverages knowledge from related tasks to improve performance on limited datasets, but it often struggles with negative transfer when source and target tasks are not closely related. Transformers, pivotal in modern \ac{AI}, excel in processing sequential data but require extensive training data, computational resources, and are prone to overfitting on smaller datasets. \ac{RL} enhances model adaptability through trial-and-error, yet suffers from sample inefficiency and exploration challenges. \ac{LLM}s face hurdles like data bias, resource intensity, and ensuring generalization across diverse tasks. Overcoming these challenges necessitates innovative strategies in data simulation, optimization, and efficient architecture design.

\item\textbf{Imbalanced data: } 
Imbalanced data, where one class significantly outweighs another, poses challenges for training \ac{DL} models, particularly in maintaining predictive accuracy across all classes. \ac{FL} can exacerbate this issue when data imbalance occurs across distributed nodes, making global model updates less effective and complicating convergence. \ac{TL} offers a solution by using pre-trained models from balanced datasets, which can be fine-tuned to improve performance on minority classes. However, the relevance of source domains remains critical. \ac{RL}, while adaptive, struggles with reward sparsity and policy optimization in imbalanced datasets, further amplifying bias toward majority classes. Transformers and \ac{LLM}s, although powerful, require balanced data for effective learning and risk significant bias amplification in imbalanced settings, particularly when fine-tuned without addressing class disparities.

\item\textbf{Interpretability of data:  }
Interpretability of data is a crucial challenge in \ac{DL} as models grow in complexity. \ac{FL}, while preserving data privacy, limits transparency due to distributed data storage, making it harder to understand how local data influences global models. \ac{TL}, though effective, raises questions about the relevance and transferability of learned features, particularly when adapting to tasks with differing data distributions. \ac{RL} models face interpretability issues in understanding decision-making processes, especially when reward structures are complex or sparsely distributed. Transformers and \ac{LLM}s present significant challenges due to their vast number of parameters and opaque attention mechanisms, making it difficult to discern why specific predictions are made. The added complexity of federated, reinforcement, and transfer learning setups, combined with the expansive architectures of transformers and \ac{LLM}s, exacerbates these interpretability concerns. 

\item\textbf{Uncertainty scaling: } 
Uncertainty scaling poses significant challenges in \ac{FL}, \ac{TL}, \ac{RL}, Transformers, and \ac{LLM}s, adding complexities that hinder reliable deployment. Confidence scores, crucial for robust predictions, are often poorly calibrated across these domains. \ac{FL} and \ac{TL} face added difficulties due to heterogeneous data distributions across clients or tasks, intensifying uncertainty quantification challenges. In \ac{RL}, the reliance on cumulative rewards makes the confidence in exploration-exploitation trade-offs unreliable, complicating policy optimization. Transformers and \ac{LLM}s, while successful, frequently produce overconfident predictions due to softmax-based probabilities, particularly when generalizing to unseen or out-of-distribution data. These challenges are further amplified in critical fields such as healthcare and autonomous systems, where poorly scaled uncertainties can result in life-threatening decisions or system failures.

\item\textbf{Overfitting: } 
Data overfitting is a prevalent issue in \ac{DL} models during training, caused by the complex correlations among a large number of parameters. This overfitting reduces the model's effectiveness on new, unseen data and is a challenge not limited to any single field but common across various tasks, including \ac{FL}, \ac{TL}, \ac{RL}, Transformers, and \ac{LLM}s. In \ac{FL}, the decentralized nature and heterogeneous client data exacerbate overfitting risks. Similarly, \ac{TL} struggles with overfitting when the source and target domains have limited similarity. In \ac{RL}, the sparse reward structure and high-dimensional state-action spaces often lead to policies overfitting the training environment. Transformers and \ac{LLM}s face overfitting challenges due to their immense parameter spaces and reliance on pretraining finetuning paradigms, particularly when adapting to domain-specific tasks.

\item\textbf{Unified assessment: }
In numerous studies, there is a notable absence of comprehensive details concerning the technical facets of the experiments conducted. The choice of measurement indicators and baseline methods often seems random, leading to a non-standardized evaluation process. Researchers primarily focus on metrics such as accuracy, sensitivity, or specificity of \ac{DL} networks. However, Unified Assessment poses additional challenges across advanced methodologies. For \ac{FL}, standardizing evaluations is difficult due to data heterogeneity, privacy concerns, and inconsistent benchmarks for data. \ac{RL} struggles with variability in reward structures, task complexity, and cross-environment benchmarking. \ac{TL} lacks clarity in reporting domain adaptation success and the influence of pre-trained models. Transformers face challenges in scalability, computational overhead, and variations in results across implementations. \ac{LLM}s complicate assessment with issues of fine-tuning effectiveness, interpretability, and resource-intensive evaluations. A standardized framework addressing these challenges is vital for cancer detection.

\item\textbf{High computational requirements: } 
Transformers, particularly \ac{ViT}s, are computationally intensive and require significant GPU/TPU resources for training and inference. This makes it difficult to deploy these models in real-time clinical settings, especially in resource-constrained environments. Similarly, \ac{FL} amplifies computational demands due to frequent communication overheads between clients and servers, as well as the need to handle non-IID data across devices. \ac{RL} is hindered by high computational costs associated with iterative policy optimization and simulation-based training, particularly in complex environments. \ac{TL} also faces challenges when fine-tuning large pre-trained models on domain-specific datasets, which often requires extensive computational resources. Transformers exacerbate the problem with their quadratic complexity concerning input sequence length, making them unsuitable for low-resource scenarios. \ac{LLM}s demand immense memory and processing power during both training and inference, further complicating their integration into clinical workflows. Addressing these challenges is critical for scalable deployment in healthcare.

\end{enumerate}

\subsection{Future directions}

As advanced \ac{DL} methods represent a significant leap in modern \ac{AI} techniques, current and future research focuses heavily on their application in cancer diagnosis. Despite extensive research efforts in this field, the use of these methods is still in its early stages, offering vast potential for future development. The complexity of cancer diagnosis arises not only from the variety of types and detection methods but also from handling complex medical data, including radiological images and genetic analyses. To overcome these challenges, \ac{AI} systems require advanced capabilities in analysis, decision-making, tool utilization, and memory management. Consequently, researchers have identified several key aspects that need to be addressed in systems based on advanced \ac{DL} methods for cancer diagnosis.

\setlength{\leftmargini}{0.4cm}
\begin{enumerate}[label={(\alph*)}]
\item \textbf{\ac{TL}: } \ac{TL} will continue to advance by addressing challenges in negative transfer, where knowledge from a pretrained model adversely impacts target tasks. Research is focusing on improving domain adaptation techniques to ensure efficient knowledge transfer across diverse datasets. TL in multitask learning is another promising area, enabling simultaneous optimization of multiple related tasks. The development of scalable and lightweight TL models for deployment on edge devices is also a key focus. Additionally, incorporating TL into continual learning frameworks will allow models to adapt incrementally to new tasks, ensuring long-term usability without catastrophic forgetting.

\item \textbf{\ac{FL}:} \ac{FL} is poised to expand in privacy-preserving applications through advancements in differential privacy and homomorphic encryption, ensuring secure model training on sensitive data like healthcare records. A major future direction includes adaptive FL frameworks that can handle \ac{non-IID} data and address client variability. FL in edge computing is also expected to grow, enabling decentralized AI systems on low-power devices for real-time applications. Additionally, the integration of FL with emerging technologies like blockchain for secure data sharing and quantum computing for accelerated training promises to redefine its scalability and efficiency in large-scale deployments.

\item \textbf{\ac{RL}: } \ac{RL} is moving toward more sample-efficient algorithms to reduce training time, making it suitable for real-time applications such as computer network \cite{gueriani2023deep},   and healthcare. A key future direction is the integration of RL with deep learning architectures, like Transformers, for solving complex, high-dimensional problems. Multi-agent RL is also gaining traction, enabling collaboration and competition among agents in dynamic environments. Furthermore, RL is anticipated to play a pivotal role in optimizing autonomous systems, such as self-driving cars and energy management. Bridging RL with unsupervised and semi-supervised learning will enhance its adaptability in scenarios with limited labeled data.

\item \textbf{ Transformers and LLM: } The future of Transformers and \acp{LLM} is closely tied to advancements in \ac{RAG} \cite{lewis2020retrieval}, which combines LLMs with external knowledge retrieval systems. \ac{RAG} enables LLMs to access and utilize vast databases, reducing the dependency on model size by dynamically fetching relevant information. Future research will focus on optimizing retrieval mechanisms for domain-specific tasks, improving both speed and accuracy. Additionally, integrating \ac{RAG} with real-time knowledge updates ensures that models remain current and relevant without extensive retraining. The development of privacy-preserving \ac{RAG} systems, particularly for sensitive domains like healthcare, will be crucial. \ac{RAG}’s ability to enhance interpretability by linking model outputs to specific knowledge sources also addresses transparency and accountability challenges, paving the way for its broader adoption in fields like personalized medicine, legal analysis, and scientific research.

\item \textbf{Hybrid approaches:}
Future hybrid methods in advanced deep learning promise to revolutionize AI by combining the strengths of techniques such as \ac{FL}, \ac{RL}, \Ac{TL}, Transformers, and \ac{RAG}. For instance, FL and TL can work together to deploy privacy-preserving, personalized models across decentralized systems, especially in healthcare and edge computing. RL integrated with Transformers can enhance sequential decision-making in applications like robotics and conversational agents, while TL combined with \ac{RAG} enables domain-specific LLMs to incorporate real-time knowledge dynamically. Additionally, FL and RL hybrids can optimize resource allocation in decentralized networks, ensuring privacy and efficiency. Transformers coupled with multimodal learning techniques will further enhance cross-modal understanding for applications in autonomous vehicles and medical diagnostics. By strategically integrating these methods, hybrid approaches will create scalable, efficient, and context-aware AI systems, addressing real-world challenges in personalized healthcare, smart cities, and knowledge-intensive fields like law and medicine.

\item \textbf{Blockchain for data security:} In the context of cancer detection, the transition from traditional models to advanced \ac{DL} methods, such as \ac{FL}, \ac{RL}, \ac{TL}, Transformers, \acp{LLM}, is driven by the need for improved prediction accuracy, scalability, and real-time data processing of heterogeneous datasets. Traditional models, which relied on centralized datasets, posed significant privacy and security risks, making them unsuitable for handling sensitive patient data. Advanced methods like \ac{FL}, integrated with blockchain, enable decentralized model training while ensuring secure, tamper-proof data sharing and patient confidentiality. \ac{RL} enhances decision-making in dynamic scenarios, such as personalized treatment plans and diagnostic workflows, with blockchain providing secure logging of decision-making processes and auditability. \ac{TL}, when paired with blockchain, enables efficient adaptation of pretrained models to cancer datasets, ensuring data provenance and reducing risks of tampered inputs. Transformers and \acp{LLM} deliver exceptional accuracy and contextual understanding for cancer diagnostics, with blockchain enhancing transparency and traceability in multimodal data integration and model updates. Together, these technologies ensure secure, distributed, and precise systems tailored to the complex challenges of cancer detection.

\item \textbf{Interpretability and scalability:} The application of advanced \ac{DL} techniques, such as \ac{FL}, \ac{RL}, \ac{TL}, Transformers, and \acp{LLM}, holds transformative potential for cancer detection. These methods can improve interpretability by providing clearer insights into patterns and anomalies in medical images or genetic data, making diagnostic results more actionable for physicians. \acp{LLM} and Transformers excel in multimodal data analysis, enabling better differentiation between malignant and normal tissues, while \ac{FL} ensures privacy-preserving collaboration across decentralized datasets. Scalability remains a critical future direction, particularly in integrating these techniques with large, heterogeneous datasets to support diverse cancer types and clinical scenarios. Advanced architectures, such as lightweight \acp{LLM} and optimized \ac{FL} frameworks, can address the challenges of computational efficiency and resource constraints. Future research should prioritize improving interpretability to enhance physician trust and understanding, alongside developing scalable models that deliver precise and personalized cancer diagnosis at population-wide levels.

\end{enumerate}

\section{Conclusion}
\label{sec7}

In this comprehensive review, we have explored the transformative potential of advanced \ac{DL} methodologies, such as \ac{RL}, \ac{FL}, \ac{TL}, Transformers, and \acp{LLM}, in enhancing cancer detection and diagnosis. These techniques represent significant progress in overcoming longstanding challenges in medical applications, such as data privacy, model scalability, and the scarcity of labeled datasets. \ac{RL} has shown promise in optimizing diagnostic and treatment pathways, offering dynamic learning capabilities that adapt to real-time clinical scenarios. By leveraging reward-based mechanisms, \ac{RL} has demonstrated its utility in tasks ranging from tumor localization to treatment decision-making. \ac{FL}, on the other hand, provides a privacy-preserving framework for collaborative model training across decentralized datasets, enabling the inclusion of diverse data while addressing privacy concerns. Its application in cancer diagnosis highlights the growing trend toward ethical \ac{AI} deployment in healthcare. \ac{TL} continues to mitigate the challenge of limited labeled data by enabling the adaptation of pretrained models to specific cancer datasets. This approach not only enhances model performance but also significantly reduces the computational costs associated with training from scratch. The introduction of Transformer-based models and \acp{LLM} has further revolutionized the field by enabling complex, context-aware analysis of medical data, including imaging, genomics, and unstructured text. Their ability to process vast datasets and deliver contextually rich predictions exemplifies the potential of cutting-edge \ac{NLP} and vision models in oncology.

The review also addresses critical technical challenges, including data imbalance and generalizability issues, by discussing advanced solutions like data augmentation, ensemble methods, and cost-sensitive learning. These approaches ensure that AI models maintain robustness across diverse cancer types and clinical conditions. Furthermore, the incorporation of evaluation metrics and datasets provides a robust foundation for assessing model performance and identifying areas for future improvement. Despite these advancements, several challenges remain. Issues such as the interpretability of complex DL models, the high computational demands of training, and the ethical implications of \ac{AI} deployment in clinical settings must be addressed to fully realize the potential of these technologies. Future research directions should focus on integrating \ac{DL} techniques with emerging technologies, such as quantum computing and edge \ac{AI}, to further enhance efficiency and scalability. In conclusion, the integration of advanced \ac{DL} techniques into cancer detection and diagnosis is reshaping the landscape of healthcare. By addressing both technical and ethical challenges, these methodologies hold the promise of improving diagnostic accuracy, personalizing treatment plans, and ultimately enhancing patient outcomes health condition. This review serves as a foundational resource for researchers and practitioners, highlighting the current state, challenges, and future directions in applying advanced DL in oncology.

\bibliographystyle{elsarticle-num}
\bibliography{refs.bib}
\end{document}

%% file: acro_list.tex
\begin{acronym}[AWGN]  
\acro{AI}{artificial intelligence}
\acro{ANN}{artificial neural networks} 
\acro{AUC}{area under the curve}
\acro{ATM}{attention transformer mechanisms}

\acro{BBQ}{bayesian binning into quantiles}
\acro{BDN}{bayesian deep network}
\acro{BraTS}{brain tumor segmentation}
\acro{BERT}{bidirectional encoder representations from transformers}
\acro{ROI}{region of interest}

\acro{CAD}{computer-aided diagnosis}
\acro{CNN}{convolutional neural networks}
\acro{CT}{computed tomography}
\acro{CBAM}{convolutional block attention module} 
\acro{CF}{catastrophic forgetting}
\acro{CBIR}{content-based image retrieval}
\acro{CE}{contrast-enhanced}
\acro{CvT}{convolutional vision transformer} 
\acro{CONAF}{convolutional neural network with attention feedback}

\acro{DL}{deep learning}
\acro{DNN}{deep neural networks}
\acro{DBN}{deep belief networks}
\acro{DBM}{deep boltzmann machines}
\acro{DCE}{dynamic contrast-enhanced}
\acro{DeiT}{data-efficient image transformers}
\acro{DWT}{discrete wavelet transform}
\acro{DQN}{deep Q-network}
\acro{DQL}{deep Q-learning}
\acro{DDPG}{deep deterministic policy gradient}
\acro{BPNN}{back-propagation neural network}
\acro{CCT}{compact convolutional Transformer}
\acro{GLaM}{generalist language model}
\acro{RAG}{retrieval-augmented generation}
\acro{CPU}{central processing unit}
\acro{EHR}{electronic health records} 
\acro{KG}{knowledge graph}
\acro{cfRNA}{plasma cell-free RNA}
\acro{EGP}{exploding gradient problem} 
\acro{EM}{ensemble model}
\acro{DBN}{deep belief network}
\acro{XAI}{explainable AI}
\acro{EL}{ensemble learning}
\acro{FC}{fully connected}
\acro{FPGA}{field-programmable gate array}
\acro{FL}{federated learning}
\acro{FNN}{feedforward neural networks}
\acro{HFL}{horizontal federated learning}
\acro{VFL}{vertical federated learning}
\acro{FSSL}{federated semi-supervised learning}
\acro{FedDis}{federated disentanglement}
\acro{FedAvg}{federated averaging}
\acro{FFN}{feed-forward network}

\acro{LF}{loss function}

\acro{BN}{batch normalization}

\acro{JUMC}{jimma university medical center}
\acro{AF}{activation function}
\acro{GAN}{generative adversarial networks} 
\acro{GRU}{gated recurrent units}
\acro{GAP}{global average pooling}
\acro{GGG}{gleason grade group}
\acro{GCPSO}{guaranteed convergence particle swarm optimization}
\acro{GRN}{gated residual network}
\acro{GPT-4V}{generative pretrained transformer 4 with vision}
\acro{GPT}{generative pretrained transformer}
\acro{GBDTMO}{gradient boosting decision tree-based mayfly optimization}

\acro{HRNet}{high-resolution network}
\acro{HBM}{high bandwidth memory}
\acro{HAM10000}{human against machine} 
\acro{ID}{imbalanced data}
\acro{IPD}{interpretability of data}
\acro{IoT}{Internet of Things}
\acro{IDC}{invasive ductal carcinoma}
\acro{IoMT}{internet of medical things}
\acro{ISIC}{international skin imaging collaboration}

\acro{LSTM}{long short-term memory} 
\acro{LCDEiT}{linear complexity data-efficient image transformer}
\acro{LLM}{large language model}
\acro{LUPI}{learning using privileged information}

\acro{KNN}{K-nearest neighbor}

\acro{MHSA}{multi-head self-attention} 

\acro{MMD}{maximum mean discrepancy}
\acro{ML}{machine learning}
\acro{MRI}{magnetic resonance imaging}
\acro{MAC}{multiply-accumulate operations}
\acro{MC}{model compression}
\acro{MDS}{medical data sharing}
\acro{MIL}{multiple instance learning}
\acro{MLP}{multilayer perceptron}
\acro{MSHT}{multi-stage hybrid transformer}
\acro{MIST}{multiple instance learning network}
\acro{MEREDITH}{medical evidence retrieval and data integration for tailored healthcare}
\acro{MTB}{Molecular Tumor Board}
\acro{AC}{actor-critic}
\acro{MCMOCL}{multi-cancer multi-omics clinical dataset laboratories}
\acro{MaOEA-IS}{stands for Many-Objective Evolutionary Algorithm based on Integrated Strategy}
\acro{MAD}{mean absolute differences}
\acro{LDA}{latent Dirichlet allocation}
\acro{NLP}{natural language processing}
\acro{NMF}{non-negative matrix factorization}

\acro{SSL}{semi-supervised learning}
\acro{SL}{supervised learning}
\acro{SVM}{support vector machine}
\acro{SGD}{stochastic gradient descent}
\acro{SIMD}{single instruction, multiple data}
\acro{SepViT}{separable vision transformer}
\acro{adam}{adaptive moment estimation}
\acro{OF}{Overfitting}
\acro{SKT}{spatiotemporal knowledge teacher student}
\acro{AE}{auto-encoder}
\acro{PD}{private data}
\acro{PLA}{privacy-embedded lightweight and efficient automated}
\acro{PPO}{proximal policy optimization}
\acro{CL}{contrastive learning}
\acro{PET}{positron emission tomography}
\acro{PiT}{pooling-based vision transformer} 
\acro{FedAvg}{federated Averaging}
\acro{FedProx}{federated proximal}
\acro{FedPDA}{federated post-deployment adaptation}
\acro{CMT}{combined topic models}
\acro{non-IID}{non-independent and identically distributed}
\acro{RNN}{recurrent neural networks}
\acro{RL}{reinforcement learning}
\acro{RvNN}{recursive neural networks} 
\acro{ReLU}{Rectified Linear Unit}
\acro{RNTN}{recursive neural tensor networks} 
\acro{RAN}{residual attention network} 
\acro{RMTF-Net}{residual mix transformer fusion network}
\acro{RBF}{radial basis function}
\acro{RAMAF}{recurrent attention model with annotation feedback}
\acro{RANet}{recurrent attention network} 
\acro{VGG}{visual geometry group}
\acro{VGP}{vanishing gradient problem}
\acro{ViT}{vision transformers}
\acro{DPG}{deterministic policy gradient}
\acro{TreeRNN}{tree-structured recursive neural networks}
\acro{DCRN}{decoupled classification reinforcement network}
\acro{MDP}{Markov decision process}
\acro{IRL}{inverse reinforcement learning}
\acro{RTL}{reinforcement transfer learning}
\acro{A3C}{asynchronous advantage actor-critic}
\acro{SAC}{Soft actor-critic}
\acro{TD}{training data}
\acro{TL}{transfer learning}
\acro{T2T-ViT}{tokens-to-token vision transformers}
\acro{TNM}{Tumor, node, and metastasis}

\acro{CAE}{convolutional autoencoder}

\acro{EGFR}{epidermal growth factor receptor}

\acro{VAE}{variational autoencoders} 

\acro{USL}{unsupervised learning} 
\acro{UCS}{uncertainty scaling}
\acro{UDS}{underspecification}
\acro{UA}{unified assessment}
\acro{UCSF}{university of california, san francisco}

\acro{OF}{overfitting}

\acro{YOLOv8}{you only look once version 8}

\end{acronym}